\documentclass[twoside,11pt]{article}

\usepackage{jmlr2e}

\usepackage{amsmath}
\usepackage{algorithmicx,algpseudocode}
\usepackage[linesnumbered,ruled]{algorithm2e}
\usepackage{bm}
\usepackage[usenames]{color}
\usepackage{enumerate}
\usepackage{float}
\usepackage{mathrsfs}
\usepackage[caption=false]{subfig}

\newtheorem{assumption}{Assumption}

\def\argmin{\mathop{\rm arg\, min\,}}

\newcommand{\pa}{\textnormal{\textsc{pa}}}
\newcommand{\an}{\textnormal{\textsc{an}}}
\newcommand{\inter}{\textnormal{\textsc{in}}}
\newcommand{\df}{\textnormal{\textsc{d}}}
\newcommand{\diag}{\text{Diag}}

\newcommand{\pto}{\overset{p}{\longrightarrow}}
\newcommand{\dto}{\overset{d}{\longrightarrow}}
\newcommand{\E}{\operatorname{\mathbb E}}
\newcommand{\PP}{\operatorname{\mathbb P}}
\newcommand{\Var}{\operatorname{Var}}
\newcommand{\Cov}{\operatorname{Cov}}
\newcommand{\Corr}{\operatorname{Corr}}
\newcommand{\I}{\operatorname{I}}

\newcommand{\bs}[1]{\boldsymbol{\mathbf{#1}}}  

\SetKwInOut{Parameter}{Parameters}

\usepackage{lastpage}
\jmlrheading{24}{2023}{1-\pageref{LastPage}}{7/21; Revised
1/23}{2/23}{21-0855}{Chunlin Li, Xiaotong Shen, and Wei Pan}

\ShortHeadings{Inference for a Large Directed Acyclic Graph with Unspecified Interventions}{Li, Shen, and Pan}
\firstpageno{1}

\begin{document}

\title{Inference for a Large Directed Acyclic Graph with Unspecified Interventions}

\author{\name Chunlin Li\thanks{To whom correspondence should be addressed.} \email li000007@umn.edu \\
       \addr School of Statistics,\\
       University of Minnesota,\\
       Minneapolis, MN 55455, USA
       \AND
       \name Xiaotong Shen \email xshen@umn.edu\\
       \addr School of Statistics,\\
       University of Minnesota,\\
       Minneapolis, MN 55455, USA
       \AND
       \name Wei Pan \email panxx014@umn.edu \\
       \addr Division of Biostatistics,\\
       University of Minnesota,\\
       Minneapolis, MN 55455, USA}

\editor{Pradeep Ravikumar}

\maketitle

\begin{abstract} 
Statistical inference of directed relations given some unspecified interventions (i.e., the intervention targets are unknown) is challenging. In this article, we test hypothesized directed relations with unspecified interventions. First, we derive conditions to yield an identifiable model. Unlike classical inference, testing directed relations requires identifying the ancestors and relevant interventions of hypothesis-specific primary variables. To this end, we propose a peeling algorithm based on nodewise regressions to establish a topological order of primary variables. Moreover, we prove that the peeling algorithm yields a consistent estimator in low-order polynomial time. Second, we propose a likelihood ratio test integrated with a data perturbation scheme to account for the uncertainty of identifying ancestors and interventions. Also, we show that the distribution of a data perturbation test statistic converges to the target distribution. Numerical examples demonstrate the utility and effectiveness of the proposed methods, including an application to infer gene regulatory networks. The R implementation is available at \url{https://github.com/chunlinli/intdag}.
\end{abstract}


\section{Introduction}

Directed relations are essential to explaining pairwise dependencies among multiple interacting units. 
In gene network analysis, regulatory gene-to-gene relations are a focus of biological investigation \citep{sachs2005causal}, while in a human brain network, scientists investigate causal influences among regions of interest to understand how the brain functions \citep{liu2017effective}. 
In such a situation, a Gaussian directed acyclic graph (DAG) is commonly employed to describe the directed relations; however, inferring the directed effects without other information is generally impossible because a Gaussian DAG often lacks model identifiability \citep{vandegeer2013ell}. 
Hence, external interventions are introduced to treat a non-identifiable situation \citep{heinze2018causal}. For instance, the genetic variants such as single-nucleotide polymorphisms (SNPs) can be, and indeed are increasingly, treated as external interventions to infer inter-trait causal relations in a quantitative trait network \citep{brown2020phenome} and gene interactions in a gene regulatory network \citep{teumer2018common,molstad2021covariance}. 
In neuroimaging analysis, scientists use randomized experimental stimuli as interventions 
to identify causal relations in a functional brain network \citep{grosse2016identification,bergmann2021inferring}.
However, the interventions in these studies often have unknown targets and off-target effects \citep{jackson2003expression,eaton2007exact}. 
Consequently, inferring directed relations while identifying useful interventions for inference is critical.
This paper focuses on the simultaneous inference of directed relations subject to unspecified interventions (i.e., the intervention targets are unknown).

In a DAG model, the research has been centered on the reconstruction of directed
relations in observational and interventional studies \citep{vandegeer2013ell,oates2016estimating,zheng2018dags,yuan2019constrained,li2023nonlinear}; see \citet{heinze2018causal} for a review. 
For uncertainty quantification, Bayesian methods \citep{friedman2003being,luo2011bayesian,viinikka2020towards} have been popular. 
Yet, statistical inference remains under-studied, especially for interventional models in high dimensions \citep{peters2016causal,rothenhausler2019causal}.
Recently, for observational data, \citet{jankova2018inference} propose a debiased test of a single directed relation, and \citet{li2020likelihood} derive a constrained likelihood ratio test for multiple directed relations.

Despite progress, challenges remain. First, inferring directed relations requires identifying a certain DAG topological order \citep{vandegeer2013ell}, while the identifiability in a Gaussian DAG with unspecified interventions remains under-explored. Second, the inferential results should agree with the acyclicity requirement for a DAG. As a result, degenerate and intractable situations can occur, making inference greatly different from the classical ones. Third, likelihood-based methods for learning the DAG topological order often use permutation search \citep{vandegeer2013ell} or continuous optimization subject to the acyclicity constraint \citep{zheng2018dags,yuan2019constrained}, 
where a theoretical guarantee of the actual estimate (instead of the global optimum) has not been established for these approaches. Recently, an important line of work \citep{ghoshal2018learning,rajendran2021structure,rolland2022score} has focused on order-based algorithms with computational and statistical guarantees. However, in Gaussian DAGs, existing methods often rely on some error scale assumptions \citep{peters2014identifiability}, which is sensitive to variable scaling like the common practice of standardizing variables. This drawback could limit their applications, especially in causal inference, as causal relations are typically invariant to scaling.

To address the above issues, we develop structure learning and inference methods for a Gaussian DAG with unspecified additive interventions. Unlike the existing approach treating structure discovery and subsequent inference separately, our proposal integrates DAG structure learning and testing of directed relations, accounting for the uncertainty of structure learning for inference. With suitable interventions called instrumental variables (IVs), the proposed approach removes the restrictive error scale assumptions and delivers creditable outcomes with theoretical guarantees in low-order polynomial time. This indicates IVs, a well-known tool in causal inference \citep{angrist1996identification}, can play important roles in structure learning even if some interventions do not meet the IV criteria. Our contributions are summarized as follows. 
\begin{itemize}
 \item For modeling, we establish the identifiability conditions for a Gaussian DAG with unspecified interventions. In particular, the conditions allow interventions on more than one target, which is suitable for multivariate causal analysis \citep{murray2006avoiding}. 

 \item For methodology, we develop likelihood ratio tests for directed edges and pathways in a super-graph of the true DAG, called the ancestral relation graph (ARG), where the ARG is formed by ancestral relations and candidate interventional relations, offering the topological order for inference. We reconstruct the ARG by the peeling algorithm, which automatically meets the acyclicity requirement. 
 On this basis, we introduce the concepts of nondegeneracy and regularity to characterize the behavior of hypothesis testing under a DAG model.
 By integrating structure learning with inference, we account for the uncertainty of ARG estimation for the proposed tests via a novel data perturbation (DP) scheme, which effectively controls the type-I error while enjoying high statistical power. 

  \item For theory, we prove that the proposed peeling algorithm based on nodewise regressions yields consistent results in $O(p \times \log\kappa_{\max}^\circ \times (q^3 + nq^2) )$ operations almost surely, where $p,q$ are the numbers of primary and intervention variables, $n$ is the sample size, and $\kappa_{\max}^\circ$ is the sparsity. Then we justify the proposed DP inference method by establishing the convergence of the DP likelihood ratio to the target distribution and desired power properties.
  
  \item The numerical studies and real data analysis demonstrate the utility and effectiveness of the proposed methods. The implementation of the proposed tests and structure learning method is available at \url{https://github.com/chunlinli/intdag}.

\end{itemize}

{The rest of the article is structured as follows.
Section \ref{section:model} establishes model identifiability and states two inference problems of interest.
Section \ref{section:method} develops the proposed methods for structure learning and statistical inference.
Section \ref{section:theory} presents statistical theory to justify the proposed methods.
Section \ref{section:simulation} performs simulation studies, followed by an application to infer gene pathways from gene expression and SNP data in Section \ref{section:real-data}. Section \ref{section:conclusion} concludes the article. 
The Appendix contains illustrative examples and technical proofs.}

\section{Gaussian Directed Acyclic Graph with Additive Interventions}\label{section:model}

To infer directed relations among $p$ primary variables (i.e., variables of primary interest) $\bm Y = (Y_1,\ldots,Y_p)^\top$, consider a structural equation model with $q$ additive interventions:  
\begin{equation}
    \label{equation:model}
    \bm Y= \bs U^{\top} \bm Y + \bs W^{\top} \bm X + \bm \varepsilon, 
    \quad \bm \varepsilon \sim N(\bm 0, \bm\Sigma), \quad \bm \Sigma = \diag(\sigma^2_1,\ldots,\sigma^2_p),
\end{equation}
where $\bm X=(X_1,\ldots, X_q)^\top$ is a vector of additive intervention variables,
$\bs U\in \mathbb R^{p \times p}$ and $\bs W \in \mathbb R^{q \times p}$ are unknown coefficient matrices, 
and $\bm \varepsilon = (\varepsilon_1,\ldots,\varepsilon_p)^\top$ is a vector of random errors with $\sigma_j^2>0$; $j=1,\ldots,p$. In \eqref{equation:model}, $\bm \varepsilon$ is independent
of $\bm X$ but the components of $\bm X$ can be dependent.
The matrix $\bs U$ specifies the directed relations among $\bm Y$, 
where $\mathrm U_{kj}\neq 0$ if $Y_k$ is a direct cause of $Y_j$, denoted by $Y_k \rightarrow Y_j$, {and $Y_k$ is called a {parent} of $Y_j$ or $Y_j$ a {child} 
of $Y_k$.}
Thus, $\bs U$ represents a directed graph, which is further required to be acyclic to ensure the validity of the local Markov property \citep{spirtes2000causation}.
The matrix $\bs W$ specifies the targets and strengths of interventions, 
where $\mathrm W_{lj}\neq 0$ indicates $X_l$ intervenes on $Y_j$, denoted by 
$X_l\to Y_j$. In \eqref{equation:model}, no directed edge from a primary 
variable $Y_j$ to an intervention variable $X_l$ is permissible. 

In what follows, we will focus on the DAG $\mathcal G = (\bm Y,\bm X; \mathcal E,\mathcal I)$ with primary variables $\bm Y$, intervention variables $\bm X$, primary edges $\mathcal E = \{ (k,j) : \mathrm U_{kj}\neq 0 \}$, and intervention edges $\mathcal I=\{ (l,j) : \mathrm W_{lj}\neq 0 \}$.
To facilitate discussion, we introduce some concepts and notations for $\mathcal G$. 
If there is a directed path $Y_k\to\cdots\to Y_j$ in $\mathcal G$, 
$Y_k$ is an ancestor of $Y_j$ or $Y_j$ is a descendant of $Y_k$.
If $Y_k\to Y_j$ and there is no other directed path from $Y_k$ to $Y_j$, then we say $Y_k$ is an unmediated parent of $Y_j$. 
Let $\pa_{\mathcal G}(j) = \{ k : Y_k \to Y_j \}$, $\an_{\mathcal G}(j) = \{ k: Y_k\to \cdots\to Y_j \}$, and $\inter_{\mathcal G}(j)=\{ l : X_l\to Y_j \}$ be the parent, ancestor, and intervention sets of $Y_j$, respectively.

\subsection{Identifiability and Instruments}\label{section:identifiability}

Model \eqref{equation:model} is generally non-identifiable without interventions 
($\bs W = \bm 0$) when errors do not meet some requirements such as the equal-variance assumption \citep{peters2014identifiability} and its variants \citep{ghoshal2018learning,rajendran2021structure}.
Moreover, the model can be identified when $\bm \varepsilon$ in \eqref{equation:model}
is replaced by non-Gaussian errors \citep{shimizu2006linear}
or linear relations are replaced by nonlinear ones \citep{peters2014causal}. 
Regardless, suitable interventions can make \eqref{equation:model} identifiable. 
When intervention targets are known, the identifiability issue has been studied 
\citep{oates2016estimating,chen2018two}. 
However, it is less so when the exact targets and strengths of interventions are unknown
as in many biological applications \citep{jackson2003expression,kulkarni2006evidence}, 
which is referred to as the case of \emph{unspecified} or \emph{uncertain} interventions 
\citep{heinze2018causal,eaton2007exact,squires2020permutation}. 

We now categorize interventions as instruments and invalid instruments.

\begin{definition}[DAG instrument] \label{definition:instrument}
An intervention variable is an instrument in $\mathcal G$ if 
\begin{enumerate}
  \item [(A)] it intervenes on at least one primary variable in $\mathcal G$; 
  \item [(B)] it does not intervene on more than one primary variable in $\mathcal G$.
\end{enumerate}
Otherwise, it is an invalid instrument in $\mathcal G$.
\end{definition}

Here, (A) requires an intervention to be active, while (B) prevents simultaneous interventions of a single intervention variable on multiple primary variables. 
This is critical to identifiability because an instrument on a (potential) cause variable $Y_1$ helps reveal its directed effect on an outcome variable $Y_2$, which breaks the symmetry in a Gaussian DAG that results in non-identifiability of directed relations $Y_1 \rightarrow Y_2$ and $Y_2 \rightarrow Y_1$.

\begin{remark}
    The conventional definition of instrumental variable differs from Definition \ref{definition:instrument}.
    In the literature \citep{angrist1996identification}, an instrument $X$ for estimating the effect from a potential cause $Y_1$ to the outcome $Y_2$ is required to satisfy  (i) $X$ is related to $Y_1$, called relevance, (ii) $X$ has no directed edge to the outcome $Y_2$, called exclusion, and (iii) $X$ is not related to unmeasured confounders, called unconfoundedness. In Definition \ref{definition:instrument}, (A) is the relevance property, (B) generalizes the exclusion property for a DAG model, and the unconfoundedness is satisfied because no confounder is present in model \eqref{equation:model}. 
\end{remark}

Next, we make some assumptions on intervention variables to yield an identifiable 
model, where dependencies among intervention variables are permissible.
 
\begin{assumption}\label{assumption:identifiability}
  Assume that model \eqref{equation:model} satisfies the following conditions.
  \begin{enumerate}
    \item [(1A)] $\E\bm X\bm X^\top$ is positive definite.
      
    \item [(1B)] $\Cov(Y_j, X_l \mid \bm X_{\{1,\ldots,q\}\setminus\{l\}})\neq 0$ if $X_l$ intervenes on any unmediated parent of $Y_j$.

    \item [(1C)] Each primary variable is intervened by at least one instrument.
  \end{enumerate}
\end{assumption}

Assumption 1A imposes mild distributional restrictions on $\bm X$, permitting discrete variables such as SNPs. Assumption 1B requires the interventional effects through unmediated parents not to cancel out, as multiple targets from an invalid instrument are permitted. 
Importantly, if either Assumption 1B or 1C fails, model \eqref{equation:model} is generally not identifiable, as shown in Example \ref{example:identifiability} of Appendix \ref{section:identifiability-example}. 
In Section \ref{section:simulation}, we empirically examine the situation when Assumption 1C is not met.

\begin{proposition}\label{proposition:identifiability}
Under Assumption \ref{assumption:identifiability}, $(\bs U, \bs W, \bm\Sigma)$ in model \eqref{equation:model} are identifiable from the distribution of $(\bm Y,\bm X)$.
\end{proposition}

Proposition \ref{proposition:identifiability} {(proved in Appendix \ref{proof:identifiability})} is derived for a DAG model with unspecified interventions. 
This is in contrast to Proposition 1 of \citet{chen2018two}, which proves the identifiability of the parameters in a directed graph with target-known instruments on each primary variable.
Moreover, the estimated graph in \citet{chen2018two} may be cyclic and lacks the local Markov property for causal interpretation \citep{spirtes2000causation}. 

\subsection{Problem Statement: Inference for a DAG}\label{section:testability}

Our goal is to perform statistical inference of directed edges and pathways in the DAG $\mathcal G$. 
Let $\mathcal{H}\subseteq \{ (k,j):k\neq j, 1\leq k,j\leq p \}$ be an edge set among primary variables $\{ Y_1,\ldots,Y_p \}$, where $(k,j) \in \mathcal{H}$ 
specifies a (hypothesized) directed edge $Y_k \to Y_j$ in \eqref{equation:model}. 
We shall focus on two types of testing with null and alternative hypotheses $H_0$ and $H_a$.
For simultaneous testing of directed edges,
\begin{eqnarray}
\label{equation:link-test}
H_0: \mathrm U_{kj}= 0; \text{ for all } (k,j) \in \mathcal{H} \quad \mbox{ versus } 
\quad H_a: \mathrm U_{kj}\neq 0 \text{ for some } (k,j) \in \mathcal{H}; 
\end{eqnarray}
for simultaneous testing of directed pathways, 
\begin{eqnarray}
\label{equation:pathway-test}
H_0: \mathrm U_{kj}= 0; \text{ for some } (k,j) \in \mathcal{H} \quad \mbox{ versus }
\quad H_a: \mathrm U_{kj} \neq  0 \text{ for all } (k,j) \in \mathcal{H},
\end{eqnarray}
where $(\bs U_{\mathcal {H}^c},\bs W, \bm \Sigma)$ 
are unspecified nuisance parameters and $^c$ is the set complement. 
Note that $H_0$ in \eqref{equation:pathway-test} is a composite hypothesis that can be decomposed into sub-hypotheses
  \begin{equation*}
  H_{0,\nu}: \mathrm U_{k_\nu,j_\nu}= 0, \quad \mbox{ versus } 
  \quad {H}_{a,\nu}: \mathrm U_{k_\nu,j_\nu}\neq 0; \quad \nu=1,\ldots,|\mathcal H|,
  \end{equation*}
where $\mathcal{H}=\{ (k_1,j_1),\ldots,(k_{|\mathcal H|},j_{|\mathcal H|}) \}$ and testing each sub-hypothesis is a directed edge test.
Thus, we treat \eqref{equation:pathway-test} as an extension of \eqref{equation:link-test}. 

We will also estimate $(\mathbf U,\mathbf W)$ as well as identify the nonzero elements in $\mathbf U$ to recover the directed edges among the primary variables $\bm Y$ in $\mathcal G$.

\section{Methodology}\label{section:method}

This section develops the main methodology, including the peeling algorithm for structure learning and the data perturbation inference for simultaneous testing of directed edges \eqref{equation:link-test} and pathways \eqref{equation:pathway-test}. 
To proceed, suppose the data matrices $\mathbf Y_{n\times p} = (\bm Y_{1},\ldots,\bm Y_n)^\top$ and $\mathbf X_{n\times q} = (\bm X_1,\ldots,\bm X_n)^\top$ are given, where the rows $\{(\bm Y_i^\top,\bm X_i^\top)\}_{1\leq i\leq n}$ are independently sampled from \eqref{equation:model}. 
Then the log-likelihood is (up to a constant)
\begin{equation}
\label{equation:likelihood}
    L(\bm\theta,\bm\Sigma)= -
   \frac{1}{2} \sum_{i=1}^n
    \big\|\bm\Sigma^{-1/2} 
    ((\bs I - \bs U^\top)\bm Y_i - \bs W^\top \bm X_i ) \big\|^2_2 - 
    n\log\sqrt{\det(\bm\Sigma)},
\end{equation}
where $\bm\theta = (\bs U,\bs W)$, $\bm \Sigma=\diag(\sigma_1^2,\ldots,\sigma_p^2)$, and $\bs U$ is subject to the acyclicity constraint \citep{zheng2018dags,yuan2019constrained} in that no directed cycle is permissible in the DAG.

One major challenge to this likelihood approach lies in the optimization of \eqref{equation:likelihood} subject to the acyclicity constraint, which imposes difficulty on not only computation but also asymptotic theory. As a result, there is a gap between the asymptotic distribution of a global maximum and that of the actual estimate which can be a local maximum \citep{jankova2018inference,li2020likelihood}. Moreover, the actual estimate may give an imprecise topological order, tending to impact adversely on inference.

To circumvent the acyclicity requirement, we propose to use the ancestral relation graph (ARG) to describe the topological order of the DAG, where the ARG can be efficiently estimated without explicitly imposing the acyclicity constraint while enjoying a statistical guarantee of the actual estimate. 

\begin{definition}[Ancestral relation graph] ~
    \begin{enumerate}[(A)]
        \item A graph $\mathcal M = (\bm Y,\bm X; \mathcal A, \mathcal C)$ is an ARG if it is acyclic and  $\mathcal A = \{ (k,j) : k\in \an_{\mathcal M}(j) \}$. 
        \item Given DAG $\mathcal G=(\bm Y,\bm X;\mathcal E,\mathcal I)$, its ARG is defined as $\mathcal G_+=(\bm Y,\bm X; \mathcal E_+,\mathcal I_+)$, where 
        \begin{equation*}
            \mathcal E_+ = \Big\{ (k,j) : k\in\an_{\mathcal G}(j) \Big\}, \quad 
            \mathcal I_+ = \Big\{ (l,j) : l\in \bigcup_{k\in \an_{\mathcal G}(j)\cup\{j\}} \inter_{\mathcal G}(k)\Big\}.
        \end{equation*}
    \end{enumerate}
\end{definition}

Given $\mathcal G_+$ (which is acyclic), we have $\bm\theta=(\bm\theta_{\mathcal G_+},\bm \theta_{\mathcal G^c_+})$, where $\bm \theta_{\mathcal G_+}=(\bs U_{\mathcal E_+}, \bs W_{\mathcal I_+})$ are the (effective) parameters and $\bm\theta_{\mathcal G^c_+} = (\bs U_{\mathcal E^c_+},\bs W_{\mathcal I^c_+}) = \bm 0$. Then the log-likelihood \eqref{equation:likelihood} becomes
\begin{equation}
\label{equation:likelihood-supergraph}
  \begin{split}
 L((\bm \theta_{\mathcal G_+},\bm 0),\bm \Sigma)
    &= -\sum_{j = 1}^p\underbrace{\sum_{i=1}^n \Big(Y_{ij} - \sum_{(k,j)\in\mathcal E_+} U_{kj}Y_{ik} - \sum_{(l,j)\in\mathcal I_+} W_{lj}X_{il} \Big)^2}_{:= \text{RSS}_j(\bm\theta)}/2\sigma_j^2 + n\log(\sigma_j),
  \end{split}
\end{equation} 
which involves $|\mathcal{E}_+|+|\mathcal I_+|$ parameters for $\bm \theta_{\mathcal G_+}$.
From \eqref{equation:likelihood-supergraph}, we can reconstruct $\mathcal G$ and conduct inference for \eqref{equation:link-test} and \eqref{equation:pathway-test}.

Our plan is as follows. In Section \ref{section:causal-discovery}, we construct $\mathcal G_+$ without the acyclicity constraint for $\bs U$. On this basis, in Section \ref{section:dp-inference} we develop likelihood ratio tests for \eqref{equation:link-test} and \eqref{equation:pathway-test}. 

\subsection{Structure Learning via Peeling}\label{section:causal-discovery}

This section develops a novel structure learning method to construct $\mathcal G_+$ in a hierarchical manner.
First, we observe an important connection between primary variables and 
intervention variables. 
Rewrite \eqref{equation:model} as
\begin{equation}
\label{equation:model2}
\bm Y = \bs V^\top \bm X + \bm \varepsilon_V, \quad \bm \varepsilon_V=(\bs I-\bs U^\top)^{-1}\bm \varepsilon \sim N(\bm 0,\bm\Omega^{-1}),
\end{equation}
where $\bm\Omega= (\bs I - \bs U)\bm\Sigma^{-1}(\bs I -\bs U^\top)$
and $\bs V = \bs W (\bs I - \bs U)^{-1}$. 

\begin{proposition} \label{proposition:peeling} 
   Suppose Assumption \ref{assumption:identifiability} is satisfied. 
    \begin{enumerate}[(A)]
        \item If $\mathrm V_{lj}\neq 0$, then $X_{l}$ intervenes on $Y_j$ or an ancestor of $Y_j$; 
        \item In $\mathcal G$, $Y_j$ is a leaf variable (having no child) if and only if there is an instrument $X_l$ such that $\mathrm V_{lj}\neq 0$ and $\mathrm V_{lj'}=0$ for $j'\neq j$. 
    \end{enumerate}
\end{proposition}

The proof of Proposition \ref{proposition:peeling} is deferred to Appendix \ref{proof:peeling}.
Intuitively, $\mathrm V_{lj}\neq 0$ implies the dependence of $Y_j$ on $X_{l}$ through a directed path $X_l\to \cdots \to Y_j$, and hence that $X_l$ intervenes on $Y_j$ or an ancestor of $Y_j$. 
Thus, the instruments on a leaf variable are independent of the other primary variables conditional on the rest of interventions. This observation suggests a method to reconstruct the DAG topological order by recursively identifying and removing (i.e., peeling) the leaf variables. 

Next, we discuss the estimation of $\bs V$ and construction of $\mathcal G_+$.

\subsubsection{Nodewise constrained regressions}

We estimate $\bs V=(\mathbf V_{\cdot 1},\ldots,\mathbf V_{\cdot p})$ via nodewise $\ell_0$-constrained regressions,
\begin{equation}
    \label{pseudo-likelihood}
    \widehat{\bs V}_{\cdot j} = \argmin_{\bs V_{\cdot j}} \ 
    \sum_{i=1}^n \Big(Y_{ij} - \bs V_{\cdot j}^\top \bm X_i\Big)^2
    \quad \text{s.t.} \quad \sum_{l=1}^q \I(\mathrm V_{lj}\neq 0) \leq \kappa_j; \quad j=1,\ldots,p,
\end{equation}
where $1 \leq \kappa_j\leq q$ is an integer-valued tuning parameter controlling the sparsity and can be chosen by BIC or cross-validation.
To solve \eqref{pseudo-likelihood}, we use $J(z; \tau_j) = \min(|z|/\tau_j,1)$ as a surrogate of $\I(z \neq 0)$ \citep{shen2012likelihood} and develop a difference-of-convex (DC) program with the $\ell_0$-projection to improve the globality of the solution of \eqref{pseudo-likelihood}. 
Specifically, at the $(t+1)$th iteration, given $\widetilde{\bs V}^{[t]}_{\cdot j}$, we solve the weighted Lasso problem, 
\begin{eqnarray}
  \label{equation:penalized-regression}
  \widetilde{\bs V}_{\cdot j}^{[t+1]}=\argmin_{\bs V_{\cdot j}} \
  \sum_{i=1}^n \Big(Y_{ij} - \bs V_{\cdot j}^\top \bm X_i\Big)^2+2n\gamma_j \tau_j \sum_{l=1}^q \I\big(|\widetilde{\mathrm V}^{[t]}_{lj}|\leq \tau_j\big) 
  |\mathrm V_{lj}|; \quad j=1,\ldots,p, 
\end{eqnarray}
where $\gamma_j > 0$ is an internal hyperparameter used by the DC program; see Remark \ref{remark:hyperparameter} below. 
The DC program terminates at $\widetilde{\bs V}_{\cdot j} = \widetilde{\bs V}_{\cdot j}^{[t]}$ such that $\|\widetilde{\bs V}_{\cdot j}^{[t+1]}-\widetilde{\bs V}_{\cdot j}^{[t]}\|_{\infty}\leq \sqrt{\text{tol}}$ or $t$ achieves the maximum iteration number, where $\text{tol}$ is the machine precision.
Then, the solution $\widehat{\bs V}_{\cdot j}$ of \eqref{pseudo-likelihood} is computed by projecting $\widetilde{\bs V}_{\cdot j}$ onto the set $\Big\{\mathbf v\in\mathbb R^q : \|\bs v\|_0 \leq \kappa_j\Big\}$. 

Algorithm \ref{algorithm:nodewise-regression} summarizes the computation method.

\begin{algorithm}[H]
\caption{Constrained estimation via DC program + $\ell_0$ projection} \label{algorithm:nodewise-regression}
\KwIn{data $\mathbf Y$ and $\mathbf X$.}
\Parameter{$\{(\kappa_j,\tau_j)\}_{1\leq j\leq p}$; candidate values $\{\gamma^{(1)} > \cdots >\gamma^{(R)}\}$ for $\gamma_j$.}
\KwOut{the estimate $\widehat{\mathbf V}$.}

\For{each $1\leq j\leq p$}{
\For{each $1\leq r\leq R$}{

Initialize $\widehat{\mathbf V}_{\cdot j}^{(r)}\leftarrow \bm 0$, and $\gamma_j\leftarrow \gamma^{(r)}$\;

Compute $\widetilde{\bs V}_{\cdot j}$ via DC program \eqref{equation:penalized-regression} with $\widetilde{\bs V}_{\cdot j}^{[0]} \leftarrow \bm 0$\;

Let $B\leftarrow\big\{ l : |\widetilde{\mathrm V}_{lj}|\neq 0 \ \text{is among the largest } \kappa_j \ \text{elements of } \{ |\widetilde{\mathrm V}_{l'j}| \}_{1\leq l'\leq q} \big\}$\; 

Compute $\widehat{\bs V}_{\cdot j}^{(r)} \leftarrow \argmin_{\{\bs v: \mathbf v_{B^c}=\bm 0\}} \|\mathbf Y_{\cdot j} - \mathbf X\mathbf v\|_2^2$\;

}

Compute $\widehat{\bs V}_{\cdot j} \leftarrow \argmin_{\{\widehat{\bs V}^{(r)}_{\cdot j}\}_{1\leq r\leq R}} \|\mathbf Y_{\cdot j} - \mathbf X\widehat{\mathbf V}^{(r)}_{\cdot j}\|_2^2$\;
}

\KwRet{$\widehat{\mathbf V} \leftarrow (\widehat{\mathbf V}_{\cdot 1},\ldots,\widehat{\mathbf V}_{\cdot p})$}\;
\end{algorithm}

\begin{remark}\label{remark:hyperparameter}
    In Algorithm \ref{algorithm:nodewise-regression}, $\gamma_j$ is chosen from a set of candidate values $\{\gamma^{(r)}\}_{1\leq r\leq R}$. In our implementation, $\gamma_j$ is not directly tuned by the user and $\{\gamma^{(r)}\}_{1\leq r\leq R}$ is provided by default. 
    When $\{(\kappa_j,\tau_j)\}_{1\leq j\leq p}$ are suitably specified by the user, for any value $\gamma^{(r)}$ lies in the proper ranges, $({\mathbf V}_{\cdot 1},\ldots,{\mathbf V}_{\cdot p})$ are the global solutions of \eqref{pseudo-likelihood} almost surely; see Theorem \ref{theorem:consistent-structure-learning} in Section \ref{section:consistency}.
    Moreover, solving the DC programs for $\gamma^{(1)} > \cdots >\gamma^{(R)}$ is efficient with the warm start trick \citep{friedman2010regularization,breheny2011coordinate}.
\end{remark}

\subsubsection{Peeling}

Now, we describe a \emph{peeling} algorithm to estimate $\mathcal G_+$ based on $\bs V$. 
Proposition \ref{proposition:peeling} suggests that the leaf variables of $\mathcal G$ (with their instruments) can be identified based on matrix $\mathbf V$.
To proceed, let $\mathcal L$ and its complement $\mathcal L^c$ be (generic) nonempty subsets of $\{1,\ldots,p\}$ such that $\bm Y_{\mathcal L^c}$ are non-descendants of $\bm Y_{\mathcal L}$.
Define a sub-DAG $\mathcal G_{\mathcal L^c} = (\bm Y_{\mathcal L^c},\bm X;\mathcal E_{\mathcal L^c},\mathcal I_{\mathcal L^c})$,
where $\mathcal E_{\mathcal L^c}\subseteq \mathcal E$ is the set of primary edges among $\bm Y_{\mathcal L^c}$ and $\mathcal I_{\mathcal L^c}\subseteq\mathcal I$ is the set of intervention edges between $\bm X$ and $\bm Y_{\mathcal L^c}$.
The following proposition offers insights into the connection between $\mathbf V$ and $\mathcal G_+$.

\begin{proposition}\label{proposition:ancestral-relation} 
    Suppose Assumption \ref{assumption:identifiability} is satisfied. 
    Let $Y_k$ be a leaf in $\mathcal G_{\mathcal L^c}$ and $Y_j$ be in $\bm Y_{\mathcal L}$.
    \begin{enumerate}[(A)]
    \item If $\mathrm V_{lj} \neq 0$ for each instrument $X_l$ of $Y_k$ in $\mathcal G_{\mathcal L^c}$,  we have $(k,j) \in \mathcal{E}_{+}$.
    \item If $Y_k$ is an unmediated parent of $Y_j$, then $\mathrm V_{lj} \neq 0$ for 
    each instrument $X_l$ of $Y_k$ in $\mathcal G_{\mathcal L^c}$.
    \end{enumerate}
\end{proposition}
 
Proposition \ref{proposition:ancestral-relation} (proved in Appendix \ref{proof:ancestral-relation}) together with Proposition \ref{proposition:peeling} indicates that $\mathcal G_+$ can be constructed from $\mathbf V$.
In particular, we can sequentially identify each leaf $Y_k$ with its instrument(s) $X_l$ in the DAG $\mathcal G$ or the sub-DAG $\mathcal G_{\mathcal L^c}$ (where $\bm Y_{\mathcal L}$ are peeled variables). Then $\mathcal E_+$ can be constructed by including all edges $(k,j)$ such that $Y_k$ is a leaf in the sub-DAG $\mathcal G_{\mathcal L^c}$, $Y_j$ is a peeled variable, and $\mathrm V_{lj} \neq 0$ for each instrument $X_l$ of leaf $Y_k$ in $\mathcal G_{\mathcal L^c}$. By Proposition \ref{proposition:ancestral-relation}, (A) confirms that all such edges are in $\mathcal E_+$ so no extra edges are included, and (B) guarantees that every directed edge from an unmediated parent must be included, which is sufficient to determine all the ancestral relations. Thus, $\mathcal E_+$ can be recovered from $\mathbf V$. Then $\mathcal I_+$ is equal to $\{ (l,j) : \mathrm V_{lk}\neq 0 \ \text{if} \ k = j \ \text{or}\ (k,j)\in\mathcal E_+ \}$.

The peeling algorithm is summarized in Algorithm \ref{algorithm:peeling} and 
a detailed illustration is presented in Example \ref{example:peeling} of Appendix \ref{section:peeling-example}. 

\begin{algorithm}[H]
\caption{Reconstruction of ARG by peeling} \label{algorithm:peeling}
\KwIn{matrix $\widehat{\mathbf V}$.}
\KwOut{estimated ARG $\widehat{\mathcal G}_+$.}

  Initialize ${\mathbf V}^{\text{work}}\leftarrow \widehat{\mathbf V}$, $\widehat{\mathcal E}_+ \leftarrow \emptyset$, $\widehat{\mathcal I}_+\leftarrow \{(l,k): \widehat{\mathrm V}_{lk}\neq 0 \}$\;
  
  Initialize $\mathcal G^{\text{work}}$ by $\mathcal V_{Y} \leftarrow \{1,\ldots,p\}$, $\mathcal V_{X} \leftarrow \{1,\ldots,q\}$, $\mathcal E_-\leftarrow \widehat{\mathcal E}_+$, $\mathcal I_-\leftarrow \widehat{\mathcal I}_+$\;

  \While{$\mathcal V_{Y}$ is not empty}{
  
    In $\mathcal{G}^{\text{work}}$, identify the instruments on leaves $\mathcal B \leftarrow \big\{l : l \text{ minimizes } \|{\mathbf V}^{\text{work}}_{l,\cdot}\|_0 \text{ and } \|{\mathbf V}^{\text{work}}_{l, \cdot}\|_0 > 0 \big\}$ and the leaves 
    $\mathcal L \leftarrow \bigcup_{l\in A} \big\{k : k \text{ maximizes } |{\mathrm V}^{\text{work}}_{l k}|\big\}$\;

    Let $\mathcal B_k \leftarrow \big\{ l\in \mathcal B : k \text{ maximizes } |{\mathrm V}^{\text{work}}_{l k}| \big\}$ be the instruments for each $k\in\mathcal L$\;

    Update $\widehat{\mathcal{E}}_{+} \leftarrow \widehat{\mathcal{E}}_{+} \cup \big\{(k,j): j \in \{1,\ldots,p\}\setminus \mathcal V_{Y}, \ k \in \mathcal{L}, \ \widehat{\mathrm V}_{lj}\neq 0 \text{ for all } l \in \mathcal{B}_k \big\}$\;
    
    Update $\mathcal{G}^{\text{work}}$ by $\mathcal V_{Y} \leftarrow \mathcal V_{Y} \setminus \mathcal{L}$ and update ${\mathbf V}^{\text{work}}$ by keeping the columns in $\mathcal V_{Y}$\;   
  }
  
  Update $\widehat{\mathcal{E}}_{+}\leftarrow \big\{(k,j): Y_k \to \cdots \to Y_j \text{ in }  \widehat{\mathcal{E}}_{+} \big\}$\;
  
  Update $\widehat{\mathcal{I}}_{+}\leftarrow \widehat{\mathcal I}_+ \cup \big\{(l,j): (l,k) \in \widehat{\mathcal{I}}_{+} \text{ and } (k,j) \in \widehat{\mathcal{E}}_{+} \big\}$\;

  \KwRet{$\widehat{\mathcal E}_+$ and $\widehat{\mathcal I}_+$}\;
\end{algorithm}

\begin{remark}
Given $\widehat{\mathcal G}_+=(\bm Y,\bm X;\widehat{\mathcal E}_+,\widehat{\mathcal I}_+)$, we estimate $(\bs U_{\widehat{\mathcal E}_+},\bs W_{\widehat{\mathcal I}_+})$ column-wise from \eqref{equation:likelihood-supergraph},
\begin{equation}\label{directed-likelihood}
    \min \  
      \sum\limits_{i=1}^n \Big(Y_{ij} - \sum\limits_{k\in \an_{\widehat{\mathcal G}_+}(j)} \mathrm U_{kj} Y_{ik} - \sum\limits_{l\in \inter_{\widehat{\mathcal G}_+}(j)} \mathrm W_{lj} X_{il}  \Big)^2 \ 
    \mathrm{s.t.} \ 
    \sum\limits_{k\in\an_{\widehat{\mathcal G}_+}(j)} \I(\mathrm U_{kj}\neq 0) \leq \kappa_j'.
\end{equation}
The final estimates are $\widehat{\bs U} = (\widehat{\bs U}_{\widehat{\mathcal E}_+},\bm 0)$
and $\widehat{\bs W} = (\widehat{\bs W}_{\widehat{\mathcal I}_+},\bm 0)$.
In \eqref{directed-likelihood}, the sparsity constraint is imposed to recover the nonzero elements of $\mathbf U$; see Appendix \ref{section:structure-learning-theory} for the technical discussion.
\end{remark}

\subsection{Likelihood Inference for a DAG}\label{section:dp-inference}

Now, we propose an inference method for testing \eqref{equation:link-test} and \eqref{equation:pathway-test}. First, we derive the likelihood ratio based on $\mathcal G_+$. Next, we perform tests via data perturbation, accounting for the uncertainty of estimating $\mathcal G_+$.

\subsubsection{Likelihood ratio, nondegeneracy, and regularity}

We commence with the likelihood inference for \eqref{equation:link-test}, since \eqref{equation:pathway-test} can be treated as an extension of \eqref{equation:link-test}; see the discussion followed by \eqref{equation:pathway-test}. 
From \eqref{equation:likelihood-supergraph}, the maximum likelihood becomes
\begin{equation*}
    \begin{split}
        \max_{\mathcal G_+,\bm\Sigma} \ \max_{\bm\theta=(\bm\theta_{\mathcal G_+},\bm 0)} \ L(\bm\theta,\bm\Sigma).
    \end{split}
\end{equation*}
Thus, we define the likelihood ratio by 
\begin{equation}\label{equation:likelihood-ratio-statistic}
 \begin{split}
     \text{Lr} = L\Big(\widehat{\bm\theta}^{(1)},\widehat{\bm\Sigma}\Big) -  L\Big(\widehat{\bm\theta}^{(0)},\widehat{\bm\Sigma}\Big)
     = \sum_{j=1}^p \Big({\text{RSS}_j(\widehat{\bm\theta}^{(0)}) - \text{RSS}_j(\widehat{\bm\theta}^{(1)})}\Big) / {2\widehat{\sigma}_j^2},
 \end{split}
 \end{equation}
 where $\widehat{\bm\theta}^{(0)} = (\widehat{\bm\theta}^{(0)}_{\widehat{\mathcal M}},\bm 0)$ and $\widehat{\bm\theta}^{(1)} = (\widehat{\bm\theta}^{(1)}_{\widehat{\mathcal M}},\bm 0)$ are MLEs (given $\widehat {\mathcal M}$) under $H_0$ and $H_a$, respectively, $\widehat{\mathcal M} = (\bm Y, \bm X; \widehat{\mathcal E}_+\cup\mathcal H, \widehat{\mathcal I}_+)$ is an estimate for $\mathcal G_+$, and $\widehat{\bm\Sigma}$ is an estimate for $\widehat{\bm \Sigma}$.
 Instead of $\widehat{\mathcal G}_+ = (\bm Y,\bm X;\widehat{\mathcal E}_+,\widehat{\mathcal I}_+)$, the graph $\widehat {\mathcal M}$ (with hypothesized edges being added) is used because we need to test the presence of any edge in $\mathcal H$.

 In many statistical models, the likelihood ratio often has a nondegenerate and tractable distribution when $H_0$ is true, for instance, a chi-squared distribution with degrees of freedom $|\mathcal{H}|$. However, since $\mathcal H$ is pre-specified by the user, $\widehat{\mathcal M}$ may not be a DAG, and thus not all edges in $\mathcal H$ could present in the DAG parameterized by $(\widehat{\bm\theta}^{(1)}_{\widehat{\mathcal M}},\bm 0)$. As a result, $\text{Lr}$ for \eqref{equation:link-test} may converge to a distribution with degrees of freedom less than $|\mathcal H|$ and the distribution may be even intractable, making inference for a DAG greatly different from the classical ones, as illustrated by Example \ref{example:testability}.

\begin{figure}[H]
  \centering
  \includegraphics[width=0.25\textwidth]{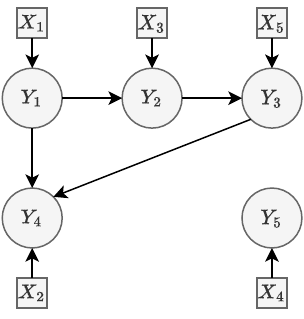}
  \caption{A DAG $\mathcal G$ of five primary variables $Y_1,\ldots,Y_5$ and five intervention variables $X_1,\ldots,X_5$, where directed edges are represented by solid arrows while dependencies among $\bm X$ are not displayed.}
\label{figure:testability}
\end{figure}

\begin{example}\label{example:testability} Consider the likelihood ratio test under null $H_0$ 
and alternative $H_a$ for the DAG $\mathcal G$ displayed in Figure \ref{figure:testability}. 
For simplicity, assume $\widehat{\mathcal G}_+ = \mathcal G_+$ and $\widehat{\mathcal M} =(\bm Y,\bm X;\mathcal E_+\cup \mathcal H,\mathcal I_+)$.
\begin{itemize}
  \item $H_0: \mathrm U_{21} = 0$ versus $H_a: \mathrm U_{21}\neq 0$, where $\mathcal H=\{ (2,1) \}$. 
  Here, $(2,1)$ forms a cycle together with the edges in $\mathcal E \setminus \mathcal H$ (namely the edges not considered by the hypothesis), and thus $\widehat{\mathcal M}$ has a directed cycle. Given $\widehat{\mathcal M}$, when a random sample is obtained under $H_0$, the likelihood tends to be maximized under the ARG $\mathcal G_+$ corresponding to the underlying DAG (which implies $\mathrm U_{21} = 0$), especially so when the asymptotics kicks in as the sample size increases. Consequently, the likelihood ratio $\text{Lr}$ becomes zero, constituting a degenerate situation.

  \item $H_0: \mathrm U_{45} = \mathrm U_{53} = 0$ versus $H_a: \mathrm U_{45}\neq 0$ or $\mathrm U_{53}\neq 0$, where $\mathcal H=\{ (4,5),(5,3)\}$.
  In this case, $\{(4,5),(5,3)\}$ forms a cycle with the edges in $\mathcal E\setminus\mathcal H$, and $\widehat{\mathcal M}$ is cyclic.
  Given $\widehat{\mathcal M}$, the likelihood tends to be maximized under either DAG $(\bm Y,\bm X;\mathcal E_+\cup\{(4,5)\},\mathcal I_+ )$ or DAG $(\bm Y,\bm X;\mathcal E_+\cup\{(5,3)\},\mathcal I_+ )$ when data is sampled under $H_0$. 
  Thus, we have 
  \begin{equation*}
  L(\widehat{\bm\theta}^{(1)},\widehat{\bm\Sigma})=\max\Big(L(\widehat{\mathrm U}_{45},\widehat{\mathrm U}_{53}=0,\widehat{\bs U}_{\mathcal H^c},\widehat{\bs W},\widehat{\bm\Sigma}), L(\widehat{\mathrm U}_{45}=0,\widehat{\mathrm U}_{53},\widehat{\bs U}_{\mathcal H^c},\widehat{\bs W},\widehat{\bm\Sigma})\Big).
  \end{equation*}
  As a result, the likelihood ratio distribution becomes complicated in this situation due to the dependence between the two components in $L(\widehat{\bm\theta}^{(1)},\widehat{\bm\Sigma})$.
\end{itemize}
\end{example}

Motivated by Example \ref{example:testability}, we introduce the concepts of nondegeneracy and regularity.

\begin{definition}[Nondegeneracy and regularity with respect to DAG]\label{definition:testability} ~
\begin{enumerate}[(A)]
  \item An edge $(k,j)\in \mathcal{H}$ is nondegenerate with respect to DAG $\mathcal G$ if $\{(k,j)\}\cup \mathcal{E}$ contains no directed cycle,
  where $\mathcal{E}$ denotes the edge set of $\mathcal G$. Otherwise, $(k,j)$ is degenerate.
  Let $\mathcal{D} \subseteq \mathcal{H}$ be the set of all nondegenerate edges with respect to $\mathcal G$.
  A null hypothesis $H_0$ is nondegenerate with respect to DAG $\mathcal G$ if $\mathcal{D}\neq\emptyset$.
  Otherwise, $H_0$ is degenerate.
  \item A null hypothesis $H_0$ is said to be regular with respect to DAG $\mathcal G$ if $\mathcal{D} \cup \mathcal{E}$ contains no directed cycle, where $\mathcal{E}$ denotes the edge set of $\mathcal G$. Otherwise, $H_0$ is called irregular.
\end{enumerate}
\end{definition}

\begin{remark}
  In practice, $\mathcal D$ is unknown and needs to be estimated from data.
  Indeed, $\widehat{\mathcal D}$ can be computed based on the estimated ARG $\widehat{\mathcal G}_+$,
  because a directed edge $(k,j)$ is nondegenerate if and only if $\{(k,j)\}\cup \mathcal{E}_+$ contains no directed cycle.
\end{remark}

Nondegeneracy ensures nonnegativity of the likelihood ratio. 
In testing \eqref{equation:link-test}, 
regularity excludes intractable situations for the null distribution.
In testing \eqref{equation:pathway-test}, if $H_0$ is irregular, 
then $\mathcal D\cup\mathcal{E}$ has a directed cycle,
which means the hypothesized directed pathway cannot exist due to the acyclicity constraint. 
Thus, regularity excludes the degenerate situations in testing \eqref{equation:pathway-test}. In what follows, we mainly focus on nondegenerate and regular hypotheses. 
For the degenerate case, the p-value is defined to be one.
{For the irregular case of edge test \eqref{equation:link-test},  
we decompose the hypothesis into regular sub-hypotheses and conduct multiple testing.
For the irregular case of pathway test \eqref{equation:pathway-test}, the p-value is defined to be one.
More discussions on the implementation in irregular cases are provided in Appendix \ref{section:appendix-irregular-hypothesis}.}

\subsubsection{Testing directed edges via data perturbation}

Assuming $H_0$ is nondegenerate and regular, then $\widehat{\bm\theta}^{(1)}$ is the MLE subject to the DAG $\widehat{\mathcal S} = (\bm Y,\bm X; \widehat{\mathcal E}_+\cup \widehat{\mathcal D},\widehat{\mathcal I}_+)$ and $\widehat{\bm\theta}^{(0)}$ is the MLE subject to an additional constraint $\mathbf U_{\mathcal H}=\bm 0$. The likelihood ratio \eqref{equation:likelihood-ratio-statistic} can be further simplified.
Let $\df_{\widehat{\mathcal S}}(j) = \{ k:(k,j)\in\widehat{\mathcal{D}} \}$ in DAG $\widehat{\mathcal S}$, where $\widehat{\mathcal D}$ is the estimated set of nondegenerate edges of $H_0$ with respect to $\mathcal G$. 
Furthermore, observe that if ${\df}_{\widehat{\mathcal S}}(j) = \emptyset$, then 
${\text{RSS}}_{j}(\widehat{\bm\theta}^{(0)}) = {\text{RSS}}_{j}(\widehat{\bm\theta}^{(1)})$.  
Hence, $\text{Lr}$ only summarizes the contributions from the primary variables with the (estimated) nondegenerate hypothesized edges,
\begin{equation}\label{equation:likelihood-ratio-edge}
  \text{Lr}
 = \sum_{\{j : {\df}_{\widehat{\mathcal S}}(j)\neq \emptyset\}}
 \Big( {{\text{RSS}}_j(\widehat{\bm\theta}^{(0)})
 - {\text{RSS}}_j(\widehat{\bm\theta}^{(1)})}\Big) / {2\widehat\sigma_j^2},
\end{equation}
where we estimate $\bm\Sigma=\diag(\sigma_j^1,\ldots,\sigma_p^2)$ by    
\begin{equation}
\label{degree-freedom}
\widehat{\sigma}^2_j=
{\text{RSS}_j(\widehat{\bm\theta}^{(1)})} / \big({n-|{\pa}_{\widehat{\mathcal S}}(j)|-|{\inter}_{\widehat {\mathcal S}}(j)|}\big),
\quad j=1,\ldots,p. 
\end{equation}

The likelihood ratio \eqref{equation:likelihood-ratio-edge} for testing directed edges \eqref{equation:link-test} requires an estimation of $\mathcal G_+$ (and $\mathcal S$), where we must account for the uncertainty of $\widehat{\mathcal G}_+$ (and $\widehat{\mathcal S}$) for finite-sample inference.
To proceed, we consider the test statistic $\text{Lr}$ based on a ``correct'' ARG $\mathcal M\supseteq \mathcal G_+$, where $\mathcal M=(\bm Y,\bm X;\mathcal A, \mathcal C) \supseteq \mathcal G_+=(\bm Y,\bm X;\mathcal{E}_+, \mathcal{I}_+)$ means that $\mathcal{A}\supseteq\mathcal E_+$ and $\mathcal C\supseteq\mathcal I_+$.  
Intuitively, a ``correct'' ARG distinguishes descendants and nondescendants, and thus can help infer the true directed relations defined by the local Markov property \citep{spirtes2000causation} without introducing model errors, yet may lead to a less powerful test when $\mathcal M$ is much larger than $\mathcal G_+$. By comparison, a ``wrong'' ARG $\mathcal M \not \supseteq \mathcal G_+$ may provide an incorrect topological order, and a test based on a ``wrong'' ARG may be biased, accompanied by an inflated type-I error.

Let $\bs{{Z}} = (\bs{Y}, \bs{X})$ denote the data matrix of primary and intervention variables and let $\bs{{e}} =(\bm\varepsilon_1,\ldots,\bm\varepsilon_n)^\top$ denote the error matrix, where the rows $\{\bs Z_{i,\cdot}=(\bm Y_i^\top,\bm X_i^\top)\}_{1\leq i\leq n}$ and $\{\bm\varepsilon_i^\top\}_{1\leq i\leq n}$ are sampled independently from \eqref{equation:model}.
From \eqref{equation:likelihood-ratio-edge}, assuming $\widehat{\mathcal G}_+\supseteq \mathcal G_+$ is a ``correct'' ARG, the likelihood ratio becomes 
\begin{equation} \label{equation:likelihood-ratio}
  \begin{split}
    \text{Lr} 
    &=  \sum_{\{j:{\df}_{\widehat{\mathcal S}}(j)\neq\emptyset\}} 
    \frac{\big\| (\bs{{P}}_{ \widehat A_j } - \bs{{P}}_{ \widehat B_j})^{1/2} 
    \bs{ Y}_{\cdot j} \big\|_2^2}
    { 2 \big\|(\bs I-\bs{{P}}_{ \widehat A_j })^{1/2}\bs{ Y}_{\cdot j}\big\|_2^2/(n-|\widehat{A}_j|)} \\ 
    &= \sum_{\{j:{\df}_{\widehat{\mathcal S}}(j)\neq\emptyset\}}
    \frac{\big\|(\bs{{P}}_{ \widehat A_j } - \bs{{P}}_{ \widehat B_j})^{1/2}
    (\bs{ Y}_{\cdot,{\df}_{\widehat{\mathcal S}}(j)}\bs U_{{\df}_{\widehat{\mathcal S}}(j),j} + \bs{ e}_{\cdot j})\big\|_2^2}{2\big\|(\bs I-\bs{{P}}_{ \widehat A_j })^{1/2}\bs{ e}_{\cdot j}\big\|_2^2/(n-|\widehat{A}_j|)},
  \end{split}
\end{equation}
where $\bs{{P}}_A = \bs{ Z}_{\cdot,A}(\bs{ Z}_{\cdot,A}^\top\bs{ Z}_{\cdot,A})^{-1}\bs{ Z}_{\cdot,A}^\top$ is the projection matrix onto the column span of $\bs{ Z}_{\cdot,A}$,
$\widehat{A}_j = {\pa}_{\widehat{\mathcal S}}(j)\cup {\inter}_{\widehat{\mathcal S}}(j)$, and 
{$\widehat{B}_j = \big({\pa}_{\widehat{\mathcal S}}(j)\cup {\inter}_{\widehat{\mathcal S}}(j)\big)\setminus{\df}_{\widehat{\mathcal S}}(j)$} for $1\leq j\leq p$.
In \eqref{equation:likelihood-ratio}, we have $\bs{ Y}_{\cdot,{\df}_{\mathcal S}(j)}\bs U_{{\df}_{\mathcal S}(j),j} = \bm 0$ for all $j$ under the null hypothesis $H_0$,
while $\bs{ Y}_{{\df}_{\mathcal S}(j)}\bs U_{{\df}_{\mathcal S}(j),j} \neq \bm 0$ for some $j$ under the alternative hypothesis $H_a$, and thus $\text{Lr}$ tends to be large under $H_a$. 

Now, we propose the data perturbation (DP) method \citep{shen2002adaptive,breiman1992little} to approximate the null distribution of \text{Lr} in \eqref{equation:likelihood-ratio}. The idea behind DP is to assess the sensitivity of the estimates through perturbed data 
$\bs Y^* = \bs Y + \bs e^*$, where the rows $\{(\bm\varepsilon_i^*)^\top\}_{1\leq i\leq n}$ of perturbation errors $\bs e^*_{n\times p}$ is sampled independently from $N(0,\widehat{\bm\Sigma})$.
Let $(\bs{ Z}^*,\bs{ e}^*) = (\bs{ X},\bs{ Y}^*,\bs{ e}^*)$ be the
DP sample. Note that the perturbation errors $\bs{ e}^*$ are only injected into $\bs Y$ and the perturbation errors $\bs{ e}^*$ are \emph{known} in the DP sample $(\bs{ Z}^*,\bs{ e}^*)$.
Given $(\bs{ Z}^*,\bs{ e}^*)$, we compute the perturbation estimate $\widehat{\mathcal G}_+^*$ (and $\widehat{\mathcal S}^*$) by Algorithms \ref{algorithm:nodewise-regression}-\ref{algorithm:peeling}. In \eqref{equation:likelihood-ratio}, under the null hypothesis $H_0$, the likelihood ratio $\text{Lr} = \Lambda(\bs Z, \bs{ e})$ is a function of observed data $\bs{Z}$ and unobserved errors $\bs{ e}$. By definition, the perturbation error $\bs{e}^*$ is accessible in the DP sample $(\bs{ Z}^*,\bs{ e}^*)$, suggesting the DP likelihood ratio $\text{Lr}^*:=\Lambda(\bs{Z}^*,\bs{e}^*)$ that is equal to
\begin{equation}\label{equation:likelihood-ratio-dp}
  \begin{split}
    \text{Lr}^* = 
      \sum_{\{j:{\df}_{\widehat{S}}(j)\neq\emptyset\}} 
          \frac{\big\| (\bs{{P}}^*_{\widehat A^*_j} - \bs{{P}}^*_{\widehat B^*_j})^{1/2}\bs{ e}^*_{\cdot j}\big\|_2^2}{2\big\|(\bs I-\bs{{P}}^*_{\widehat A^*_j})^{1/2}\bs{ e}^*_{\cdot j}\big\|_2^2/(n-|\widehat A^*_j|)}.
  \end{split}
\end{equation}
Note that \eqref{equation:likelihood-ratio-dp} mimics \eqref{equation:likelihood-ratio} when $\bs{ Y}_{\cdot,{\df}_S(j)}\bs U_{{\df}_S(j)} = \bm 0$. 
As a result, when $\widehat{\mathcal G}_+^*\supseteq \mathcal G_+$, the conditional distribution of $\text{Lr}^*$ given the data $\bs{ Z}$ well approximates the null distribution of $\text{Lr}$, 
where the model selection effect is accounted for by assessing the variability of $\{\widehat A^*_j, \widehat{B}^*_j\}_{1\leq j\leq p}$ over different realizations of $(\bs{ Z}^*,\bs{ e}^*)$.

In practice, we use Monte-Carlo to approximate the distribution
of $\text{Lr}^*$ given $\bs{ Z}$. We generate $M$ perturbed samples $\{(\bs{{Z}}^*_m,\bs{ e}_m^*)\}_{1\leq m \leq M}$ independently and compute $\{\text{Lr}^*_m\}_{1\leq m\leq M}$, respectively.
Then, we examine the condition $\widehat{\mathcal G}^*_{+,m}\supseteq \mathcal G_+$ by checking its empirical counterpart $\widehat{\mathcal G}_{+,m}^*\supseteq \widehat{\mathcal G}_+$.
The DP p-value of the edge test in \eqref{equation:link-test} is defined as 
\begin{equation}\label{pval:link}
  \text{Pval} = \Big(\sum_{m=1}^{M} 
  \I(\text{Lr}_m^*\geq \text{Lr},\widehat{\mathcal G}_{+,m}^*\supseteq\widehat{\mathcal G}_+)\Big)/
  \Big( \sum_{m=1}^{M} \I(\widehat{\mathcal G}^*_{+,m}\supseteq \widehat{\mathcal G}_+)\Big),
\end{equation}
where $\I(\cdot)$ is the indicator function.

\begin{remark}
Instead of \eqref{equation:likelihood-ratio-dp}, a naive approach is to recompute the likelihood ratio by treating the perturbed sample $\bs Z^*$ as $\bs Z$ while not using the information of $\bs e^*$.
However, this is infeasible.
For explanation, assuming $\widehat{\mathcal G}_+^*\supseteq \mathcal G_+$ is a ``correct'' ARG, then this naive likelihood ratio is equal to 
\begin{equation*}
\begin{split} 
  \sum_{\{j:{\df}_{\widehat{\mathcal S}^*}(j)\neq \emptyset\}} \frac{\big\|
  (\bs{{P}}^*_{\widehat{A}^*_j} - \bs{{P}}^*_{\widehat{B}^*_j})^{1/2}(\bs{ Y}_{\cdot,j} + \bs{ e}^*_{\cdot j})\big\|_2^2 }{ 2\big\|(\bs I-\bs{{P}}^*_{ \widehat A^*_j })^{1/2}(\bs{ Y}_{\cdot j} + \bs{e}^*_{\cdot j})\big\|_2^2/(n-|\widehat{A}^*_j|)}.
\end{split}
\end{equation*} 
Note that $\{\bs{ Y}_{\cdot,j}\}_{1\leq j\leq p}$ given $\bs{ Z}$ are deterministic and do not vanish under either $H_0$ or $H_a$. Thus, the conditional distribution of this naive likelihood ratio given $\bs{Z}$ does not approximate the null distribution of $\text{Lr}$, in contrast to the DP likelihood ratio in \eqref{equation:likelihood-ratio-dp}.
\end{remark}

\subsubsection{Extension to hypothesis testing for a directed pathway}

Next, we extend the DP inference for \eqref{equation:pathway-test}. 
Denote $\mathcal{H}=\{ (k_1,j_1),\ldots,(k_{|\mathcal{H}|},j_{|\mathcal{H}|}) \}$. 
Then the test of pathways in \eqref{equation:pathway-test} can be reduced to testing sub-hypotheses
\begin{equation*}
  H_{0,\nu}: \mathrm U_{k_\nu,j_\nu}= 0, \quad \mbox{ versus } 
  \quad {H}_{a,\nu}: \mathrm U_{k_\nu,j_\nu}\neq 0; \quad \nu=1,\ldots,|\mathcal H|,
\end{equation*}
where testing each sub-hypothesis is a directed edge test. 
Given $(\widehat{\mathcal S},\widehat{\bm\Sigma})$, the likelihood ratio 
for ${H}_{0,\nu}$ is 
$ \text{Lr}_\nu
= L(\widehat{\bm\theta}^{(1)},\widehat{\bm \Sigma}) 
- L(\widehat{\bm \theta}^{(0,\nu)},\widehat{\bm\Sigma})$,
where $\widehat{\bm\theta}^{(0,\nu)}$ is the MLE under 
the constraint that $\mathrm U_{k_\nu,j_{\nu}}=0$. 
When $\widehat{\mathcal G}_+\supseteq \mathcal G_+$, we have 
\begin{equation}\label{equation:likelihood-ratio-dp-pathway}
  \begin{split}
    \text{Lr}_\nu = 
  \frac{\big\| (\bs{{P}}_{\widehat A_{j_\nu}} - \bs{{P}}_{\widehat B_{j_\nu}})^{1/2}
  (\bs{ Y}_{k_\nu} \mathrm U_{k_\nu, j_\nu} + \bs{ e}_{\cdot j_\nu}) \big\|_2^2}{2\big\|(\bs I-\bs{{P}}_{\widehat A_{j_\nu}})^{1/2}\bs{ e}_{\cdot j_\nu}\big\|_2^2/(n-|\widehat A_{j_\nu}|)}, 
  \
  \text{Lr}_\nu^* =
  \frac{\big\| (\bs{{P}}^*_{\widehat A^*_{j_\nu}} - \bs{{P}}^*_{\widehat B^*_{j_\nu}})^{1/2}
  \bs{ e}^*_{\cdot j_\nu}\big\|_2^2}{2\big\|(\bs I-\bs{{P}}^*_{\widehat A^*_{j_\nu}})^{1/2}\bs{ e}^*_{\cdot j_\nu}\big\|_2^2/
  (n-|\widehat A^*_{j_\nu}|)},
  \end{split}
\end{equation}
where the distributions of $\text{Lr}_\nu^*$ given $\bs{ Z}$
approximates the null distributions of $\text{Lr}_\nu$. 
Finally, define the p-value of the pathway test in \eqref{equation:pathway-test} as
\begin{equation}\label{pval:path}
  \text{Pval} = \max_{1\leq \nu\leq |\mathcal{H}|} 
  \Big(\sum_{m=1}^{M} \I(\text{Lr}_{\nu,m}^*\geq \text{Lr}_{\nu},\widehat{S}^*_{m}\supseteq\widehat{S})\Big)
  /\Big( \sum_{i=1}^{M} \I(\widehat{S}^*_m\supseteq \widehat{S})\Big).
\end{equation}
Note that if any ${H}_{0,\nu}$ is degenerate, then $\text{Pval} = 1$.

Algorithm \ref{algorithm:data-perturbation} summarizes the DP method for hypothesis testing.

\begin{algorithm}[H]
\caption{DP likelihood ratio test} \label{algorithm:data-perturbation}

\KwIn{hypothesis $H_0$; data $\bs{Z}$}
\Parameter{Monte Carlo size $M$; parameters for Algorithm \ref{algorithm:nodewise-regression}.}
\KwOut{the p-value Pval.}

Compute $(\widehat{\sigma}^2_1,\ldots,
\widehat{\sigma}^2_p)$ and $\mathrm{Lr}$ with data $\bs{ Z}$\;

\For{each $1\leq m\leq M$ in parallel}{

Generate perturbed data $(\bs{ Z}^*_m,\bs{ e}_{m}^{*})$\;

For \eqref{equation:link-test}, compute $\text{Lr}^*_m$ based on \eqref{equation:likelihood-ratio-dp}\;

For \eqref{equation:pathway-test}, 
compute $\{\text{Lr}^*_{\nu,m}\}_{1\leq\nu\leq|\mathcal H|}$ based on \eqref{equation:likelihood-ratio-dp-pathway}\;

}

Compute $\mathrm{Pval}$ as \eqref{pval:link} or \eqref{pval:path} accordingly\;

\KwRet{$\mathrm{Pval}$}\;
\end{algorithm}

\begin{remark}
  For acceleration, we parallelize Step 2 in Algorithm \ref{algorithm:data-perturbation}. Additionally, we use the estimate $\widehat{\bm\theta}$ as a warm-start initialization for the DP estimates,  effectively reducing the computing time. 
\end{remark}

\begin{remark}[Connection with bootstrap]
  One may consider parametric or nonparametric bootstrap for $\text{Lr}$. The parametric bootstrap requires a good initial estimate of $(\bs U,\bs W)$. Yet, it is rather challenging to correct the bias of this estimate because of the acyclicity constraint. By comparison, DP does not rely on such an estimate. On the other hand, nonparametric bootstrap resamples the original data with replacement. In a bootstrap sample, only about 63\% distinct observations in the original data are used in model selection and fitting, leading to deteriorating performance \citep{kleiner2012big}, especially in a small sample. As a result, nonparametric bootstrap may not well-approximate the distribution of $\text{Lr}$, while DP provides a better approximation of $\text{Lr}$, taking advantage of a full sample. 
\end{remark}

\section{Theory}\label{section:theory}

This section provides the theoretical justification for the proposed methods.

\subsection{Convergence and Consistency of Structure Learning}\label{section:consistency}

First, we introduce some technical assumptions to derive statistical 
and computational properties of Algorithms \ref{algorithm:nodewise-regression} and \ref{algorithm:peeling}. 
Let $\bm\zeta$ be a generic vector and $\bm\zeta_A$ be the subvector of $\bm \zeta$ with coordinates in $A$. Let 
$\kappa^\circ_j = \|\bs V_{\cdot j}\|_0$ and 
$\kappa^\circ_{\max}=\max_{1\leq j\leq p} \kappa_j^\circ$.

\begin{assumption} \label{assumption:intervention}
For constants $c_1,c_2 > 0$, 
\begin{enumerate}[(A)]
    \item $\min\limits_{ \{A : |A|\leq 2 \kappa^\circ_{\max} \} }\min\limits_{\{ \bm\zeta : \|\bm\zeta_{A^c}\|_1\leq 3\|\bm\zeta_{A}\|_1 \}} {\|\bs{ X} \bm\zeta \|_2^2}/{n\| \bm\zeta \|_2^2} \geq c_1$ almost surely.
    \item $\max\limits_{1\leq l\leq q} \ (\bs{ X}^\top\bs{ X})_{ll}/n\leq c_2^2$ almost surely.
\end{enumerate}
\end{assumption}

\begin{assumption}\label{assumption:signal}
  $\min\limits_{\mathrm V_{lj} \neq 0}|\mathrm V_{lj}|
  \geq 100 c_1^{-1}c_2\max\limits_{1 \leq j \leq p}\big(\Omega_{jj}^{-1/2}\big)\sqrt{\log(q)/n + \log(n)/n}$.
\end{assumption}

Assumption \ref{assumption:intervention} is a common condition for proving the convergence rate of the Lasso \citep{bickel2009simultaneous,zhang2014lower}. 
As a replacement of Assumption \ref{assumption:identifiability}A, it can be satisfied for isotropic subgaussian or bounded $\bm X$ \citep{rudelson2013reconstruction}. 
Assumption \ref{assumption:signal}, as an alternative to Assumption 1B, specifies the minimal signal strength over candidate interventions. Such a signal strength requirement  is used for establishing the high-dimensional variable selection consistency \citep{fan2014strong,loh2017support,zhao2018pathwise}. Moreover, Assumption \ref{assumption:signal} can be further relaxed to a less intuitive condition, Assumption \ref{assumption:relax-signal}; see Appendix \ref{section:relax-signal} for details.

\begin{theorem} \label{theorem:consistent-structure-learning}
Suppose Assumptions \ref{assumption:identifiability}-\ref{assumption:signal} are met with constants $c_1 < 6c_2$, and the machine precision $\mathrm{tol}\ll 1/n$ is negligible. For $1\leq j \leq p$, if the tuning parameters $(\kappa_j,\tau_j)$ of Algorithm \ref{algorithm:nodewise-regression} are suitably chosen such that
\begin{equation*}
    \kappa_j = \kappa_j^\circ, \qquad 
    \frac{36c_2}{c_1}\sqrt{\Omega_{jj}^{-1}\Big(\frac{\log (q)}{n} + \frac{\log(n)}{n}\Big)} \leq \tau_j \leq 
  \frac{2}{5} \min_{\mathrm V_{lj} \neq 0}|\mathrm V_{lj}|,
\end{equation*}
then for any $\gamma_j$ such that $\tau_j^{-1}({32c_2^2\Omega_{jj}^{-1}n^{-1}(\log(q)+\log(n))})^{1/2}\leq \gamma_j \leq c_1/6$, almost surely we have Algorithm \ref{algorithm:nodewise-regression} yields a global solution $\widehat{\bs V}_{\cdot j}$ of \eqref{pseudo-likelihood} in at most $1 + \lceil \log (\kappa_{\max}^\circ) / \log(4)\rceil$ DC iterations when $n$ is sufficiently large, where $\lceil \cdot\rceil$ is the ceiling function. Moreover, almost surely we have
Algorithm \ref{algorithm:peeling} recovers $\mathcal E_+$ and $\mathcal I_+$ when $n$ is sufficiently large.
\end{theorem}

In view of Theorem \ref{theorem:consistent-structure-learning} {(proved in Appendix \ref{proof:consistent-structure-learning})}, it suffices to specify the maximum number of DC iterations as $1 + \lceil \log (\kappa_{\max}^\circ) / \log(4)\rceil$. Then the time complexity of Algorithm \ref{algorithm:nodewise-regression} is $p\times O(\log \kappa^\circ_{\max}) \times O(q^3+n q^2)$, where $O(q^3+n q^2)$ is that of solving a weighted Lasso \citep{efron2004least}. Note that Algorithm \ref{algorithm:peeling} does not involve heavy computation, so the overall time complexity for estimating the ARG (Algorithms \ref{algorithm:nodewise-regression}-\ref{algorithm:peeling}) is $O(p\times \log \kappa^\circ_{\max} \times (q^3+n q^2) )$.
Finally, the peeling method does not apply to observational data ($\bs W = \bm 0$). In a sense, interventions are essential.

Theorem \ref{theorem:consistent-structure-learning} establishes the consistent reconstruction by the peeling algorithm for the ARG. Yet, it does not provide any uncertainty measure for the presence of some directed edges in the true DAG. In what follows, we will develop an asymptotic theory for hypothesis tests concerning directed edges of interest.

\subsection{Inferential Theory}\label{section:inference-theory}

Given $H_0$, let $\mathcal S = (\bm Y,\bm X; \mathcal E_+\cup\mathcal D,\mathcal I_+)$.

\begin{assumption}\label{assumption:dimension}
$\max\limits_{\{j : \df_{\mathcal S}(j)\neq \emptyset\}} \big({|\pa_{\mathcal S}(j)| + {|\inter_{\mathcal S}(j)|}}\big)/{n} \leq \rho$ as $n \to \infty$, for a constant $0<\rho<1$.
\end{assumption}

Assumption \ref{assumption:dimension} is a hypothesis-specific condition restricting the underlying dimension of the testing problem. Usually, $|\pa_{\mathcal S}(j)| \asymp |\an_{\mathcal G}(j)| \asymp |\inter_{\mathcal S}(j)| \asymp \kappa_j^\circ \ll p$; $1\leq j \leq p$, which relaxes the condition $n\gg  p\log(p){\sqrt{|\mathcal D|}}$ for the constrained likelihood ratio test \citep{li2020likelihood}.

\begin{theorem}[Empirical p-values]
  \label{theorem:dp-null-distribution}
Suppose Assumptions \ref{assumption:identifiability}-\ref{assumption:dimension} are met and $H_0$ is regular.  
Assume the tuning parameters in Algorithm \ref{algorithm:nodewise-regression}
satisfy the requirements in Theorem \ref{theorem:consistent-structure-learning}.
  \begin{enumerate}[(A)]
    \item For the test of directed edges \eqref{equation:link-test}, 
    \begin{itemize}
        \item [] $\lim\limits_{
       \substack{n\to\infty\\
       \bm \theta \textnormal{ satisfies }H_0 \textnormal{ in } \eqref{equation:link-test}}
       } 
       \lim\limits_{M \rightarrow \infty} \PP_{\bm\theta}(\mathrm{Pval} < \alpha) = \alpha$, if $H_0$ is nondegenerate.
        \item [] $\lim\limits_{
       \substack{n\to\infty\\
       \bm \theta \textnormal{ satisfies }H_0 \textnormal{ in } \eqref{equation:link-test}}
       } 
      \lim\limits_{M \rightarrow \infty} \PP_{\bm\theta}(\mathrm{Pval} = 1) = 1$, if $H_0$ is degenerate.
    \end{itemize}
    \item For the test of directed pathways \eqref{equation:pathway-test},
    \begin{itemize}
        \item [] 
        $\limsup\limits_{
       \substack{n\to\infty\\
       \bm \theta \textnormal{ satisfies }H_0 \textnormal{ in } \eqref{equation:pathway-test}}
       }  
     \lim\limits_{M \rightarrow \infty} \PP_{\bm\theta}(\textnormal{Pval} < \alpha) = \alpha$,
     if $H_0$ is nondegenerate with $|\mathcal D|=|\mathcal H|$.
        \item [] $\limsup\limits_{
      \substack{n\to\infty\\
      \bm \theta \textnormal{ satisfies H}_0 \textnormal{ in } \eqref{equation:pathway-test}}
      }  
      \lim\limits_{M \rightarrow \infty} \PP_{\bm\theta}(\textnormal{Pval} = 1) = 1$, if $|\mathcal D|<|\mathcal H|$.
    \end{itemize}
  \end{enumerate}
\end{theorem} 

By Theorem \ref{theorem:dp-null-distribution} (proved in Appendix \ref{proof:dp-null-distribution}), the DP likelihood ratio test yields a valid p-value for \eqref{equation:link-test} and \eqref{equation:pathway-test} under appropriate conditions.
Note that $|\mathcal D|$ is permitted to depend on $n$. 
Moreover, Proposition \ref{proposition:asymptotic} (proved in Appendix \ref{proof:asymptotic}) summarizes the asymptotics for directed edge test \eqref{equation:link-test}. 

\begin{proposition} [Asymptotics of edge test] \label{proposition:asymptotic}
  Suppose the assumptions of Theorem \ref{theorem:dp-null-distribution} are met. Under a nondegenerate and regular $H_0$, as $n\to\infty$, 
  \begin{enumerate}[(A)]
      \item $2 \textnormal{Lr}  \dto \chi^2_{|\mathcal D|}$, if $|\mathcal D|>0$ is fixed.
      \item $(2 \textnormal{Lr} - |\mathcal D|)/\sqrt{2|\mathcal D|} \dto N(0,1)$, if $|\mathcal D|\log (|\mathcal D|)/{n} \rightarrow 0$.
  \end{enumerate}
\end{proposition}

\begin{remark}\label{remark:local-structures}
  As opposed to the entire $\mathcal G_+$ (or $\mathcal S$), Theorem \ref{theorem:dp-null-distribution} and Proposition \ref{proposition:asymptotic} only require correct identification of the local structures $\big\{\an_{\mathcal G_+}(j),\inter_{\mathcal G_+}(j),\df_{\mathcal S}(j)\big\}_{\{ j : \df_{\mathcal S}(j)\neq\emptyset \}}$. This requirement can be reasonably satisfied when the sample size is moderately large, as illustrated in Section \ref{section:simulation}.
\end{remark}

Next, we analyze the local limit power of the proposed tests for \eqref{equation:link-test} and \eqref{equation:pathway-test}. 
Assume $\bm \theta^\circ = (\bs U^\circ,\bs W^\circ)$ satisfies $H_0$. 
Let $\bm \Delta\in\mathbb R^{p\times p}$ satisfy
$\bm \Delta_{\mathcal D^c}=\bm 0$ so that $\bs U^\circ+\bm \Delta$ represents a DAG. 
For nondegenerate and regular $H_0$, consider an alternative $H_a$: $\bs U_{\mathcal H} = \bs U_{\mathcal H}^\circ + \bm \Delta_{\mathcal H}$,
and define the power function as 
\begin{equation}
  \beta(\bm\theta^\circ,\bm \Delta) = \PP_{H_a}( \text{Pval} < \alpha ).
\end{equation}

\begin{proposition}[Local power of edge test]
\label{proposition:power-edge}
Suppose $H_0$ is nondegenerate and regular. 
Let $\| \bm \Delta\|_{F} = 
\|\bm \Delta_{\mathcal H}\|_{F} = n^{-1/2} \delta$ when $|\mathcal D| > 0$ is fixed 
and $\|\bm \Delta\|_{F} = |\mathcal D|^{1/4} n^{-1/2}h$ when $|\mathcal D|\to\infty$, 
where $\delta>0$ and $\|\cdot\|_{F}$ is the matrix Frobenius norm. 
If the assumptions of Theorem \ref{theorem:dp-null-distribution} are met, then under $H_a$, as $n,M\to\infty$,
\begin{equation*}
    \beta(\bm\theta^\circ,\bm \Delta) \geq
    \begin{cases}
      \PP_{\bm Z\sim N(\bm 0,\bs I_{|\mathcal D|\times|\mathcal D|})}\Big(\|\bm Z + c_l\sqrt{n}\bm \Delta \|_2^2 > \chi^2_{|\mathcal D|,1-\alpha} \Big) & \text{if } |\mathcal D|>0 \text{ is fixed},\\
      \PP_{Z\sim N(0,1)}\Big(Z > z_{1-\alpha} - {c_l \|\bm \Delta\|^2_2}/{\sqrt{2|\mathcal D|}} \Big) & \text{if } |\mathcal D|\to\infty, \frac{|\mathcal D|\log|\mathcal D|}{n}\to 0,
    \end{cases}
\end{equation*}
where $\chi^2_{|\mathcal D|,1-\alpha}$ and $z_{1-\alpha}$ denote the $(1-\alpha)$th quantile of distributions $\chi^2_{|\mathcal D|}$ and $N(0,1)$, respectively.
Hence, $\lim\limits_{\delta\to\infty} \lim\limits_{n\to\infty} \beta(\bm\theta^\circ,\bm \Delta) = 1$.
\end{proposition}

\begin{proposition} [Local power of pathway test]
  \label{proposition:power-path}
Suppose $H_0$ is nondegenerate and regular with $|\mathcal D| = |\mathcal H|$. 
Let $\min_{(k,j)\in \mathcal H}|\mathrm U^\circ_{kj} + \Delta_{kj}| = n^{-1/2}\delta$ when $|\mathcal H|>0$ is fixed and  $\min_{(k,j)\in \mathcal H}|\mathrm U^\circ_{kj} + \Delta_{kj}| = n^{-1/2}\delta\sqrt{\log|\mathcal H|}$ when $|\mathcal H|\to\infty$.
If the assumptions of Theorem \ref{theorem:dp-null-distribution} are met, then under $H_a$, as $n,M\to\infty$,
  \begin{equation*}
      \beta(\bm\theta^\circ,\bm \Delta) \geq 1 - \frac{|\mathcal H|}{\sqrt{2\pi}} 
      \exp\Big( -\frac{1}{2}\Big(\delta\sqrt{\log|\mathcal H|}/\max_{1\leq j\leq p}\Omega_{jj} - \sqrt{\chi^2_{1,1-\alpha}}\Big)^2   \Big),
  \end{equation*}
  where 
  $\chi^2_{1,1-\alpha}$ is the $(1-\alpha)$th quantile of distribution $\chi^2_{1}$.
  Hence, $\lim\limits_{\delta\to\infty} \lim\limits_{n\to\infty} \beta(\bm\theta^\circ,\bm \Delta) = 1$.
\end{proposition}

{The proofs of Propositions \ref{proposition:power-edge} and \ref{proposition:power-path} are deferred to Appendix \ref{proof:power-edge} and \ref{proof:power-path}.}

\section{Simulations}\label{section:simulation}

This section investigates the operating characteristics of the proposed tests and the peeling algorithm via simulations.  In simulations, we consider two setups for generating $\bs U\in\mathbb{R}^{p\times p}$, representing random and hub DAGs, respectively.
\begin{itemize}
  \item \textbf{Random graph.} The upper off-diagonal entries $\mathrm U_{kj}$; $k<j$ are sampled independently from $\{0,1\}$ according to Bernoulli$(1/p)$, while other entries are zero.
  This generates a random graph with a sparse neighborhood. 

  \item \textbf{Hub graph.} Set $\mathrm U_{1,2j+1} = 1$ and $\mathrm U_{2,2j+2} = 1$ for $j= 1,\ldots, \lfloor p/2\rfloor -2$, while other entries are zero. This generates a hub graph, where nodes 1 and 2 are hub nodes with a dense neighborhood.
\end{itemize}

Moreover, we consider three setups for intervention matrix $\bs W\in\mathbb{R}^{q\times p}$, representing different scenarios. Setups (A) and (B) are designed for inference, whereas
Setup (C) in Section \ref{section:structure-learning-simulation} is designed to compare with the method of 
\citet{chen2018two} for structure learning.
Let $\bs W = (\bs A^\top, \bs B^\top, \bm 0^\top)^\top$,
where $\bs A,\bs B\in\mathbb{R}^{p\times p}$ and $\bm 0\in\mathbb{R}^{(q-2p)\times p}$.
\begin{itemize}
  \item \textbf{Setup (A).} Set $\mathrm A_{jj} = \mathrm B_{jj} = \mathrm B_{j,j+1} = 1$; $j = 1,\ldots,p-1$, $\mathrm A_{pp}= 1$, while other entries of $\bs A$, $\bs B$ are zero. Then, $X_1,\ldots,X_p$ are instruments for $Y_1,\ldots,Y_p$, respectively, $X_{p+1},\ldots,X_{2p-1}$ are invalid instruments with two targets, and $X_{2p},\ldots,X_q$ represent inactive interventions. 

  \item \textbf{Setup (B).} Set $\mathrm A_{jj} = \mathrm A_{j,j+1} = \mathrm B_{jj} = \mathrm B_{j,j+1}=1$; $j=1,\ldots,p-1$, $\mathrm A_{pp} = 1$, while other entries of $\bs A$, $\bs B$ are zero. Here, the only valid instrument is $X_{p}$ on $Y_p$, and the other intervention variables either have two targets or are inactive. Importantly, Assumption 1C is not met.
    
\end{itemize}

To generate $(\bm Y,\bm X)$ for each setup, we sample $\bm X\sim N(\bm 0,\bm\Sigma_X)$ with $(\Sigma_X)_{ll'} = 0.5^{|l-l'|}$; $1\leq l,l'\leq q$ and sample $\bm Y$ according to \eqref{equation:model} with $(\bs U,\bs W,\sigma_1^2,\ldots,\sigma_p^2)$, where $\sigma_1^2,\ldots,\sigma_p^2$ are set to be equally spaced from $0.5$ to $1$.

\subsection{Inference}\label{section:inference-simulation}

We compare three tests in empirical type-I errors and powers in simulated examples, 
namely, the DP likelihood ratio test (DP-LR) in Algorithm \ref{algorithm:data-perturbation},
the asymptotic likelihood ratio test (LR), and the oracle likelihood ratio test (OLR). Here LR uses $\text{Lr}$, while 
OLR uses $\text{Lr}(\mathcal S,\widehat{\bm\Sigma})$ assuming that $\mathcal S$ were known in advance.
The p-values of LR and OLR are computed via Proposition \ref{proposition:asymptotic}. 
The implementation details of these tests are in Appendix \ref{section:implementation}.

For the empirical type-I error of a test, we compute the percentage of times 
rejecting $H_0$ out of 500 simulations when $H_0$ is true. 
For the empirical power of a test, we report the percentage of times
rejecting $H_0$ out of 100 simulations when $H_a$ is true under alternative hypotheses $H_a$. 
\begin{itemize}
  \item \textbf{Test of directed edges.} 
  For \eqref{equation:link-test}, we examine two different hypotheses: 
  
  (i) $H_0: \mathrm U_{1, 20} = 0$ versus $H_a: \mathrm U_{1,20} \neq  0$. In this case, $|\mathcal D|=1$.
  
  (ii) $H_0: \bs U_{\mathcal H} = \bm 0$ versus $H_a: \bs U_{\mathcal H} \neq  \bm 0$, where $\mathcal{H} = \{ (k,20): k=1,\ldots,15 \}$. In this case, $|\mathcal D|=15$.

  Moreover, five alternatives 
  $H_a: \mathrm U_{1,20} = 0.1 l$ and $\bs U_{\mathcal H\setminus\{ (1,20)\}}= \bm 0$; $l=1,2,3,4,5$,
  are used for the power analysis in (i) and (ii).
  The data are generated by modifying $\bs U$ accordingly.

  \item \textbf{Test of directed pathways.} 
  We test the directed path 
  $Y_1\to Y_5\to Y_{10}\to Y_{15}\to Y_{20}$, namely
  $\mathcal H = \{ (1,5),(5,10),(10,15),(15,20) \}$ in \eqref{equation:pathway-test}.
  Since \eqref{equation:pathway-test} is a test of composite null hypothesis, the data are generated under a graph with parameters $(\bs U,\bs W,\bm \Sigma)$ satisfying $H_0$, 
  where $\bs U_{\mathcal H} = \bm 0$. Five hypotheses $H_a: \bs U_{\mathcal H} = \bm{0.1} l $; $l=1,2,3,4,5$ are used for the power analysis.
\end{itemize}

\begin{figure}
  \centering
  \includegraphics[width=\textwidth]{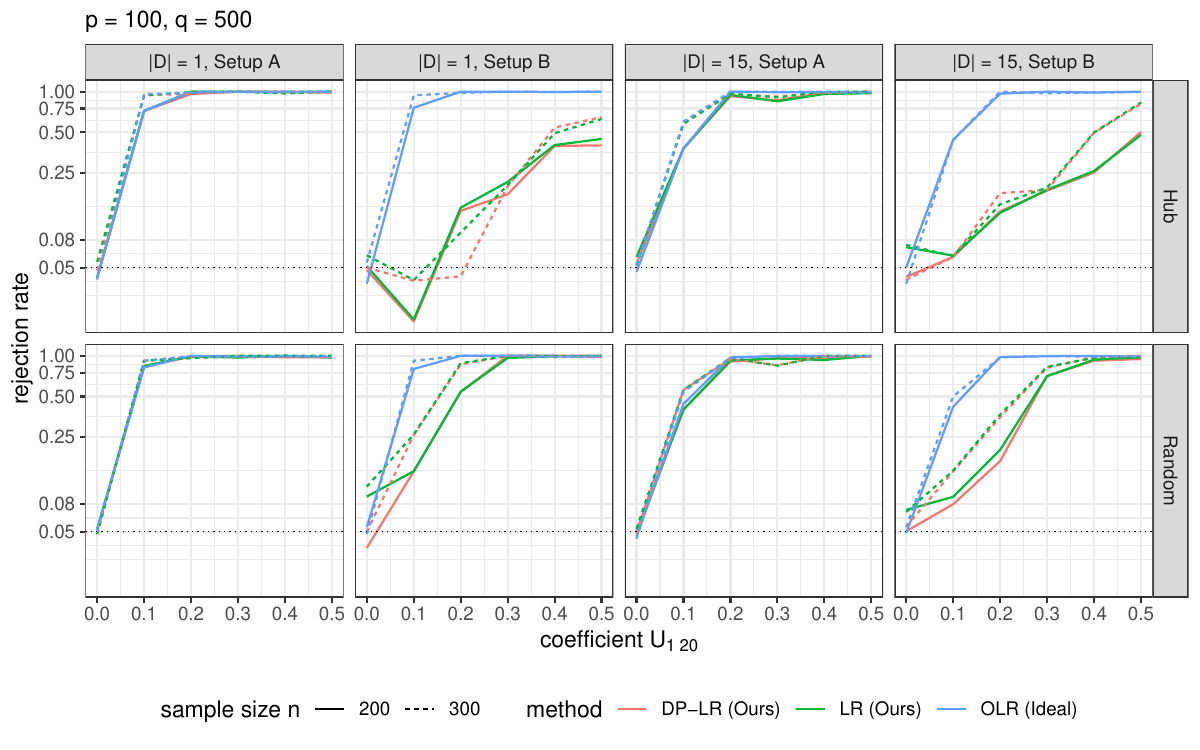}
  \caption{Empirical type-I errors and powers of tests of directed edges.
  The black dotted line marks the nominal level of significance $\alpha=0.05$.}
  \label{figure:edge-inference}
\end{figure}

\begin{figure}
  \centering
  \includegraphics[width=\textwidth]{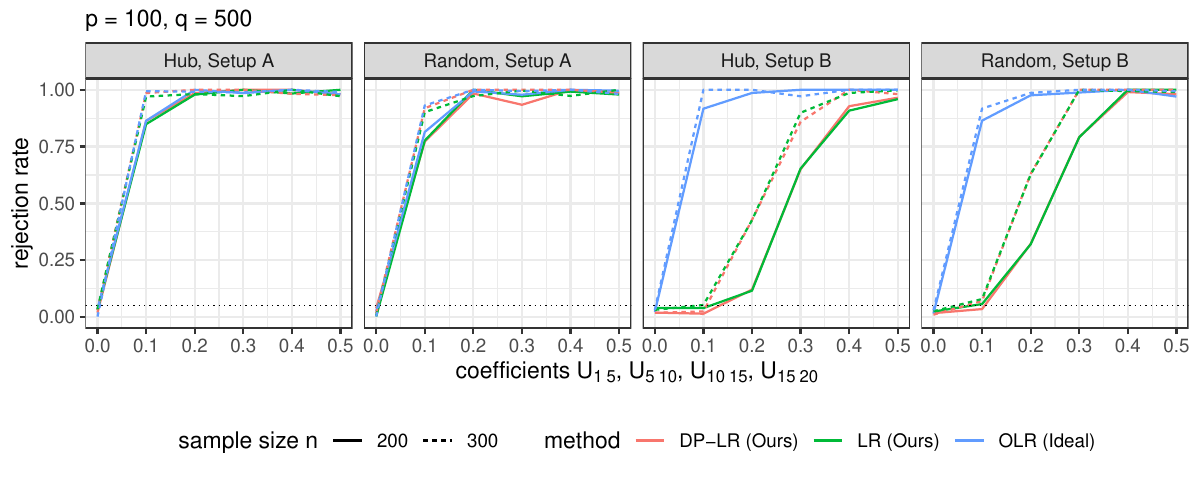}
  \caption{Empirical type-I errors and powers of tests of a directed pathway.
  The black dotted line marks the nominal level of significance $\alpha=0.05$.}
  \label{figure:path-inference}
\end{figure}

For testing directed edges, as displayed in Figure \ref{figure:edge-inference}, DP-LR and LR perform well compared to the ideal test OLR in Setup (A) with Assumption \ref{assumption:identifiability} satisfied. In Setup (B) with Assumption 1C not fulfilled, DP-LR appears to have control of 
type-I error, whereas LR has an inflated empirical type-I error compared to
the nominal level $\alpha = 0.05$. This discrepancy is attributed to the data
perturbation scheme accounting for the uncertainty of identifying 
$S$. However, both suffer a loss of power compared to the oracle test OLR in this 
setup. This observation suggests that without Assumption 1C, the peeling algorithm 
tends to yield an estimate $\widehat{\mathcal G}_+\supseteq \mathcal G_+$, which overestimates $\mathcal G_+$, resulting in a power loss. 

For testing directed pathways, as indicated in Figure \ref{figure:path-inference},
we observe similar phenomena as in the previous directed edge tests.
Of note, both LR and DP-LR are capable in controlling type-I error of directed path tests.

In summary, DP-LR has a suitable control of type-I error when there are invalid instruments and Assumption 1C is violated. 
Concerning the power, DP-LR and LR are comparable in all scenarios and 
their powers tend to one as the sample size $n$ or the signal strength 
of tested edges increases. Moreover, DP-LR and LR perform nearly as well
as the oracle test OLR when Assumption 1 is satisfied. These empirical findings
agree with our theoretical results.

\subsection{Structure Learning}\label{section:structure-learning-simulation}

This subsection compares the peeling algorithm with the Two-Stage Penalized Least Squares (2SPLS, \citet{chen2018two}) in terms of the structure learning accuracy. For peeling, we consider Algorithm \ref{algorithm:peeling} with an additional step \eqref{directed-likelihood} for structure learning of $\bs U$. 
For 2SPLS, we use the R package \texttt{BigSEM}.

\begin{figure}
  \centering
  \includegraphics[width=.9\textwidth]{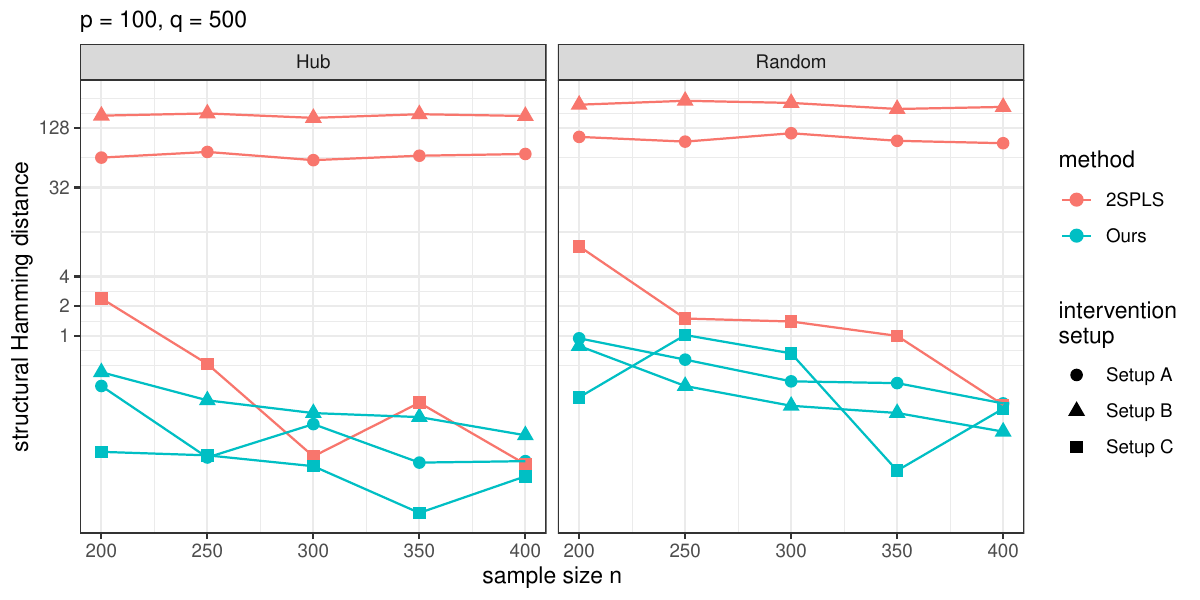}
  \caption{SHDs for the reconstructed DAG by the peeling algorithm 
and 2SPLS, where a smaller value of SHD indicates a better result.}
  \label{figure:structure-learning}
\end{figure}

2SPLS requires that all the intervention variables to be target-known instruments in addition to Assumption 1C. Thus, we consider an additional Setup (C).
\begin{itemize}
  \item \textbf{Setup (C).} Let $\bs W = (\bs I_{p\times p},\bm 0)^\top\in\mathbb{R}^{q\times p}$. Then $X_1,\ldots,X_p$ are valid instruments for $Y_1,\ldots,Y_p$, respectively, and other intervention variables are inactive.
\end{itemize}
For 2SPLS, we assign each active intervention variable to its most correlated primary variable in Setups (A)-(C). 
In Setup (C), this assignment yields a correct identification of valid instruments, meeting all the requirements of 2SPLS.

For each scenario, we compute the structural Hamming distance (SHD)
\begin{equation*}
    \text{SHD}(\widehat{\bs U}, \bs U) = \sum_{k,j}|\I(\widehat{\mathrm U}_{kj}\neq 0)-\I(\mathrm U_{kj}\neq 0)|,
\end{equation*}
averaged over 100 runs. As shown in Figure \ref{figure:structure-learning}, the peeling algorithm outperforms 2SPLS, especially when there are invalid instruments and Assumption 1C is violated. 

Appendix \ref{section:additional-simulation} contains additional numerical experiments on structure learning, including the results of different sparsity settings, SHD transition curves, and different numbers of interventions.

\subsection{Comparison of Inference and Structure Learning}\label{section:comparison-simulation}

This subsection compares the proposed DP testing method against the proposed structure learning method in \eqref{directed-likelihood} in terms of inferring the true graph structure.
To this end, we consider Setup (A) in Section \ref{section:inference-simulation} with $p = 30, q = 100$, and the hypotheses 
\begin{center}
  $H_0: \mathrm U_{1,20} = 0$ versus $H_a: \mathrm U_{1,20} = 1/\sqrt{n}$.
\end{center}
For DP inference, we use $\alpha = 0.05$ and choose the tuning parameters by BIC as in previous experiments; see Appendix \ref{section:implementation} for details. 
For structure learning, we reject the null hypothesis when $\widehat{\mathrm U}_{1,20}\neq 0$. 

\begin{figure}[H]
  \centering
  \includegraphics[width=.9\textwidth]{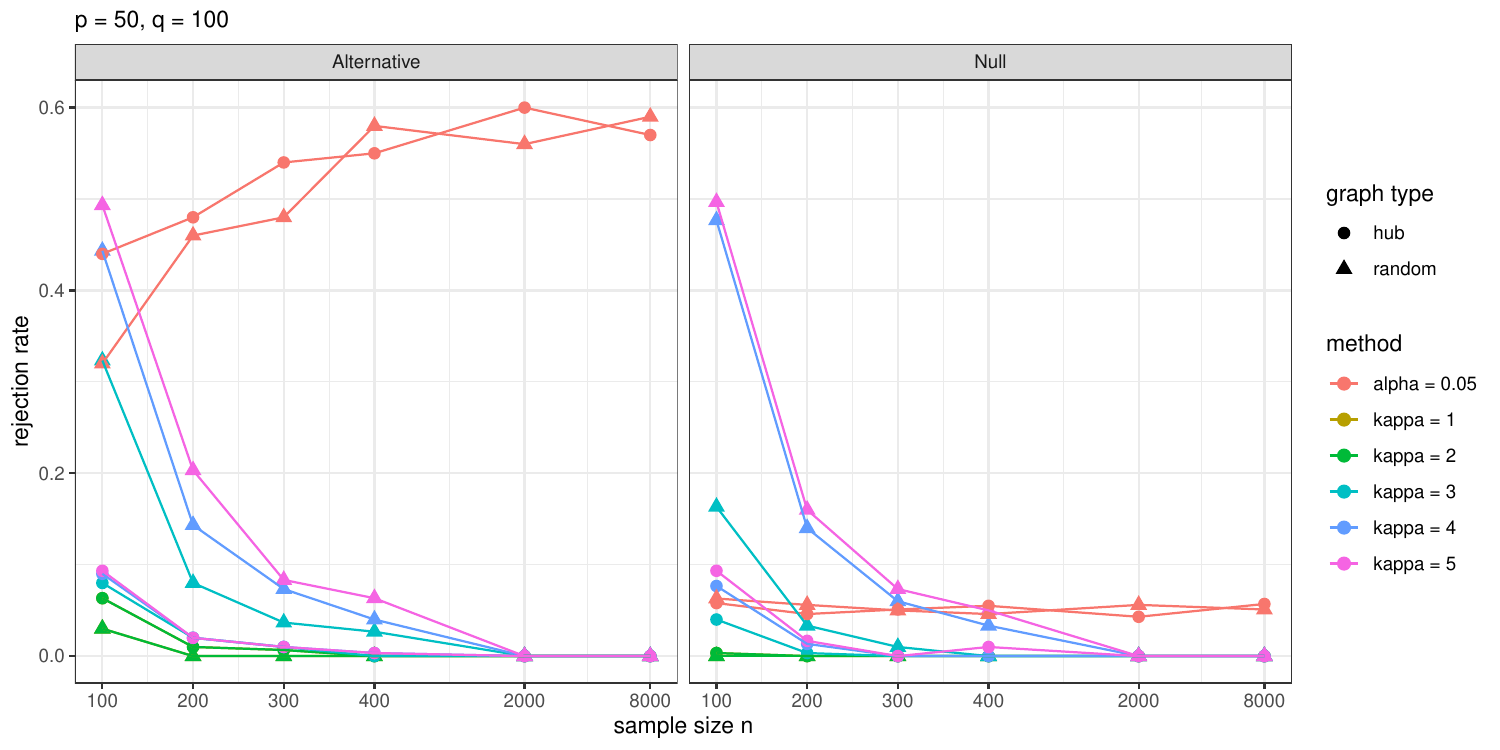}
  \caption{{Rejection rates for $H_0:\mathrm U_{1,20}=0$ versus $H_a:\mathrm U_{1,20}=1/\sqrt{n}$ by DP inference and structure learning. For structure learning, $H_0$ is rejected if $\widehat{\mathrm U}_{1,20}\neq 0$. The red lines indicate the results of DP inference using the significance level $\alpha=0.05$. The other colored lines display the results of structure learning using different sparsity parameter values $\kappa = 1, 2, 3, 4, 5$.
  The simulation is repeated for 500 times and $\kappa = 2$ is chosen by BIC in over 90\% cases.}}
  \label{figure:comparison-inference-learning}
\end{figure}

As displayed in Figure \ref{figure:comparison-inference-learning},
when the null hypothesis $H_0$ is true, the DP testing method controls type-I error very close to the nominal level of 0.05, whereas the type-I error of the structure learning varies greatly depending on the tuning parameter selection. Under the alternative hypothesis $H_a$, the DP inference enjoys high statistical power than structure learning methods when $n\geq 200$. Interestingly, the power of structure learning diminishes as $n$ increases. This observation is in agreement with our theoretical results in Theorem \ref{theorem:consistent-structure-learning-U}, suggesting that consistent reconstruction requires the smallest size of nonzero coefficients to be of order $\gtrsim\sqrt{\log(n)/n}$ with the tuning parameter $\tau$ of the same order (fixing $p,q$).
In this case, the edge $\mathrm U_{1,20}$ is of order $1/\sqrt{n}$, which is less likely to be reconstructed as $n$ increases. In contrast, Proposition \ref{proposition:power-edge} indicates that a DP test has a non-vanishing power when the hypothesized edges are of order $1/\sqrt{n}$.

Figure \ref{figure:comparison-inference-learning} demonstrates some important distinctions between inference and structure learning. When different tuning parameters are used, the structure learning results correspond to different points on an ROC curve. Although it is asymptotically consistent when optimal tuning parameters are used, structure learning lacks an uncertainty measure of graph structure identification. As a result, it is nontrivial for structure learning methods to trade-off the false discovery rate and detection power in practice. This makes the interpretation of such results hard, especially when they heavily rely on hyperparameters as in Figure \ref{figure:comparison-inference-learning}. By comparsion, DP inference aims to maximize {statistical power while controlling} type-I error at a given level, offering a clear interpretation of its result. 
This observation agrees with the discussions in the literature on variable selection and inference \citep{wasserman2009high,meinshausen2010stability,lockhart2014significance,candes2018panning} and it justifies the demand for inferential tools for directed graphical models.

\section{ADNI Data Analysis}\label{section:real-data}

This section applies the proposed tests to analyze an Alzheimer's Disease Neuroimaging Initiative (ADNI) dataset. 
In particular, we infer gene pathways related to Alzheimer's Disease (AD) to highlight some gene-gene interactions differentiating patients with AD/cognitive impairments and healthy individuals.

The raw data are available in the ADNI database (\url{https://adni.loni.usc.edu}), including gene expression, whole-genome sequencing, and phenotypic data. After cleaning and merging, we have a sample size of 712 subjects. 
From the KEGG database \citep{kanehisa2000kegg}, we extract the AD reference pathway (hsa05010, \url{https://www.genome.jp/pathway/hsa05010}), including 146 genes in the ADNI data.

For data analysis, we first regress the gene expression levels on five covariates 
-- {gender}, {handedness}, {education level}, {age}, and {intracranial volume}, and 
then use the residuals as gene expressions in the following analysis. 
Next, we extract the genes with at least one SNP at a marginal significance level below $10^{-3}$, yielding $p=63$ genes as primary variables.
For these genes, we further extract their marginally most correlated two SNPs, resulting in $q = 63\times2 = 126$ SNPs as unspecified intervention variables for subsequent data analysis.
All gene expression levels are normalized. 

The dataset contains individuals in four groups, namely, Alzheimer's Disease (AD), Early Mild Cognitive Impairment (EMCI), Late Mild Cognitive Impairment (LMCI), and Cognitive Normal (CN). 
For our purpose, we treat 247 CN individuals as controls while the remaining 465 individuals as cases (AD-MCI). Then, we use the gene expressions and the SNPs to reconstruct the ancestral relations and infer gene pathways for 465 AD-MCI and 247 CN control cases, respectively.

\begin{figure}
  \centering
  \subfloat[AD-MCI]{\includegraphics[width=0.3\textwidth]{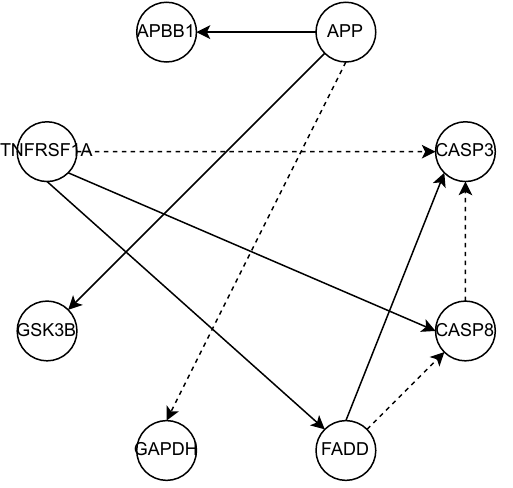}\label{figure:ad-mci-subnet}}
  \hfill
  \subfloat[CN]{\includegraphics[width=0.3\textwidth]{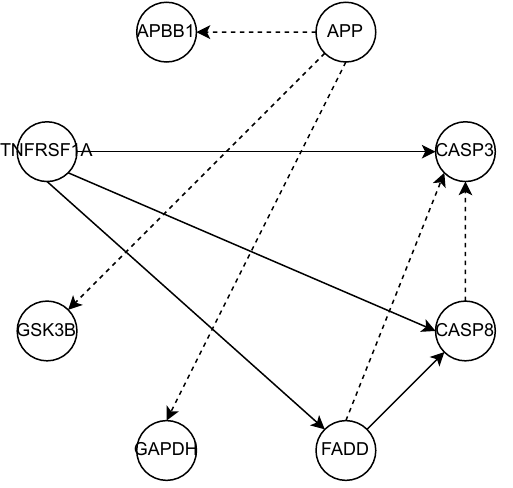}\label{figure:cn-subnet}}\\
  \subfloat[AD-MCI]{\includegraphics[width=0.3\textwidth]{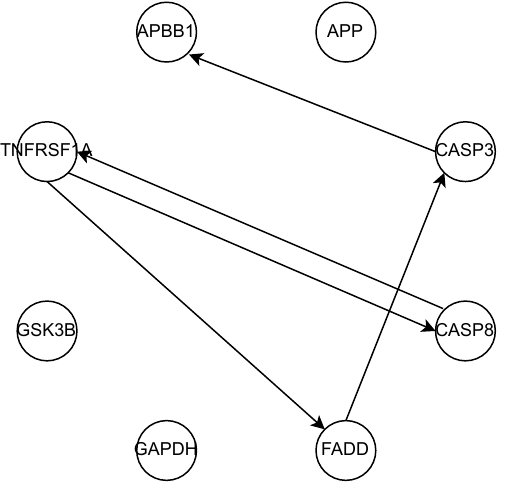}\label{figure:2sls-ad-mci-subnet}}
  \hfill
  \subfloat[CN]{\includegraphics[width=0.3\textwidth]{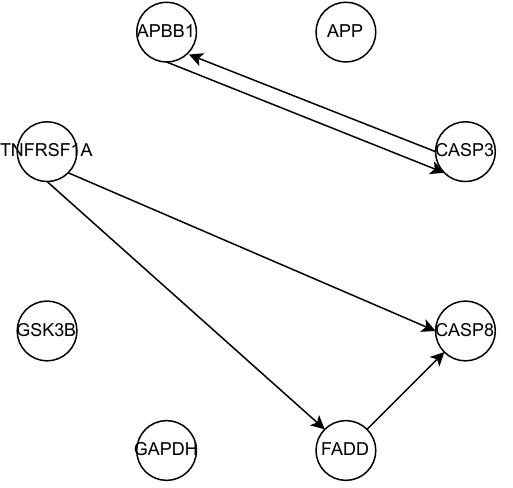}\label{figure:2sls-cn-subnet}}
  \caption{
    Display of the subnetworks associated with genes APP and CASP3.
    (a) and (b): Solid/dashed arrows indicate significant/insignificant edges at $\alpha=0.05$
    after adjustment for multiplicity by the Bonferroni-Holm correction.
    (c) and (d): Solid arrows indicate the reconstructed edges using 2SPLS \citep{chen2018two}.}
  \label{figure:network}
\end{figure}

\begin{figure}
  \centering
  \subfloat[AD-MCI]{\includegraphics[width=0.3\textwidth]{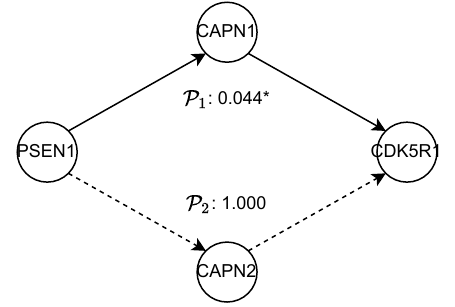}\label{figure:ad-mci-path}}
  \hfill
  \subfloat[CN]{\includegraphics[width=0.3\textwidth]{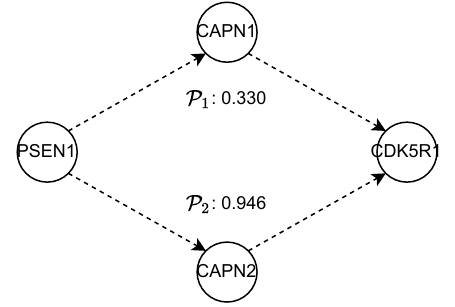}\label{figure:cn-path}}
  \caption{The p-values of pathway tests \eqref{equation:pathway-test} by the proposed tests for the AD-MCI and CN groups, where p-values are adjusted for multiplicity by the Bonferroni-Holm correction and solid/dashed arrows indicate significant/insignificant pathways at $\alpha=0.05$.}
  \label{figure:PSEN1}
\end{figure}

In the literature, genes APP, CASP3, and PSEN1 are well-known to be associated with AD, reported to play different roles in AD patients and healthy subjects \citep{julia2017genetics,su2001activated,kelleher2017presenilin}.
For this dataset, we conduct hypothesis testing on edges and pathways related to genes APP, CASP3, and PSEN1 in the KEGG AD reference (hsa05010) to evaluate the proposed DP inference by checking if DP inference can discover the differences that are reported in the biomedical literature.
First, we consider testing  
$H_0: \mathrm U_{kj}=0$ versus 
$H_a: \mathrm U_{kj}\neq0$,
for each edge $(k,j)$
as shown in Figure \ref{figure:network} (a) and (b).
Moreover, we consider two hypothesis tests of pathways
$H_0: \mathrm U_{kj}=0$ for some $(k,j)\in\mathcal{P}_\ell$ versus 
$H_a: \mathrm U_{kj}\neq 0$ for all $(k,j)\in\mathcal{P}_\ell$; $\ell=1,2$,
where the two pathways are specified by 
$\mathcal{P}_1 = \{ \text{PSEN1}\to\text{CAPN1}\to\text{CDK5R1} \}$,
and $\mathcal{P}_2 = \{ \text{PSEN1}\to \text{CAPN2}\to\text{CDK5R1} \}$.
See Figure \ref{figure:PSEN1}.
Of note, for clear visualization, Figure \ref{figure:network} (a)-(b), and Figure \ref{figure:PSEN1} only display the edges related to hypothesis testing. Also notice that the ancestral relations are reconstructed using $p=63$ genes and $q = 126$ SNPs for AD-MCI and CN groups separately.

\begin{figure}
  \centering
  \includegraphics[width=0.7\textwidth]{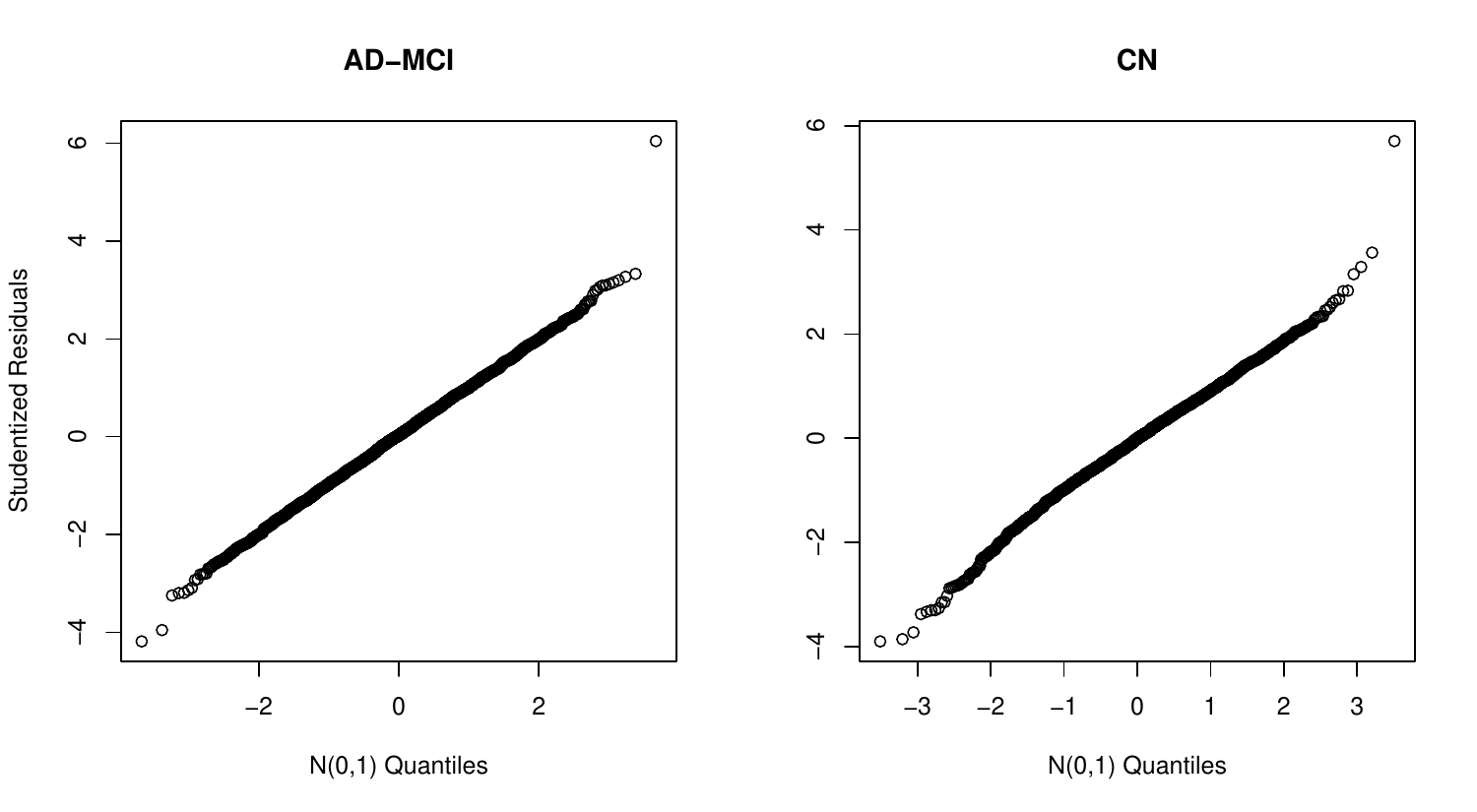}
  \caption{Normal quantile-quantile plots of studentized residuals of the AD-MCI and CN groups. }
  \label{figure:residuals}
\end{figure}

{In Figures \ref{figure:network}-\ref{figure:PSEN1}, 
the significant results under the level $\alpha=0.05$ after the Holm-Bonferroni adjustment for 
$2\times (9 + 2)= 22$ tests
are displayed.} In Figures \ref{figure:network},
the edge test in \eqref{equation:link-test} exhibits a strong evidence for the 
presence of directed connectivity $\{$APP $\to$ APBB1, APP $\to$ GSK3B, FADD $\to$ CASP3$\}$ 
in the AD-MCI group, but no evidence in the CN group. 
Meanwhile, this test suggests the presence of connections 
$\{$TNFRSF1A $\to$ CASP3, FADD $\to$ CASP8$\}$ in the CN group but not so in the AD-MCI group. 
In both groups, we identify directed connections $\{$TNFRSF1A $\to$ FADD, TNFRSF1A $\to$ CASP8$\}$. 
In Figure \ref{figure:PSEN1},
the pathway test \eqref{equation:pathway-test} supports the presence of a pathway 
PSEN1 $\to$ CAPN1 $\to$ CDK5R1 in the AD-MCI group with a p-value of $0.044$ 
but not in the CN group with a p-value of $0.33$. 
The pathway PSEN1 $\to$ CAPN2 $\to$ CDK5R1 appears insignificant at $\alpha=0.05$ for
both groups. Also noted is that some of our discoveries agree with the literature
according to the AlzGene database (\url{alzgene.org}) and the AlzNet database (\url{https://mips.helmholtz-muenchen.de/AlzNet-DB}).
Specifically, GSK3B differentiates AD patients from normal subjects;
as shown in Figures \ref{figure:network},
our result indicates the presence of connection APP $\to$ GSK3B for the AD-MCI group,
but not for the CN group, the former of which is confirmed by 
Figure 1 of \citet{kremer2011gsk3}.
The connection APP $\to$ APBB1 also differs in AD-MCI and CN groups, 
which appears consistent with Figure 3 of \citet{bu2009apolipoprotein}.
Moreover, the connection CAPN1 $\to$ CDK5R1, in the pathway PSEN1 $\to$ CAPN1 $\to$ CDK5R1 discovered in AD-MCI group, 
is found in the AlzNet database (interaction-ID 24614, \url{https://mips.helmholtz-muenchen.de/AlzNet-DB/entry/show/1870}).
Finally, as suggested by Figure \ref{figure:residuals}, the normality assumption in \eqref{equation:model} is adequate for both groups.

By comparison, as shown in Figure \ref{figure:network} (c) and (d), gene APP in the reconstructed networks by 2SPLS \citep{chen2018two} is not connected with other genes, 
{indicating no regulatory relation of APP with other genes in the AD-MCI and CN groups. However, as a well-known gene associated with AD, APP is reported to play different roles in controlling the expressions of other genes for AD patients and healthy people \citep{matsui2007expression,julia2017genetics}. Our results in Figure \ref{figure:network} (a) and (b) are congruous with the studies: the connections of APP with other genes are different in our estimated networks for AD-MCI and CN groups.}

In summary, our findings seem to agree with those in the literature \citep{julia2017genetics,su2001activated,kelleher2017presenilin}, {where the subnetworks of genes APP, CASP3 in Figure \ref{figure:network} and PSEN1 in  Figure \ref{figure:PSEN1} differentiate the AD-MCI from the CN groups}. Furthermore, the pathway PSEN1 $\to$ CAPN1 $\to$ CDK5R1 in Figure \ref{figure:PSEN1} seems to differentiate these groups, which, however, requires validation in biological
experiments.

\section{Summary}\label{section:conclusion}

This article proposes structure learning and inference methods for a Gaussian DAG with interventions, where the targets and strengths of interventions are unknown.
A likelihood ratio test is derived based on an estimated ARG formed by ancestral relations and candidate interventional relations. 
This test accounts for the statistical uncertainty of the construction of the ARG based on a novel data perturbation scheme. 
Moreover, we develop a peeling algorithm for the ARG construction. 
The peeling algorithm allows scalable computing and yields a consistent estimator.
The numerical studies justify our theory and demonstrate the utility of our methods.

The proposed methods can be extended to many practical situations beyond biological applications with independent and identically distributed data. An instance is to infer directed relations between multiple autoregressive time series \citep{pamfil2020dynotears}, where the lagged variables and covariates can serve as interventions for each time series.

The current work has two limitations. First, the inferential theory requires (asymptotically) correct recovery of the local DAG structures (Remark \ref{remark:local-structures}) to produce valid p-values, similar to \citet{shi2019linear} and \citet{zhu2020high}. As illustrated in numerical studies, the graph structures are reasonably recovered when $n$ is moderately large, and the DP scheme empirically alleviates the issue of inference after the ARG reconstruction. 
However, whether valid p-values can be obtained without the exact reconstruction of nuisance graph structures remains unclear in theory. Second, the proposed methods do not treat hidden confounding, which often arises in practice and can bias the results of both inference and learning. One future research direction is to extend the framework of unspecified interventions to allow unmeasured confounders.


\acks{The authors would like to thank the action editor and three anonymous reviewers for their helpful comments and suggestions. The research is supported by NSF grants DMS-1712564, DMS-1952539, and NIH grants R01GM126002, R01HL116720, R01HL105397, R01AG069895, R01AG065636, R01AG074858, and U01AG073079.}


\appendix

\section{Illustrative Examples and Discussions}\label{section:illustrative-example}

\subsection{Identifiability of Model \texorpdfstring{\eqref{equation:model}}{(1)} and Assumption \ref{assumption:identifiability}} \label{section:identifiability-example}

The parameter space for model \eqref{equation:model} is  
\begin{equation*}
  \{ (\bs U,\bs W,\bm\Sigma) :
    \bs U\in \mathbb{R}^{p\times p} \text{ represents a DAG}, \ 
    \bs W\in \mathbb{R}^{p\times q}, \
    \bm \Sigma = \text{diag}(\sigma_1^2,\ldots,\sigma^2_p) \}.
\end{equation*}
As suggested by Proposition \ref{proposition:identifiability}, 
Assumption \ref{assumption:identifiability} (1A-1C) suffices for identification of every parameter value
in the parameter space. Next, we show by examples
that if Assumption 1B or 1C is violated 
then model \eqref{equation:model} is no longer identifiable. 
{To proceed, we rewrite \eqref{equation:model} as
\begin{equation*}
\bm Y = \bs V^\top \bm X + \bm \varepsilon_V, \quad \bm \varepsilon_V=(\bs I-\bs U^\top)^{-1}\bm \varepsilon \sim N(\bm 0,\bm\Omega^{-1}),
\end{equation*}
where $\bm\Omega= (\bs I - \bs U)\bm\Sigma^{-1}(\bs I -\bs U^\top)$ is a precision matrix 
and $\bs V = \bs W (\bs I - \bs U)^{-1}$.}

\begin{figure}
  \centering
  \includegraphics[width=0.8\textwidth]{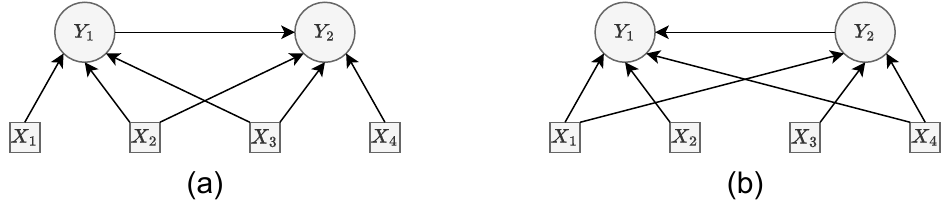}
  \caption{(a) Display of DAG defined by \eqref{example-appendix:1B-SEM1}.
(b) Display of DAG defined by \eqref{example-appendix:1B-SEM2}.}
\label{example-appendix:identifiability-1B}
\end{figure}

\begin{example}[Identifiability]\label{example:identifiability}
  In model \eqref{equation:model}, consider
two non-identifiable bivariate situations: (1) $p=2$ and $q=4$
and (2) $p=2$ and $q=2$.  
\begin{enumerate}
  \item[(1)] \textit{Model \eqref{equation:model} is non-identifiable when Assumption 
1B breaks down.}
Consider two different models with different parameter values:
  \begin{eqnarray}
    \bm \theta:  &Y_1 = X_1 + X_2 + X_3 + \varepsilon_1, \quad
      &Y_2 = Y_1 - X_2 + X_3 + X_4 + \varepsilon_2, \label{example-appendix:1B-SEM1} \\
    \widetilde{\bm \theta}: 
      &Y_1 = 0.5 Y_2 + 0.5 X_1 + X_2 - 0.5X_4 + \widetilde{\varepsilon}_1, \quad
      &Y_2 = X_1 + 2X_3 + X_4 + \widetilde{\varepsilon}_2, \label{example-appendix:1B-SEM2}
  \end{eqnarray}
  where $\varepsilon_1,\varepsilon_2\sim N(0,1)$ are independent, and
$\widetilde{\varepsilon}_1\sim N(0,0.5)$, $\widetilde{\varepsilon}_2
\sim N(0,2)$ are independent.
As depicted in Figure \ref{example-appendix:identifiability-1B},
\eqref{example-appendix:1B-SEM1} satisfies Assumption 1C. 
However, Assumption 1B is violated given that $\Cov(Y_2, X_2 \mid 
\bm X_{\{1,3,4\}}) = 0$ and $X_2$ is an intervention variable of $Y_2$. 
This is because the direct interventional effect of $X_2$ on $Y_2$ 
are canceled out by its indirect interventional effect through $Y_1$.
Similarly, \eqref{example-appendix:1B-SEM2} satisfies Assumption 1C but 
$\Cov(Y_1, X_4 \mid \bm X_{\{1,2,3\}}) = 0$ violating Assumption 1B. 
In this case, it can be verified that $\bm\theta$ and
$\widetilde{\bm\theta}$ correspond to the same distribution $\PP(\bm Y\mid \bm X)$, 
because they share the same $(\bs V,\bm \Omega)$ even with different values of $(\bs U,\bs W,\bm\Sigma)$. 
Hence, it is impossible to infer the directed relation between $Y_1$ and $Y_2$.

  \item[(2)] \textit{Model \eqref{equation:model} is non-identifiable when Assumption 1C 
breaks down.}
Consider two different models with different parameter values:
 \begin{eqnarray}
    \bm \theta: &Y_1 = X_1 + X_2 + \varepsilon_1, \quad
      &Y_2 = Y_1 + X_2 + \varepsilon_2, \label{example-appendix:1C-SEM1} \\
    \widetilde{\bm \theta}: 
      &Y_1 = 0.5 Y_2 + 0.5 X_1 + \widetilde{\varepsilon}_1, \quad
      &Y_2 = X_1 + 2X_2 + \widetilde{\varepsilon}_2, \label{example-appendix:1C-SEM2}
  \end{eqnarray}
where $\varepsilon_1,\varepsilon_2\sim N(0,1)$ are independent, and 
$\widetilde{\varepsilon}_1\sim N(0,0.5)$, $\widetilde{\varepsilon}_2
\sim N(0,2)$ are independent. Note that 
\eqref{example-appendix:1C-SEM1} and \eqref{example-appendix:1C-SEM2}
satisfy Assumption 1B. In \eqref{example-appendix:1C-SEM1}, $Y_2$ does not 
have any instrumental intervention although it has an invalid instrument $X_2$. Similarly, in \eqref{example-appendix:1C-SEM2}, neither
does $Y_1$ have any instrumental intervention while having an invalid instrument $X_1$. As in the previous case,
$\bm\theta$ and 
$\widetilde{\bm\theta}$ yield the same distribution $\PP(\bm Y\mid\bm X)$ because they share the same $(\bs V,\bm\Omega)$ even with different values of $(\bs U,\bs W,\bm\Sigma)$.
In this case, it is impossible to infer the directed relation between 
$Y_1$ and $Y_2$.
\end{enumerate}
\end{example}

  

\begin{figure}
  \centering
  \includegraphics[width=0.7\textwidth]{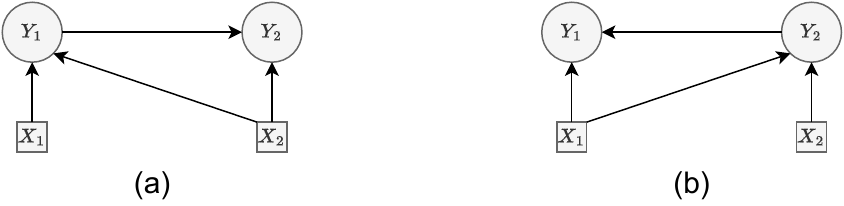}
  \caption{(a) Display of DAG defined by \eqref{example-appendix:1C-SEM1}.
(b) Display of DAG defined by \eqref{example-appendix:1C-SEM2}.}
\label{example-appendix:identifiability-1C}
\end{figure}

\subsection{Illustration of Algorithm \ref{algorithm:peeling}} \label{section:peeling-example}

We now illustrate Algorithm \ref{algorithm:peeling} by Example \ref{example:peeling}. 

\begin{example} \label{example:peeling}
  Consider model \eqref{equation:model} with $p=q=5$, 
\begin{equation}\label{equation:peeling-alg}
  \begin{aligned}
    Y_1 &= X_1 + \varepsilon_1,\\
    Y_3 &= 0.5Y_2 + X_5 + \varepsilon_3,\\
  \end{aligned}
  \qquad
  \begin{aligned}
    Y_2 &= 0.5Y_1 + X_3 + \varepsilon_2,\\
    Y_4 &= 0.5Y_3 - 0.1Y_1 + X_2 + \varepsilon_4,\\
  \end{aligned}
  \qquad
  \begin{aligned}
    Y_5 &= X_4 + \varepsilon_5,\\
        &
  \end{aligned}
\end{equation}
where $\varepsilon_1,\ldots,\varepsilon_5\sim N(0,1)$ independently. 
Then \eqref{equation:peeling-alg} defines a DAG as displayed in Figure \ref{example:testability}.
For illustration, we generate a random sample of size $n = 40$ and compute $\widehat{\bs V}$ by Algorithm \ref{algorithm:nodewise-regression}. 
In particular, 
\begin{equation*}
  \bs V = \begin{pmatrix}
    1 & 0.5 & 0.25 & 0.025 & 0\\
    0 & 0   & 0    & 1     & 0\\
    0 & 1   & 0.5  & 0.25  & 0\\
    0 & 0   & 0    & 0     & 1\\
    0 & 0   & 1    & 0.5   & 0  
  \end{pmatrix}, \quad
  \widehat{\bs V} = \begin{pmatrix}
    0.92 & 0.48   & 0.27  & 0        & 0\\
    0    & 0      & 0     & 1.08     & 0\\
    0    & 1.03   & 0.52  & 0.21     & 0\\
    0    & 0      & 0     & 0        & 1.06\\
    0    & 0      & 0.98  & 0.55     & 0
  \end{pmatrix}. \quad
\end{equation*}
Algorithm \ref{algorithm:peeling} proceeds as follows. 
\begin{itemize}
\item \textbf{$\mathcal V_Y = \{ 1,2,3,4,5 \}$:} 
The interventions indexed by $\mathcal B = \{ 2,4\}$ are instruments on the leaves that are indexed by $\mathcal L = \{ 4,5 \}$.
\begin{itemize}
  \item $X_2$ is identified as an instrument of leaf node $Y_4$ ($X_2\to Y_4$) because $\widehat{\mathrm V}_{24}\neq 0$ is the only nonzero in the row 2 with the smallest (positive) row $\ell_0$-norm.
  \item $X_4$ is identified as an instrument of leaf node $Y_5$ ($X_4\to Y_5$) because $\widehat{\mathrm V}_{45}\neq 0$ is the only nonzero in row 4 with the smallest (positive) row $\ell_0$-norm. 
\end{itemize}
Then $\{ Y_4,Y_5 \}$ are removed.

\item \textbf{$\mathcal V_Y = \{ 1,2,3 \}$:} The intervention indexed by $\mathcal B = \{ 5\}$ is an instrument on the leaf indexed by $\mathcal L=\{ 3 \}$.
\begin{itemize}
  \item $X_5$ is identified as an instrument of a leaf node $Y_3$ ($X_5\to Y_3$) in $\mathcal G^{\mathrm{work}}$ given that $\widehat{\mathrm V}_{53}\neq 0$ is the only nonzero element in the row with the smallest (positive) row $\ell_0$-norm of the submatrix for $Y_1,Y_2,Y_3$. 
\end{itemize}
Since $Y_3$ is a leaf in $\mathcal G^{\mathrm{work}}$, $Y_4$ has been removed, $X_5$ is the only instrument on $Y_3$, and $\widehat{\mathrm V}_{54}\neq 0$, we have $(3,4)\in \widehat{\mathcal E}_+$ by Proposition \ref{proposition:ancestral-relation}. Then $\{Y_3\}$ is removed.

\item \textbf{$\mathcal V_{Y}=\{ 1,2 \}$:} 
The intervention indexed by $\mathcal B = \{ 3\}$ is an instrument on the leaf indexed by $\mathcal L=\{ 2 \}$.
\begin{itemize}
  \item $X_3$ is identified as an instrument of a leaf node $Y_2$ ($X_3\to Y_2$)
  similarly in $\mathcal G^{\mathrm{work}}$ given that $\widehat{\mathrm V}_{32}\neq 0$ is the largest nonzero element in its row of the submatrix. 
\end{itemize}
Since $Y_2$ is a leaf in $\mathcal G^{\mathrm{work}}$, $Y_3$ has been removed, $X_3$ is the only instrument on $Y_2$, and $\widehat{\mathrm V}_{33}\neq 0$, we have $(2,3)\in\widehat{\mathcal E}_+$.   
Then $\{Y_2\}$ is removed.
  
\item \textbf{$\mathcal V_Y = \{1\}$:} 
The intervention indexed by $\mathcal B = \{ 1\}$ is an instrument on the leaf indexed by $\mathcal L=\{ 1 \}$.
\begin{itemize}
  \item $X_1$ is an instrument of $Y_1$ ($X_1\to Y_1$). 
  
\end{itemize}
Since $Y_1$ is a leaf node in $\mathcal G^{\mathrm{work}}$, $Y_2$ has been removed, $X_1$ is the only instrument on $Y_1$, and $\widehat{\mathrm V}_{12}\neq 0$, we have $(1,2)\in\widehat{\mathcal E}_+$.
Then $\{Y_1\}$ is removed, and the peeling process is terminated.
\end{itemize}
Finally, Steps 9 and 10 identify 
\begin{equation*}
\begin{split}
    \widehat{\mathcal E}_+ &= \{ (1,2),(2,3),(3,4),(1,3),(1,4),(2,4) \},\\
    \widehat{\mathcal I}_+ &= \{ (1,1),(1,2),(1,3),(1,4),(2,4),(3,2),(3,3),(3,4),(4,5),(5,3),(5,4) \},
\end{split}
\end{equation*}
which are equal to $\mathcal E_+$ and $\mathcal I_+$, respectively.
\end{example}

In Example \ref{example:peeling}, $\{(l,j):\widehat{\mathrm V}_{lj}\neq 0\}\neq\{(l,j):\mathrm V_{lj}\neq 0\}$, suggesting that the selection consistency of $\widehat{\bs V}$ is unnecessary for Algorithm \ref{algorithm:peeling} to correctly reconstruct the ARG $\mathcal G_+$; see also Section \ref{section:relax-signal} for the theoretical justification.

\subsection{Relaxation of Assumption \ref{assumption:signal}} \label{section:relax-signal}

Assumption \ref{assumption:signal} in Theorem \ref{theorem:consistent-structure-learning} leads to consistent identification for $\bs V$. Now, we discuss when Algorithm \ref{algorithm:peeling} correctly reconstructs $\mathcal G_+$ without requiring Assumption \ref{assumption:signal}.

\begin{assumption}\label{assumption:relax-signal}
For $1\leq j\leq p$, there exists $\tau^*_j$ such that 
\begin{enumerate}[(A)]
    \item $\Big\{l : {X_l \textnormal{ intervenes on } Y_j\\ \textnormal{ or its unmediated parents}}\Big\} \subseteq\Big\{ l: |\mathrm V_{lj}| \geq \tau_j^* \Big\}$.
    \item $\tau^*_j \geq 100 c_1^{-1}c_2 \big(\Omega_{jj}^{-1/2}\big) \sqrt{ (\kappa^\circ_j-|\{ l: |\mathrm V_{lj}|\geq \tau_j^* \}| + 1)( {\log(q)}/{n} + {\log(n)}/{n} )}$.
\end{enumerate}
\end{assumption}

  Assumption \ref{assumption:relax-signal} requires 
  the effects of intervention variables on $Y_j$ or its unmediated parents to exceed a certain signal strength $\tau_j^*$, while imposing no restrictions on the other intervention variables, for $1\leq j\leq p$. 
  These signals enable us to reconstruct $\mathcal G_+$. 
  Assumption \ref{assumption:signal} implies Assumption \ref{assumption:relax-signal} with $\tau_j^*=\min_{\mathrm V_{lj}\neq 0}|\mathrm V_{lj}|$, so Assumption \ref{assumption:relax-signal} is weaker.

\begin{theorem}\label{theorem:relax-signal}
  Suppose Assumptions \ref{assumption:identifiability}-\ref{assumption:intervention} and \ref{assumption:relax-signal} are met with constants $c_1 < 6c_2$, and the machine precision $\mathrm{tol}\ll 1/n$ is negligible. For $1\leq j \leq p$, there exist some suitable choice of tuning parameters $(\kappa_j,\tau_j)$ in Algorithm \ref{algorithm:nodewise-regression} such that
  \begin{equation*}
      |\{ l: |\mathrm V_{lj}|\geq \tau_j^* \}|\leq \kappa_j\leq \kappa_j^\circ, 
      \qquad 
      \frac{36c_2}{c_1}\sqrt{\Omega_{jj}^{-1}\Big(\frac{\log (q)}{n} + \frac{\log(n)}{n}\Big)} \leq \tau_j \leq \frac{2\tau_j^*}{5},
  \end{equation*}
  then for any $\gamma_j$ such that 
  \begin{equation*}
      \tau_j^{-1}({32c_2^2\Omega_{jj}^{-1}n^{-1}(\log(q)+\log(n))})^{1/2}\leq \gamma_j \leq c_1/6\sqrt{\kappa^\circ_j- |\{ l: |\mathrm V_{lj}|\geq \tau_j^* \}| +1},
  \end{equation*}
  almost surely we have Algorithm \ref{algorithm:nodewise-regression} terminates in at most $1 + \lceil \log (\kappa_{\max}^\circ) / \log(4)\rceil$ DC iterations when $n$ is sufficiently large. Moreover, almost surely we have Algorithm \ref{algorithm:peeling} recovers $\mathcal E_+$ and $\mathcal I_+$ when $n$ is sufficiently large.
\end{theorem}

The proof of Theorem \ref{theorem:relax-signal} is given in Appendix \ref{proof:relax-signal}.

\subsection{Comparison of Strong Faithfulness and Assumption \ref{assumption:signal} (or \ref{assumption:relax-signal})}

In the literature, a faithfulness condition is usually assumed for identifiability up to Markov equivalence classes \citep{spirtes2000causation}. For discussion, we formally introduce the concepts of faithfulness and strong faithfulness. 

Consider a DAG $\mathcal G$ with node variables $(Z_1,\ldots,Z_{p+q})^\top$. 
Nodes $Z_i$ and $Z_j$ are adjacent if $Z_i\to Z_j$ or $Z_j\to Z_i$.
A \emph{path} (\emph{undirected}) between $Z_i$ and $Z_j$ in $\mathcal G$ is a sequence of distinct nodes $(Z_i,\ldots,Z_j)$ such that all pairs of successive nodes in the sequence are adjacent. 
A nonendpoint node $Z_k$ on a path $(Z_i,\ldots,Z_{k-1},Z_k,Z_{k+1},\ldots,Z_j)$ is a \emph{collider} if $Z_{k-1}\to Z_k\leftarrow Z_{k+1}$. Otherwise it is a \emph{noncollider}.
Let $A\subseteq \{1,\ldots,p+q\}$, where $A$ does not contain $i$ and $j$.
Then $\bm Z_A$ \emph{blocks} a path $(Z_i,\ldots,Z_j)$ if at least one of the following holds: (i) the path contains a noncollider that is in $\bm Z_A$, or (ii) the path contains a collider that is not in $\bm Z_A$ and has no descendant in $\bm Z_A$.
A node $Z_i$ is d-separated from $Z_j$ given $\bm Z_A$ if $\bm Z_A$ block every path between $Z_i$ and $Z_j$; $i\neq j$ \citep{pearl2009causality}.

According to \citet{uhler2013geometry}, a multivariate 
Gaussian distribution of $(Z_1,\ldots,Z_{p+q})^{\top}$ is said to be $\varsigma$-strong faithful 
to a DAG with node set $\mathcal{V}=\{1,\ldots,p+q\}$ if
\begin{equation}
\label{equation:strong-faithfulness}
\min_{A \subseteq
\mathcal V\backslash\{i,j\}} \Big\{|{\Corr}(Z_i,Z_j \mid \bm Z_A)|: 
Z_i \text{ is not d-separated from }Z_j \text{ given }\bm Z_A \Big\}  >\varsigma,
\end{equation}
for $1 \leq i \neq j \leq p+q$,
where $\varsigma\in [0,1)$, ${\Corr}$ denotes the correlation. When $\varsigma = 0$, \eqref{equation:strong-faithfulness} is equivalent to faithfulness. For consistent structure learning (up to Markov equivalence classes), it often requires that $\varsigma \gtrsim \sqrt{s_0 \log (p+q)/n}$, where $s_0$ is a sparsity measure; see \citet{uhler2013geometry} for a survey. For a pair  $(i,j)$, the number of possible sets for $A$ is $2^{(p+q-2)}$. If $Z_i\to Z_j$, then $\text{Corr}(Z_i,Z_j \mid\bm  Z_A)\neq 0$ for any $A$. Therefore, for this $(i,j)$ pair alone, \eqref{equation:strong-faithfulness} could require exponentially many conditions. Indeed, \eqref{equation:strong-faithfulness} is very restrictive in high-dimensional situation \citep{uhler2013geometry}.

By comparison, Algorithm \ref{algorithm:peeling} yields consistent structure learning based on
Assumption \ref{assumption:signal} or \ref{assumption:relax-signal} instead of strong faithfulness. 
In some sense, Assumption \ref{assumption:signal} or \ref{assumption:relax-signal} requires sufficient signal strength that is analogous to the condition for consistent feature selection \citep{shen2012likelihood}.
This assumption may be thought of as an alternative to strong faithfulness. 
As illustrated in Example \ref{example:faithfulness}, 
Assumption \ref{assumption:signal} or \ref{assumption:relax-signal} is less stringent than 
strong faithfulness. 

\begin{example}[Faithfulness]\label{example:faithfulness}
Assume $\bm X\sim N(\bm 0,\bs I)$. 
Consider model \eqref{equation:model} with $p=q=3$,
\begin{eqnarray*}
Y_1=\mathrm W_{11}X_1 + \varepsilon_1, \quad Y_2=\mathrm U_{12}Y_1 + \mathrm W_{22}X_2 +\varepsilon_2, \quad
Y_3=\mathrm U_{13} Y_1+ \mathrm U_{23} Y_2 + \mathrm W_{33}X_3 +\varepsilon_3, 
\end{eqnarray*}
where $\varepsilon_1,\varepsilon_2,\varepsilon_3\sim N(0,1)$ are independent and $\mathrm U_{12},\mathrm U_{13},\mathrm U_{23}, \mathrm W_{11}, \mathrm W_{22}, \mathrm W_{33}\neq 0$. Denote $\bm Z=(Y_1,Y_2,Y_3,X_1,X_2,X_3)^{\top}$. 
Since the directed relations among $\bm X$ are not of interest, \eqref{equation:strong-faithfulness} becomes 
\begin{equation}\label{equation:faithfulness}
  \begin{split}
  \min_{ \substack{\bm Z_A = (\bm Y_{A_1},\bm X_{A_2}) : \\ A_1 \subseteq\{i,j\}^c, A_2}  } & \Big\{|{\Corr}(Y_i,Y_j \mid \bm Z_{A})|: 
  Y_i, Y_j \text{ are not d-separated given } \bm Z_A \Big\}  >\varsigma,\\
  \min_{ \substack{\bm Z_A = (\bm Y_{A_1},\bm X_{A_2}) : \\A_1 \subseteq
  \{j\}^c, A_2\subseteq \{ l\}^c }} &\Big\{|{\Corr}(Y_j,X_l \mid \bm Z_{A})|: 
  Y_j, X_l \text{ are not d-separated given } \bm Z_A \Big\}  >\varsigma,
  \end{split}
\end{equation}
for each pair $(i,j)$ with $i\neq j$ and each pair $(l,j)$.
Then strong-faithfulness in \eqref{equation:faithfulness} assumes 
$152$ conditions for the correlations. 
By comparison, Assumption \ref{assumption:signal} requires the absolute values of 
$\mathrm V_{11},\mathrm V_{12}, \mathrm V_{13}, \mathrm V_{22}, \mathrm V_{23}, \mathrm V_{33} \gtrsim \sqrt{\log(q)/n}$,
which in turn requires the minimum absolute value of 
six correlations $\gtrsim \sqrt{\log(q)/n}$, \\
(i) ${\Corr}(Y_1,X_1\mid X_2,X_3)$, \quad 
(ii) ${\Corr}(Y_2,X_1\mid X_2,X_3)$, \quad 
(iii) ${\Corr}(Y_3,X_1\mid X_2,X_3)$,\quad \\
(iv) ${\Corr}(Y_2,X_2\mid X_1,X_3)$, \quad 
(v) ${\Corr}(Y_3,X_2\mid X_1,X_3)$, \quad 
(vi) ${\Corr}(Y_3,X_3\mid X_1,X_2)$.\\
Importantly,
(i)-(vi) are required in \eqref{equation:faithfulness}, 
suggesting that the strong-faithfulness is more stringent than Assumption \ref{assumption:signal}.
\end{example}

\subsection{Irregular Hypothesis}\label{section:appendix-irregular-hypothesis}

Assume, without loss of generality, that $\widehat{\mathcal{D}} = \mathcal{D}$ and $\widehat{\mathcal{G}}_+=\mathcal{G}$ are correctly reconstructed in the following discussion.

\begin{itemize}
  \item For testing of directed edges \eqref{equation:link-test}, 
suppose $H_0$ is irregular, namely, $\mathcal{D}\cup \mathcal{E}$ contains a directed cycle. This implies that a directed cycle exists in $\widehat{\mathcal D}\cup\widehat{\mathcal{E}}_+$. 
In this situation, we decompose $H_0$ into sub-hypotheses $H_0^{(1)},\ldots,H_0^{(\nu)}$, each of which is regular. 
Then testing $H_0$ is equivalent to multiple testing for $H_0^{(1)},\ldots,H_0^{(\nu)}$.
{For instance, in Example \ref{example:testability}, $H_0:\mathrm U_{45}=\mathrm U_{53}=0$ is irregular, 
and $H_0$ can be decomposed into $H_{0}^{(1)}:\mathrm U_{45}=0$ and $H_0^{(2)}:\mathrm U_{53}=0$.}

  \item For testing of directed pathways \eqref{equation:pathway-test}, if $H_0$ is irregular, 
  then $\widehat{\mathcal D}\cup\widehat{\mathcal{E}}_+$ has a directed cycle. 
  The p-value is defined to be one in this situation since no evidence supports the presence of the pathway.
\end{itemize}

\subsection{Theoretical Results on Structure Learning}\label{section:structure-learning-theory}

The regression \eqref{directed-likelihood} can be solved by Algorithm \ref{algorithm:nodewise-regression} with the input 
$\mathbf Y_{\cdot j}$ as the response variable and the input $(\mathbf Y_{\cdot,\an_{\widehat{\mathcal G}_+}(j)},\mathbf X_{\cdot, \inter_{\widehat{\mathcal G}_+}(j)})$ as the covariates for $1\leq j\leq p$. The tuning parameters for solving \eqref{directed-likelihood} by Algorithm \ref{algorithm:nodewise-regression} are denoted by $\{ (\kappa_j',\tau_j') \}_{1\leq j\leq p}$.

Let $\overline{\kappa} = \max\limits_{1\leq j\leq p} |\an_{\mathcal G_+}(j)| + |\inter_{\mathcal G_+}(j)|$ and $\bs Z = (\bs Y,\bs X)$.

\begin{assumption}\label{assumption:design}
For constants $c_3,c_4>0$,
  \begin{enumerate}[(A)]
      \item $\min\limits_{\{ A : |A|\leq 2 \overline{\kappa}\} }
\min\limits_{\{ \bm\zeta : \|\bm\zeta_{A^c}\|_1\leq 3\|\bm\zeta_{A}\|_1 \}}
{\|\bs{ Z} \bm\zeta \|_2^2}/{n\| \bm\zeta \|_2^2} \geq c_3$ almost surely.
      \item $\max\limits_{1\leq k\leq p+q} n^{-1}(\bs{ Z}^\top\bs{ Z})_{kk}
    \leq c_4^2$ almost surely.
  \end{enumerate}
\end{assumption}

\begin{assumption}\label{assumption:signal-U}
  $\min\limits_{\mathrm U_{kj} \neq 0}|\mathrm U_{kj}|
  \geq 100 c_3^{-1}c_4 \max\limits_{1\leq j\leq p}(\sigma_j) \sqrt{{\log(p)}/{n} + {\log(n)}/{n}}$.
\end{assumption}

\begin{theorem} \label{theorem:consistent-structure-learning-U}
Suppose the assumptions in Theorem \ref{theorem:consistent-structure-learning} are satisfied. In addition, suppose Assumptions \ref{assumption:design}-\ref{assumption:signal-U} are met with constants $c_3<6c_4$. For $1\leq j \leq p$, assuming $|\an_{\mathcal G}(j)|\ll n$ and $|\inter_{\mathcal G}(j)| \ll n$, if the tuning parameters $(\kappa_j',\tau_j')$ are suitably chosen such that
\begin{equation*}
    \kappa_j' = |\pa_{\mathcal G}(j)|, \qquad 
    \frac{36c_4}{c_3}\sigma_j\sqrt{\frac{\log (p)}{n} + \frac{\log(n)}{n}} \leq \tau_j' \leq 
  \frac{2}{5} \min_{\mathrm U_{kj} \neq 0}|\mathrm U_{kj}|,
\end{equation*}
then for any $\gamma_j$ such that $(\tau_j')^{-1}({32c_4^2\sigma_{j}^{2}n^{-1}(\log(p)+\log(n))})^{1/2}\leq \gamma_j \leq c_3/6$, almost surely we have 
$\widehat{\mathcal E} = \mathcal E$, when $n$ is sufficiently large. 
\end{theorem}

\section{Technical Proofs}\label{section:proof}

\subsection{Proof of Proposition \ref{proposition:identifiability}} \label{proof:identifiability}

Suppose that $\bm \theta=(\bs U,\bs W, \bm \Sigma)$ and $\tilde{\bm\theta}=(\tilde{\bs U},\tilde{\bs W}, \tilde{\bm\Sigma})$ render the same distribution of $(\bm Y,\bm X)$. We will prove that $\bm \theta=\tilde{\bm\theta}$.

Denote by $\mathcal G(\bm\theta)$ and $\mathcal G(\tilde{\bm\theta})$ the DAGs corresponding to $\bm\theta$ and $\tilde{\bm\theta}$, respectively. First, consider $\mathcal G(\bm\theta)$. Without loss of generality, assume $Y_1$ is a leaf node in $\mathcal G(\bm\theta)$. By Assumption 1C, there exists an instrumental intervention with respect to $\mathcal G(\bm\theta)$, say $X_1$. Then, 
\begin{eqnarray}
  &\Cov(Y_j,X_1\mid \bm X_{\{2,\ldots,q\}}) = 0, &\quad j = 2,\ldots,p, \label{proposition:identifiability-proof1} \\
  &\Cov(Y_1,X_1\mid \bm Y_{A}, \bm X_{\{2,\ldots,q\}}) \neq 0, &\quad \text{for any } A\subseteq \{2,\ldots,p\}.
  \label{proposition:identifiability-proof2}
\end{eqnarray}
By the local Markov property \citep{spirtes2000causation}, \eqref{proposition:identifiability-proof2} implies that $X_1\to Y_1$ in $\mathcal G(\tilde{\bm\theta})$. 
Suppose $Y_1$ is not a leaf node in $\mathcal G(\tilde{\bm\theta})$.
Without loss of generality, assume that $Y_1$ is an unmediated parent of $Y_2$. 
Then $\Cov(Y_2,X_1\mid \bm X_{\{2,\ldots,q\}}) = 0$ but $X_1\to Y_1$ and $Y_1$ is an unmediated parent of $Y_2$, which contradicts to Assumption 1B.
This implies that if $Y_1$ is a leaf node in $\mathcal G(\bm\theta)$ then it must be a leaf node in $\mathcal G(\tilde{\bm\theta})$. In both $G(\bm\theta)$ and $\mathcal G(\tilde{\bm\theta})$, the parents and interventions of $Y_1$ can be identified by 
\begin{equation*}
  \begin{split}
    \E (Y_1\mid \bm Y_{\{2,\ldots,p\}}, \bm X) &= \E(Y_1\mid \bm Y_{\pa_{\mathcal G(\bm\theta)}(1)},\bm X) = \E(Y_1\mid \bm Y_{\pa_{\mathcal G(\bm\theta)}(1)},\bm X_{\inter_{\mathcal G(\bm\theta)}(1)}),\\
    \E (Y_1\mid \bm Y_{\{2,\ldots,p\}}, \bm X) &= \E(Y_1\mid \bm Y_{\pa_{\mathcal G(\tilde{\bm\theta})}(1)},\bm X) = \E(Y_1\mid \bm Y_{\pa_{\mathcal G(\tilde{\bm\theta})}(1)},\bm X_{\inter_{\mathcal G(\tilde{\bm\theta})}(1)}).
  \end{split}
\end{equation*}
Consequently, $Y_1$ has the same parents and interventions in $\mathcal G(\bm\theta)$ and $\mathcal G(\tilde{\bm\theta})$.

The forgoing argument is applied to other nodes sequentially. First, we remove $Y_1$ with any directed edges to $Y_1$, which does not alter the joint distribution of $(\bm Y_{\{2,\ldots,p\}},\bm X)$ and the sub-DAG of nodes $Y_2,\ldots,Y_p$. By induction, we remove the leafs in $\mathcal G(\bm\theta)$ until it is empty, leading to $\mathcal G(\bm\theta) = \mathcal  G(\tilde{\bm\theta})$. Finally, $\bm\theta = \tilde{\bm\theta}$ because they have the same locations for nonzero elements and these model parameters (or regression coefficients) are uniquely determined under Assumption 1A \citep{shojaie2010penalized}. This completes the proof. 

\subsection{Proof of Proposition \ref{proposition:peeling}} \label{proof:peeling}

\subsubsection*{Proof of (A)}

Note that the maximal length of a path in a DAG of $p$ nodes is at most $p-1$. Then it can be verified that $\bs U$ is nilpotent in that $\bs U^p = \bm 0$. An application of the matrix series expansion yields that $(\bs I -\bs U)^{-1}=\bs I + \bs U +\cdots+\bs U^{p-1}$. Using the fact that $\bs V= \bs W (\bs I -\bs U)^{-1}$ from \eqref{equation:model2}, we have, for
any $1 \leq l,j \leq p$, 
\begin{equation*}
  \mathrm V_{lj}=\sum_{k=1}^p \mathrm W_{lk}\Big(\mathrm I_{kj} + \mathrm U_{kj} + \cdots + (\bs U^{p-1})_{kj}\Big),
\end{equation*}
where $\mathrm U_{kj}$ is the $(k,j)$th entry of $\bs U$. 
If $\mathrm V_{lj}\neq 0$, then there exists $k$ such that $W_{lk}\neq 0$ and $(\bs U^{r})_{kj}\neq 0$ for some $0\leq r \leq p-1$. If $r=0$, then we must have $k = j$, and $X_l\to Y_j$. If $r>0$, then $X_l\to Y_{k}$ and $Y_{k}$ is an ancestor of $Y_j$.

\subsubsection*{Proof of (B)}

First, for any leaf node variable $Y_j$, by Assumption 1, there exists an instrument $X_l\to Y_j$. 
If $\mathrm V_{lj'}\neq 0$ for some $j'\neq j$, then $Y_j$ must be an ancestor of $Y_{j'}$, which contradicts the fact that $Y_j$ is a leaf node variable.

Conversely, suppose that $\mathrm V_{lj}\neq 0$ and $\mathrm V_{lj'}=0$ for $j'\neq j$. If $Y_j$ is not a leaf node variable, then there exists a variable $Y_{j'}$ such that $Y_j$ is an unmediated parent of $Y_{j'}$, that is $\mathrm U_{jj'}\neq 0$ and $(\bs U^r)_{jj'}=0$ for $r>1$. 
Then $\mathrm V_{lj'} = \mathrm W_{lj} \mathrm U_{jj'}\neq 0$, a contradiction. 

\subsection{Proof of Proposition \ref{proposition:ancestral-relation}} \label{proof:ancestral-relation}

Suppose $Y_k$ is an unmediated parent of $Y_j$. Let $X_l$ be an instrument of $Y_k$ in $\mathcal G_{\mathcal L^c}$. Then there are two cases: (1) $X_l$ intervenes on $Y_k$ but does not intervene on $Y_j$, namely $X_l\to Y_k$ but $X_l\not\to Y_j$; (2) $X_l$ intervenes on $Y_k$ and $Y_j$ simultaneously, namely $X_l\to Y_k$ and $X_l\to Y_j$. For (1), $\mathrm V_{lj} = \mathrm W_{lk}\mathrm U_{kj}\neq 0$. For (2), Assumption 1B implies that $\mathrm V_{lj}\neq 0$. This holds for every instrument of $Y_k$ in $\mathcal G_{\mathcal L^c}$, and the desired result follows. 

Conversely, suppose for each instrument $X_l$ of $Y_k$ in $\mathcal G_{\mathcal L^c}$, we have $\mathrm V_{lj}\neq 0$. Let $X_{l'}$ be an instrument of $Y_k$ in $\mathcal G$, which is also an instrument in $\mathcal G_{\mathcal L^c}$. Then $0\neq \mathrm V_{l'j} = \mathrm W_{l'k} \mathrm U_{kj}$, which implies $\mathrm U_{kj}\neq 0$. This completes the proof.

\subsection{Proof of Theorem \ref{theorem:consistent-structure-learning}} \label{proof:consistent-structure-learning}

The proof proceeds in two steps: (A) we show that $\{l :\widehat{\mathrm V}_{l j}\neq 0 \} = \{l: \mathrm V_{lj} \neq 0\}$ almost surely for $1 \leq j \leq p$; and (B) we show that $\widehat{\mathcal G}_+ = \mathcal G_+$ if $\widehat{\bs V}$ satisfies the property in (A). 

\subsubsection*{Proof of (A)}

Let $A^\circ_j = \Big\{ l: \mathrm V_{lj}\neq 0 \Big\}$
and $A^{[t]}_j = \Big\{ l : |\widetilde{\mathrm V}^{[t]}_{lj}|\geq \tau_j \Big\}$ be the estimated support of penalized solution at the
$t$-th iteration of Algorithm \ref{algorithm:nodewise-regression}. 
For the penalized solution, define 
the false negative set 
 $\text{FN}^{[t]}_j = A_j^\circ\setminus A_j^{[t]}$
and the false positive set 
$\text{FP}^{[t]}_j = A_j^{[t]}\setminus A_j^\circ$ for $t\geq 0$.  
Consider a ``good'' event 
\begin{equation*}
    \mathscr{E}_j =\Big\{ \|\bs{ X}^\top\widehat{\bm \xi}_j/n\|_\infty\leq 0.5\gamma_j\tau_j \Big\} \cap \Big\{ \|\widehat{\bs V}^\circ_{\cdot j} - \bs V_{\cdot j}\|_\infty \leq 0.5\tau_j \Big\},
\end{equation*}
where $\widehat{\bm\xi}_j = \bs{ Y}_j - \bs{ X}\widehat{\bs V}^\circ_{\cdot j}$ is the residual of the oracle least squares estimate $\widehat{\bs V}^\circ_{\cdot j}$ such that $A_j=\Big\{ l: \widehat{\mathrm V}^\circ_{lj}\neq 0\Big\}$; $1\leq j \leq p$. We shall show that $\text{FN}^{[t]}_j$ and $\text{FP}^{[t]}_j$ are eventually empty on event $\mathscr{E}_j$ which has a probability tending to one.

First, we will show that if $|A_j^\circ\cup A^{[t-1]}_j|\leq 2\kappa_{\max}^\circ$ on $\mathscr E_j$ for $t \geq 1$, then $|A_j^\circ\cup A^{[t]}_j|\leq 2\kappa^\circ_{\max}$, to be used in Assumption \ref{assumption:intervention}A. To proceed, suppose $|A_j^\circ\cup A^{[t-1]}_j|\leq 2 \kappa^\circ_{\max}$ on $\mathscr E_j$ for $t \geq 1$. By the optimality condition of \eqref{equation:penalized-regression},
\begin{equation*}
      \begin{split}
        \Big(\widehat{\bs V}^\circ_{\cdot j} - \widetilde{\bs V}^{[t]}_{\cdot j}\Big)^\top 
        \Big(-\bs{ X}^\top (\bs{ Y}_j - \bs{ X}\widetilde{\bs V}^{[t]}_{\cdot j})/n 
        + \gamma_j\tau_j 
        \nabla \|\widetilde{\bs V}_{(A^{[t-1]}_j)^c, j}^{[t]}\|_1 \Big) \geq 0,
      \end{split}
\end{equation*} 
where $\widetilde{\bs V}^{[t]}$ is defined in \eqref{equation:penalized-regression}. 
Plugging $\widehat{\bm \xi}_j = \bs{ Y}_j - \bs{ X}\widehat{\bs V}^\circ_{\cdot j}$ into the inequality and rearranging it, we have $\|\bs{ X}(\widetilde{\bs V}_{\cdot j}^{[t]}-\widehat{\bs V}_{\cdot j}^\circ)\|_2^2/n$ is no greater than 
\begin{equation}
\begin{split}
        &\Big(\widetilde{\bs V}_{\cdot j}^{[t]}-\widehat{\bs V}_{\cdot j}^\circ\Big)^\top
        \Big( \bs{ X}^\top\widehat{\bm \xi}_j/n - \gamma_j\tau_j 
        \nabla \|\widetilde{\bs V}^{[t]}_{(A_j^{[t-1]})^c, j}\|_1 \Big)\\
  =\ & \Big(\widetilde{\bs V}_{A_j^\circ \setminus A_j^{[t-1]}, j}^{[t]} - \widehat{\bs V}_{A_j^\circ \setminus A_j^{[t-1]}, j}^\circ\Big)^\top 
        \Big( \bs{ X}^\top_{A_j^\circ\setminus A^{[t-1]}_j}\widehat{\bm \xi}_j/n - \gamma_j\tau_j 
        \nabla \|\widetilde{\bs V}^{[t]}_{A_j^\circ\setminus A^{[t-1]}_j, j}\|_1 \Big)  \\
 & + \Big(\widetilde{\bs V}_{(A_j^\circ\cup A^{[t-1]}_j)^c, j}^{[t]} - \widehat{\bs V}_{(A_j^\circ\cup A^{[t-1]}_j)^c, j}^\circ\Big)^\top 
        \Big( \bs{ X}^\top\widehat{\bm \xi}/n - \gamma_j\tau_j 
        \nabla \|\widetilde{\bs V}^{[t]}_{(A_j^\circ\cup A^{[t-1]}_j)^c, j}\|_1 \Big) \\
  & + \Big(\widetilde{\bs V}_{A^{[t-1]}_j\setminus A_j^\circ, j}^{[t]} - \widehat{\bs V}_{A^{[t-1]}_j \setminus A_j^\circ, j}^\circ\Big)^\top
  \bs{ X}^\top_{A_j^{[t-1]}\setminus A_j^\circ}\widehat{\bm \xi}/n,
\label{equation:basic-inequality}
\end{split}
\end{equation}
where $\bs X_{A^\circ_j}^\top \widehat{\bs\xi}_j/n = \bs 0$ has been used.
Note that 
    \begin{equation*}
      \Big(\widetilde{\bs V}_{(A_j^\circ\cup A_j^{[t-1]})^c, j}^{[t]} - \widehat{\bs V}_{(A_j^\circ\cup A_j^{[t-1]})^c, j}^\circ\Big)^\top \nabla \Big\|\widetilde{\bs V}^{[t]}_{(A_j^\circ\cup A_j^{[t-1]})^c, j}\Big\|_1
    = \Big\|\widetilde{\bs V}_{(A_j^\circ\cup A_j^{[t-1]})^c, j}^{[t]} - \widehat{\bs V}_{(A_j^\circ\cup A_j^{[t-1]})^c, j}^\circ\Big\|_1.
    \end{equation*} 
    Then \eqref{equation:basic-inequality} is no greater than 
\begin{equation}\label{equation:l1-inequality}
      \begin{split}
& \Big\| \widetilde{\bs V}_{A_j^\circ\triangle A^{[t-1]}_j, j}^{[t]} - \widehat{\bs V}_{A_j^\circ\triangle A^{[t-1]}_j, j}^\circ \Big\|_1 
\Big(\|\bs{ X}^\top\widehat{\bm \xi}_j/n\|_\infty + \gamma_j\tau_j \Big)\\
& 
+ \Big\|\widetilde{\bs V}_{(A_j^\circ\cup A_j^{[t-1]})^c, j}^{[t]} - \widehat{\bs V}_{(A_j^\circ\cup A_j^{[t-1]})^c, j}^\circ \Big\|_1 
\Big(\|\bs{ X}^\top\widehat{\bm \xi}_j/n\|_\infty - \gamma_j\tau_j \Big),
      \end{split}
\end{equation}
where $\triangle$ denotes the symmetric difference of two sets.
Note that $\|\bs{ X}(\widetilde{\bs V}_{\cdot j}^{[t]}-\widehat{\bs V}_{\cdot j}^\circ)\|_2^2/n\geq 0$. 
Rearranging the inequality yields that
\begin{equation*}
  \begin{split}
   \Big( \gamma_j\tau_j - \|\bs{ X}^\top\widehat{\bm \xi}_j/n\|_\infty \Big) \Big\|\widetilde{\bs V}_{(A_j^\circ\cup A_j^{[t-1]})^c, j}^{[t]} - \widehat{\bs V}_{(A_j^\circ\cup A_j^{[t-1]})^c, j}^\circ \Big\|_1 \\
  \leq \Big(\|\bs{ X}^\top\widehat{\bm \xi}_j/n\|_\infty + \gamma_j\tau_j \Big) 
    \Big\| \widetilde{\bs V}_{A_j^\circ\triangle A^{[t-1]}_j, j}^{[t]} - \widehat{\bs V}_{A_j^\circ\triangle A^{[t-1]}_j, j}^\circ \Big\|_1.
  \end{split}
\end{equation*}
On event $\mathscr E_j$, $\|\bs{ X}^\top\widehat{\bm \xi}_j/n\|_\infty
\leq \gamma_j\tau_j/2$, implying that 
\begin{equation*}
\begin{split}
    \Big\|\widetilde{\bs V}_{(A_j^\circ\cup A_j^{[t-1]})^c, j}^{[t]} - \widehat{\bs V}_{(A_j^\circ\cup A_j^{[t-1]})^c, j}^\circ\Big\|_1 
    &\leq 3 \Big\|\widetilde{\bs V}_{A_j^\circ\triangle A^{[t-1]}_j, j}^{[t]} - \widehat{\bs V}_{A_j^\circ\triangle A^{[t-1]}_j, j}^\circ\Big\|_1\\
    &\leq 3 \Big\|\widetilde{\bs V}_{A_j^\circ\cup A^{[t-1]}_j, j}^{[t]} - \widehat{\bs V}_{A_j^\circ\cup A^{[t-1]}_j, j}^\circ\Big\|_1.
\end{split}
\end{equation*}
Note that $|A_j^\circ\cup A^{[t-1]}_j|\leq 2\kappa^\circ_{\max}$. 
By Assumption \ref{assumption:intervention}A and \eqref{equation:l1-inequality},
\begin{equation}\label{equation:error-bound}
      \begin{split}
  c_1\|\widetilde{\bs V}_{\cdot j}^{[t]}-\widehat{\bs V}_{\cdot j}^\circ\|_2^2 
        \leq & \Big( \|\bs{ X}^\top\widehat{\bm \xi}_j/n\|_{\infty} + \gamma_j\tau_j \Big) \Big\|
        \widetilde{\bs V}_{A_j^\circ\triangle A_j^{[t-1]}, j}^{[t]} - \widehat{\bs V}_{A_j^\circ\triangle A_j^{[t-1]}, j}^\circ\Big\|_1  \\
        & +
        \Big(\|\bs{ X}^\top\widehat{\bm \xi}_j/n\|_\infty - \gamma_j\tau_j \Big) \Big\|\widetilde{\bs V}_{(A_j^\circ\cup A_j^{[t-1]})^c,j}^{[t]} - \widehat{\bs V}_{(A_j^\circ\cup A_j^{[t-1]})^c, j}^\circ \Big\|_1\\
        \leq & \Big( \|\bs{ X}^\top\widehat{\bm \xi}_j/n\|_{\infty} + \gamma_j\tau_j \Big) \Big\|
        \widetilde{\bs V}_{A_j^\circ\triangle A_j^{[t-1]}, j}^{[t]} - \widehat{\bs V}_{A_j^\circ\triangle A_j^{[t-1]}, j}^\circ\Big\|_1.
      \end{split}
\end{equation}
By the Cauchy-Schwarz inequality, 
\begin{equation*}
    \Big\|\widetilde{\bs V}_{A_j^\circ\triangle A_j^{[t-1]}, j}^{[t]} - \widehat{\bs V}_{A_j^\circ\triangle A_j^{[t-1]}, j}^\circ\Big\|_1
\leq \sqrt{A_j^\circ \triangle A^{[t-1]}_j} \Big\|\widetilde{\bs V}_{\cdot j}^{[t]} - \widehat{\bs V}^\circ_{\cdot j}\Big\|_2.
\end{equation*}
Thus, $c_1\|\widetilde{\bs V}_{\cdot j}^{[t]}-\widehat{\bs V}_{\cdot j}^\circ\|_2
\leq 1.5 \gamma_j\tau_j\sqrt{2\kappa^\circ_{\max}}$, since
$|A_j^\circ\triangle A_j^{[t-1]}|\leq |A_j^\circ\cup A_j^{[t-1]}| \leq 2\kappa^\circ_{\max}$ and $\|\bs{ X}^\top\widehat{\bm \xi}_j/n\|_\infty \leq 0.5\gamma_j \tau_j$.
By the condition of Theorem \ref{theorem:consistent-structure-learning}, 
$\|\widetilde{\bs V}_{\cdot j}^{[t]} - \widehat{\bs V}_{\cdot j}^\circ\|_2/\tau_j \leq \sqrt{\kappa^\circ_{\max}}$ and $\gamma_j \leq c_1/6$. On the other hand, 
for any $l\in \text{FP}^{[t]}_j= A^{[t]}_j \setminus A_j^\circ$, 
we have $|\widetilde{\mathrm V}_{lj}^{[t]} - \widehat{\mathrm V}_{lj}^\circ| = 
|\widetilde{\mathrm V}_{lj}^{[t]}|>\tau_j$.
Thus, $\sqrt{|\text{FP}^{[t]}_j|}\leq \|\widetilde{\bs V}_{\cdot j}^{[t]} - \widehat{\bs V}_{\cdot j}^\circ\|_2/\tau_j \leq \sqrt{\kappa^\circ_{\max}}$, implying $|A_j^\circ \cup A^{[t]}_j|= |A_j^\circ| + |\text{FP}^{[t]}_j|
\leq 2\kappa^\circ_{\max}$ on $\mathscr E_j$ for $t\geq 1$.
  
Next, we estimate the number of iterations required for termination. Note that the machine precision is negligible. The termination criterion is met when $A^{[t]}_j = A^{[t-1]}_j$ since the weighted Lasso problem \eqref{equation:penalized-regression} remains same at the $t$th and $(t-1)$th iterations.
To show that $A^{[t]}_j = A^\circ_j$ eventually, we
prove that $|\text{FN}^{[t]}_j| + |\text{FP}^{[t]}_j| < 1$ eventually.

Now, suppose $|\text{FN}^{[t]}_j|+|\text{FP}_j^{[t]}|\geq 1$. For any $l\in \text{FN}^{[t]}_j\cup\text{FP}_j^{[t]}$, 
by Assumption \ref{assumption:signal}, 
\begin{equation*}
    |\widetilde{\mathrm V}^{[t]}_{lj} - \widehat{\mathrm V}^\circ_{lj}| \geq |\widetilde{\mathrm V}^{[t]}_{lj} - \mathrm V_{lj}|-|\widehat{\mathrm V}^\circ_{lj}-\mathrm V_{lj}| \geq \tau_j - 0.5\tau_j,
\end{equation*}
so $\sqrt{|\text{FN}^{[t]}_j| + |\text{FP}^{[t]}_j|} \leq \|\widetilde{\bs V}_{\cdot j}^{[t]}
-\widehat{\bs V}_{\cdot j}^\circ\|_2 /0.5\tau_j$. 
By \eqref{equation:error-bound} and the Cauchy-Schwarz inequality,  
$c_1\|\widetilde{\bs V}_{\cdot j}^{[t]} - \widehat{\bs V}_{\cdot j}^\circ\|_2 \leq 
1.5\gamma_j \tau_j \sqrt{|\text{FN}_j^{[t-1]}| + |\text{FP}^{[t-1]}_j|}.
$
By conditions (1) and (2) for $(\tau_j,\gamma_j)$ in Theorem \ref{theorem:consistent-structure-learning}, we have 
\begin{equation*}
    \begin{split}
        \sqrt{|\text{FN}^{[t]}_j| + |\text{FP}^{[t]}_j|} \leq \frac{\|\widetilde{\bs V}_{\cdot j}^{[t]}-\widehat{\bs V}_{\cdot j}^\circ\|_2}
  {0.5\tau_j}
  &\leq \frac{3\gamma_j}
  {c_1}
  \sqrt{|\text{FN}_j^{[t-1]}| + |\text{FP}_j^{[t-1]}|}\\
  &\leq 0.5 \sqrt{|\text{FN}_j^{[t-1]}| + |\text{FP}_j^{[t-1]}|}.
    \end{split}
\end{equation*}
Hence, $\sqrt{|\text{FN}^{[t]}_j| + |\text{FP}^{[t]}_j| }\leq (1/2)^t\sqrt{|A^\circ_j| + |A^{[0]}_j|}$.
In particular, for $t \geq 1 + \lceil\log \kappa^\circ_j /\log 4\rceil$, 
$|\text{FN}_j^{[t]}| + |\text{FP}^{[t]}_j| < 1$ implying that $\text{FN}_j^{[t]} = \emptyset$
and $\text{FP}_j^{[t]} = \emptyset$.

Let $t_{\max} = 1 + \lceil\log \kappa^\circ_j /\log 4\rceil$. Then 
$\text{FN}_j^{[t_{\max}]} = \text{FP}^{[t_{\max}]}_j = \emptyset$ on event $\mathscr{E}_j$. 
By the condition of Theorem \ref{theorem:consistent-structure-learning}, we have 
$\Big\{l :\widetilde{\mathrm V}^{[t_{\max}]}_{l j}\neq 0 \Big\} = \Big\{l: \mathrm V_{lj} \neq 0\Big\}$.
To bound $\PP\Big(\bigcup_{j=1}^p\mathscr E_j^c\Big)$, let 
$\bm\eta = \bs{ X}^\top (\bs I - \bs{ P}_{A^\circ_j}) {\bm \xi}_j$
and $\bm \eta' = n(\bs{ X}^\top_{A^\circ_j} \bs{ X}_{A^\circ_j})^{-1}\bs{ X}^\top_{A_j^\circ}{\bm \xi}_j$, where $\bm\xi_j = ((\varepsilon_V)_{1j},\ldots,(\varepsilon_V)_{nj})^\top$. Then $\bm\eta\in\mathbb{R}^q$ and $\bm\eta'\in\mathbb{R}^{|\kappa^\circ_j|}$ are Gaussian vectors with $\Var(\eta_l)\leq \Omega_{jj}^{-1} (\bs{ X}^\top \bs{ X})_{ll}\leq n\Omega_{jj}^{-1}c_2^2$ and $\Var(\eta'_l) \leq \Omega_{jj}^{-1}((\bs{ X}^\top_{A^\circ_j} \bs{ X}_{A^\circ_j})^{-1})_{ll}\leq n\Omega^{-1}_{jj} c_2^2$.
Then 
\begin{equation*}
\begin{split}
\PP\Big(\|\bs{ X}^\top\widehat{\bm \xi}_j/n\|_\infty > 0.5\gamma_j\tau_j\Big) 
    &= \PP\Big(\|\bm\eta/n\|_\infty>0.5\gamma_j\tau_j\Big)\\
    &\leq \sum_{l=1}^q \PP\Big(|\eta_l| \geq 0.5n\gamma_j\tau_j\Big) \\
    &\leq q \int_{\frac{0.5\sqrt{n \Omega_{jj}}\gamma_j \tau_j  }{c_2}}^\infty \frac{e^{-t^2/2}}{\sqrt{2\pi}} dt\leq q\sqrt{\frac{2}{\pi}}  \exp\Big( -\frac{ n \Omega_{jj}\gamma^2_j\tau^2_j}{8c^2_2 } \Big), 
\end{split}
\end{equation*}
and similarly,
\begin{equation*}
  \begin{split}
    \PP\Big(\|\widehat{\bs V}_{\cdot j}^\circ-\bs V_{\cdot j}^\circ\|_\infty > 0.5\tau_j\Big) 
    \leq \sum_{l\in A^\circ_j} \PP\Big( |\eta'_l| >0.5\tau_j\Big)
          \leq \kappa^\circ_{\max} \sqrt{\frac{2}{\pi}} \exp\Big(- \frac{n\Omega_{jj}\tau^2_j}{8c^2_2 }\Big). 
  \end{split}
\end{equation*}
Under conditions for $(\tau_j,\gamma_j)$ in Theorem \ref{theorem:consistent-structure-learning}, 
\begin{equation*}
  \begin{split}
   & \PP\Big( \Big\{l :\widetilde{\mathrm V}^{[t_{\max}]}_{l j}\neq 0 \Big\} = \Big\{l: \mathrm V_{lj} \neq 0\Big\};\ 1\leq j\leq p \Big) \\
&\geq 1 - \PP\Big(\bigcup_{j=1}^p\mathscr E_j^c\Big) \\
&\geq 1 - p \sqrt{\frac{2}{\pi}} \Big( e^{-4(\log(q) + \log(n)) + \log(q)} + e^{-144 c_1^{-2}c_2^2 ( \log(q)+\log(n)) + \log(\kappa^\circ_j)}  \Big)
\geq 1 - \sqrt{\frac{2}{\pi}} p q^{-3} n^{-4},
  \end{split}
\end{equation*}
where $c_1<6c_2$ is used in the last inequality.
By Borel-Cantelli lemma, almost surely we have $\Big\{l :\widetilde{\mathrm V}^{[t_{\max}]}_{lj}\neq 0 \Big\} = \Big\{l:\mathrm V_{lj} \neq 0\Big\}$; $1\leq j \leq p$ when $n$ is sufficiently large.

So far, we have $\widetilde{\bs V}_{\cdot j} = \widetilde{\bs V}^{[t_{\max}]}_{\cdot j} = \widehat{\bs V}^\circ_{\cdot j}$; $1 \leq j \leq p$, almost surely.
It remains to show that $\widehat{\bs V}^\circ_{\cdot j}$ is a global minimizer 
of \eqref{pseudo-likelihood}; $1\leq j \leq p$, with a high probability. 
Note that Assumptions \ref{assumption:intervention}A and \ref{assumption:signal} imply the degree of separation condition \citep{shen2013constrained}. By Theorem 2 of \citet{shen2013constrained},
\begin{equation*}
  \PP\Big(\widehat{\bs V}_{\cdot j}^\circ \text{ is not a global minimizer of }\eqref{pseudo-likelihood}; 1\leq j \leq p\Big) 
  \leq 3 p \exp\Big( -2(\log(q) + \log(n)) \Big).
\end{equation*} 
implying that almost surely $\widehat{\bs V}_{\cdot j}=\widetilde{\bs V}^{[t_{\max}]}_{\cdot j} = \widehat{\bs V}^\circ_{\cdot j}$ is a global minimizer of \eqref{pseudo-likelihood} when $n$ is sufficiently large.
This completes the proof.

\subsubsection*{Proof of (B)}

Assume $\widehat{\bs V}$ satisfies the properties in (A).
Then Propositions \ref{proposition:peeling} and \ref{proposition:ancestral-relation} holds true for $\widehat{\bs V}$.
Clearly, we have $\widehat{\mathcal I}_+ := \{ (l,j) : \widehat{\mathrm V}_{lk}\neq 0 \text{ if } k=j \text{ or } (k,j)\in\widehat{\mathcal E}_+ \} = \mathcal I_+$ whenever $\widehat{\mathcal E}_+=\mathcal E_+$.
Thus, we only need to show $\widehat{\mathcal{E}}_+ = \mathcal{E}_+$.
We shall prove this by induction. 

Given $\mathcal G^{\text{work}}$ and $\bs V^{\text{work}}$, the set of instruments on leaves is 
\begin{equation*}
    \mathcal B = \Big\{ l : l \text{ minimizes } \|\widehat{\bs V}^{\text{work}}_{l,\cdot}\|_0 \text{ and } \|\widehat{\bs V}^{\text{work}}_{l,\cdot}\|_0 > 0 \Big\} = \Big\{  l: \|\bs V^{\text{work}}_{l,\cdot}\|_0 = 1 \Big\}.
\end{equation*}
Hence, $X_l$ is an instrument of leaf variable $Y_k$ in $\mathcal G^{\text{work}}$, when $l\in \mathcal{B}$ and $k$ maximizes $|\mathrm V_{lk}^{\text{work}}|$.
Hence, $\mathcal L$ is the index set of leaves in $\mathcal G^{\text{work}}$.
By Proposition \ref{proposition:ancestral-relation}, all $(k,j)$ such that $Y_k$ is an unmediated parent of a peeled variable $Y_j$ are in $\widehat{\mathcal E}_+$ and can be identified by Assumption \ref{assumption:identifiability}B. 
In model \eqref{equation:model}, after removing $\bm Y_{\mathcal L}$, the local Markov property of the rest variables in $\mathcal G^{\text{work}}$ remain intact. Then we repeat the procedure until all primary variables are removed.

As a result, Algorithm \ref{algorithm:peeling} 
correctly identifies a subset of ${\mathcal E}_+$ that contains all edges from an unmediated parent, so it suffices to recover $\mathcal E_+$. 
Consequently, $\mathcal E_+$ can be reconstructed by Step 9 and $\mathcal I_+$ can be recovered by Step 10.
This completes the proof of (B). 

\begin{lemma}\label{lemma:inference}
   Let 
   \begin{equation}
    T_j = 
    \frac{\bs{ e}_j^\top( \bs{ P}_{A_j} 
    - \bs{ P}_{B_j})\bs{ e}_j}{\bs{ e}_j^\top(\bs I - \bs{ P}_{A_j} )\bs{ e}_j/(n - |A_j|)},
    \quad 
    T_j^* = 
    \frac{(\bs{ e}^*_j)^\top( \bs{ P}^*_{A_j} 
    - \bs{ P}^*_{B_j})\bs{ e}^*_j}{(\bs{ e}^*_j)^\top(\bs I - \bs{ P}^*_{A_j} )\bs{ e}^*_j/(n - |A_j|)},
\end{equation}
where $A_j = \pa_{\mathcal S}(j)\cup\inter_{\mathcal S}(j)$ and 
$B^\circ_j=(\pa_{\mathcal S}(j)\cup\inter_{\mathcal S}(j))\setminus\df_{\mathcal S}(j)$.
Then $(T_1,\ldots,T_p)$ are independent and $(T_1^*,\ldots,T_p^*)$ 
are conditionally independent given $\bs{ Z}$.
Moreover, 
\begin{equation*}
  T^*_j/(|A_j|-|B_j|) \mid \bs{Z} \sim T_j/(|A_j|-|B_j|)  
  \sim F_{|A_j|-|B_j|,n-|A_j|};
  \quad 1\leq j\leq p,
\end{equation*}
where $F_{d_1,d_2}$ denotes the F-distribution with $d_1$ and $d_2$ degrees of
freedom.
\end{lemma}

\subsection*{Proof of Lemma \ref{lemma:inference}}

Let $\bs{ Z} = (\bs{ Y},\bs{ X})$ as in \eqref{equation:likelihood-ratio}.
Given data submatrix $\bs{{Z}}_{A_j}$, 
\begin{equation*}
  \frac{\bs{{e}}_j^\top( \bs{ P}_{A_j} - \bs{{P}}_{B_j})\bs{{e}}_j}{\sigma^2_j} \mid \bs{{Z}}_{A_j}\sim \chi^2_{|A_j|-|B_j|}, \quad \frac{\bs{{e}}_j^\top(\bs I - \bs{{P}}_{A_j} )\bs{{e}}_j}{\sigma^2_j} \mid \bs{{Z}}_{A_j} \sim \chi^2_{n-|A_j|},
\end{equation*}
and they are independent, because $\bs{{P}}_{A_j}-\bs{{P}}_{B_j}$ and $\bs I - \bs{{P}}_{A_j}$ are orthonormal projection matrices, $\bs{{e}}_j\mid \bs{{Z}}_{A_j}\sim N(\bm 0, \sigma_j^2\bs I_n)$, and $( \bs{{P}}_{A_j} - \bs{{P}}_{B_j})(\bs I - \bs{{P}}_{A_j} )=\bm 0$. Then, for any real number $t$, the characteristic function $t\mapsto \E \exp(\iota t T_j/(|A_j|-|B_j|) )$ 
is ($\iota$ is the imaginary unit)
\begin{equation*}
    \E \left(\E\left(  \exp(\iota t T_j/(|A_j|-|B_j|)  ) \mid \bs{{Z}}_{A_j} \right) \right)
    = \E \psi_{|A_j|-|B_j|,n-|A_j|}(t) = \psi_{|A_j|-|B_j|,n-|A_j|}(t),
\end{equation*}
where $\psi_{|A_j|-|B_j|,n-|A_j|}$ is the characteristic function 
of F-distribution with degrees of freedom $|A_j|-|B_j|,n-|A_j|$.
Hence, $T_j/(|A_j|-|B_j|) \sim F_{|A_j|-|B_j|,n-|A_j|}$. 
Similarly, we also have $T_j^*/(|A_j|-|B_j|) \sim F_{|A_j|-|B_j|,n-|A_j|}$; $j=1,\ldots,p$.

Next, we prove independence for $\bm T$ and $\bm T^*$ via a peeling argument. 
Let $\bm t =(t_1,\ldots,t_p)$. Let $Y_j$ be a leaf node of the graph $\mathcal G$. Then $\bm T_{-j}\mid \bs{ Y}_{-j},\bs{ X}$ is deterministic, where $\bm T_{-j}$ is the subvector of $\bm T$ with the $j$th component removed. The characteristic function of $\bm t^\top\bm T$ is 
\begin{equation}
  \begin{split}
    \E \exp(\iota\bm t^\top\bm T) 
    &= \E \Big( 
      \E\Big( \exp(\iota t_{j} T_{j}) 
      \mid \bs{ Y}_{-j},\bs{ X}\Big) \exp(\iota\bm t^\top_{-j}\bm T_{-j} ) \Big)\\
    &= \psi_{|A_{j}|-|B_{j}|,n-|A_{j}|}(t_{j}) \E\exp(\iota\bm t_{-j}^\top\bm T_{-j}),
  \end{split}
\end{equation}
where $\psi_{|A_{j}|-|B_{j}|,n-|A_{j}|}$ is the characteristic function of 
the F-distribution with degrees of freedom $(|A_{j}|-|B_{j}|,n-|A_{j}|)$.
Next, let $Y_{j'}$ be a leaf node of the graph $\mathcal G'$ and apply 
the law of iterated expectation again, where $\mathcal G'$ is the subgraph of $\mathcal G$ without node $Y_j$.
Repeat this procedure and after $p$ steps 
$\E\exp(\iota\bm t^\top\bm T) = \prod_{j=1}^p \psi_{|A_{j}|-|B_{j}|,n-|A_{j}|}(t_j)$,
which implies $(T_1,\ldots,T_p)$ are independent.
Similarly, $\bm T^*$ also has independent components given $\bs{ Z}$ and
has the same distribution as $\bm T$. This completes the proof.

\subsection{Proof of Theorem \ref{theorem:dp-null-distribution}} \label{proof:dp-null-distribution}

Let $\mathrm{Lr}(\mathcal S,\bm\Sigma)$ denote the likelihood ratio given $\mathcal S$ and $\bm\Sigma$. Then $\mathrm{Lr} = \mathrm{Lr}(\widehat{\mathcal S},\widehat{\bm\Sigma})$.

\subsubsection*{Proof of (A)}

Without loss of generality, assume $M$ is sufficiently large. For any real number $x$,
\begin{equation}
  \begin{split}
     & |\PP(\textnormal{Lr}\leq x) - \PP(\textnormal{Lr}(\mathcal S,\widehat{\bm\Sigma})\leq x)|\\
    \leq\ & \E|\I(\textnormal{Lr}\leq x) - \I(\textnormal{Lr}(\mathcal S,\widehat{\bm\Sigma})\leq x)|  \\
    =\ & \E|\I(\textnormal{Lr}\leq x, \widehat{\mathcal S}\neq \mathcal S) + \I(\textnormal{Lr}(\mathcal S,\widehat{\bm\Sigma})\leq x)(\I(\widehat{\mathcal S}=\mathcal S)-1)|\\
    \leq\ & 2\PP(\widehat{\mathcal S}\neq \mathcal S). 
  \end{split}
\label{equation:inference-bound-unconditional}
\end{equation}
From \eqref{equation:likelihood-ratio-dp}, $\textnormal{Lr}^*(\mathcal S,\widehat{\bm\Sigma}^*) = \sum_{\{j:\df_{\mathcal S}(j)\neq\emptyset\}} T_j^*$.
By Lemma \ref{lemma:inference}, 
\begin{equation*}
    \PP(\textnormal{Lr}(\mathcal S,\widehat{\bm\Sigma})\leq x) = \PP(\textnormal{Lr}^*(\mathcal S,
\widehat{\bm\Sigma}^*)\leq x\mid \bs{ Z}).
\end{equation*}
Note that $\textnormal{Lr}^*=\textnormal{Lr}^*(\widehat{\mathcal S}^*,
\widehat{\bm\Sigma}^*)$. Then for any real number $x$,  
\begin{equation}
 \begin{split}
    & |\PP(\textnormal{Lr}^*\leq x\mid \bs{ Z}) - \PP(\textnormal{Lr}(\mathcal S,\widehat{\bm\Sigma})\leq x)| \\
    =\ &  |\PP(\textnormal{Lr}^*\leq x, \widehat{\mathcal S}^*\neq \mathcal S\mid\bs{ Z}) + 
    \PP(\textnormal{Lr}^*\leq x, \widehat{\mathcal S}^*=\mathcal S\mid \bs{ Z})-\PP(\textnormal{Lr}^*(\mathcal S,\widehat{\bm\Sigma}^*)\leq x\mid \bs{ Z})| \\ 
    \leq\ & |\PP(\textnormal{Lr}^*\leq x, \widehat{\mathcal S}^*\neq \mathcal S\mid\bs{ Z})|+
|\PP(\textnormal{Lr}^*(\mathcal S,\widehat{\bm\Sigma}^*)\leq x, \widehat{\mathcal S}^*\neq \mathcal S\mid \bs{ Z})|\\
\leq\ & 2\PP(\widehat{\mathcal S}^*\neq \mathcal S \mid \bs{ Z}). 
  \end{split}
\label{equation:inference-bound-conditional}
\end{equation}
Since \eqref{equation:inference-bound-unconditional} and \eqref{equation:inference-bound-conditional} hold uniformly in $x$, we have 
\begin{equation*}
    \sup_{x\in\mathbb R}|\PP(\textnormal{Lr}\leq x)-\PP(\textnormal{Lr}^*\leq x\mid\bs{ Z})|
    \leq 2\PP(\widehat{\mathcal S}\neq \mathcal S) + 2\PP(\widehat{\mathcal S}^*\neq \mathcal S \mid \bs{ Z}).
\end{equation*}
By Theorem \ref{theorem:consistent-structure-learning}, we have $\PP(\widehat{\mathcal S}\neq \mathcal S)\to 0$, which holds 
uniformly for all $\bm \theta$ which satisfy the Assumptions \ref{assumption:identifiability}-\ref{assumption:dimension}.
For $\PP(\widehat{\mathcal S}^*\neq \mathcal S \mid \bs{ Z})$, 
the error terms of \eqref{pseudo-likelihood} in the perturbed data $\bs{ Z}^*$ are rescaled with $\Omega_{jj}^{-1} + \widehat{\sigma}^2_j\leq 2\Omega_{jj}^{-1}$. 
By Theorem \ref{theorem:consistent-structure-learning}, 
$\PP(\widehat{\mathcal S}^*\neq \mathcal S) = \E \PP( \widehat{\mathcal S}^*\neq \mathcal S \mid \bs{Z}) \to 0$, which implies $\PP( \widehat{\mathcal S}^*\neq \mathcal S \mid \bs{ Z}) \pto 0$ as $n\rightarrow \infty$ 
by the Markov inequality.
Consequently, $\sup_{x\in\mathbb R}|\PP(\textnormal{Lr}\leq x)-\PP(\textnormal{Lr}^*\leq x\mid\bs{ Z})|\to0$. 
For $|\mathcal D|=0$, $\PP(\textnormal{Lr} = 0)\to 1$, 
$\PP(\textnormal{Lr}^* = 0 \mid \bs{ Z}) \to 1$, and $\PP(\textnormal{Pval}=1)\to 1$. 
For $|\mathcal D|>0$, $\PP(\textnormal{Lr}^*\geq \textnormal{Lr}\mid \bs{ Z}) \to \textnormal{Unif}(0,1)$
and $\PP(\textnormal{Pval}<\alpha)\to \alpha$.
This completes the proof of (A).

\subsubsection*{Proof of (B)}

Let $\textnormal{Pval}_k = M^{-1}\sum_{m=1}^M \I(\text{Lr}^*_{k,m}\geq \text{Lr})$,
the p-value of sub-hypothesis $\text{H}_{0,k}$.  
For $|\mathcal D| < |\mathcal H|$, there exists an edge $(i_k,j_k)\in \mathcal H$ but $(i_k,j_k)\notin \mathcal D$. 
Then by (A), $\PP(\textnormal{Pval}= \textnormal{Pval}_k = 1)\to 1$.
For $|\mathcal D| = |\mathcal H|$, note that as $n,M\to\infty$,
\begin{equation*}
   \PP\Big(\textnormal{Pval}<\alpha\Big) = \PP\Big(\textnormal{Pval}_1<\alpha,\ldots,\textnormal{Pval}_{|\mathcal H|}<\alpha\Big)\leq
   \PP\Big(\textnormal{Pval}_1<\alpha\Big)\to \alpha.
\end{equation*}
Now, define a sequence $\{\bs U_{\mathcal H}^{(r)}\}_{r\geq 1}$ such that 
$\bs U_{(i_1,j_1)}^{(r)}= 0$ and $\min_{2\leq k\leq |\mathcal H|} |\bs U_{(i_k,j_k)}^{(r)}|\geq c > 0$. 
Thus, $\{\bs U_{\mathcal H}^{(r)}\}_{r\geq 1}$ satisfy $H_0$.
By Proposition \ref{proposition:power-edge}, $\text{Pval}_k\pto 0$ for $k\geq 2$ as $r\to\infty$ .
Hence, 
\begin{equation*}
  \limsup_{
  \substack{n\to\infty\\
  \bs U_{\mathcal H}^{(r)} \text{ satisfies H}_0
  } }\PP_{\bm \theta^{(r)}}\Big(\text{Pval}<\alpha\Big) \geq 
  \sup_{r} \lim_{n\to\infty} \PP_{\bm \theta^{(r)}}\Big(\text{Pval}<\alpha\Big) = \alpha.
\end{equation*}
This completes the proof.

\subsection{Proof of Proposition \ref{proposition:asymptotic}} \label{proof:asymptotic}

Since $\PP(\widehat{\mathcal S}=\mathcal S)\to 1$, it suffices to consider $\text{Lr}(\mathcal S,\widehat{\bm\Sigma})$.
For $|\mathcal D|=0$, we have $\PP(\text{Lr} = 0)\to 1$. 
Now, assume $|\mathcal D|>0$. Then 
\begin{equation*}
  \begin{split}
      2 \text{Lr}(\mathcal S,\widehat{\bm\Sigma}) 
      = \underbrace{\sum_{\{j: \df_{\mathcal S}(j)\neq\emptyset\} }
      \frac{\bs{ e}_j^\top( \bs{ P}_{A_j} 
      - \bs{ P}_{B_j})\bs{ e}_j}{\sigma^2_j} }_{R_1 }
      + \underbrace{ \sum_{\{j: \df_{\mathcal S}(j)\neq \emptyset\}}
      \left(  \frac{\sigma^2_j}{\widehat{\sigma}^2_j} - 1\right)  \frac{\bs{ e}_j^\top( \bs{ P}_{A_j} 
      - \bs{ P}_{B_j})\bs{ e}_j}{\sigma_j^2} }_{R_2}.
  \end{split}
\end{equation*}
To derive the asymptotic distribution of $R_1$, we apply the strategy with 
law of iterated expectation as in the proof of Lemma \ref{lemma:inference}. 
Then we have $\Big\{\bs{ e}_j^\top( \bs{ P}_{A_j} 
- \bs{ P}_{B_j})\bs{ e}_j/\sigma^2_j\Big\}_{\{ j : \df_{\mathcal S}(j)\neq\emptyset \}}$ are independent.
Therefore, $R_1 \sim \chi^2_{|\mathcal D|}$.
To bound $R_2$, we apply Lemma 1 of \citet{laurent2000adaptive}. 
By Assumption 5, 
\begin{equation*}
    \begin{split}
        \PP\left(\max_{\{j:\df_{\mathcal S}(j)\neq\emptyset\}} \left|\frac{\widehat{\sigma}^2_j}{\sigma^2_j} - 1\right|
        \geq  4\sqrt{\frac{\log |\mathcal D|}{(1-\rho)n}} + 8\frac{\log |\mathcal D|}{(1-\rho)n} \right)
        &\leq 2\exp( -\log |\mathcal D| ).
    \end{split}
\end{equation*}
Hence, $\max\limits_{\{j:\df_{\mathcal S}(j)\neq \emptyset\}} \left|\frac{\sigma^2_j}{\widehat{\sigma}_j^2} - 1 \right| \leq 8 \sqrt{{\log(|\mathcal D|)}/{(1-\rho)n}}$ with probability tending one. Thus
\begin{equation*}
    |R_2| 
\leq |R_1|\max_{\{j:\df_{\mathcal S}(j)\neq \emptyset\}} \left| \frac{\sigma^2_j}{\widehat{\sigma}_j^2} - 1 \right| 
\leq O_{\PP}\left( |\mathcal D| \sqrt{\frac{\log |\mathcal D|}{n}}  \right).
\end{equation*}
Consequently, the desired result follows. 

\subsection{Proof of Proposition \ref{proposition:power-edge}} \label{proof:power-edge}

Let $\bm \theta^{(n)} = (\bs U^\circ+\bm \Delta,\bs W^\circ)$.
Then 
\begin{equation*}
    L(\bm\theta^{(n)},\bm\Sigma)-L(\bm\theta^\circ,\bm\Sigma)
    = \sum_{\{j:\df_{\mathcal S}(j)\neq\emptyset\}} \Big( \sqrt{n}\bm \eta_j^\top \bm\Delta_{\cdot j} - \frac{1}{2} \sqrt{n}\bm\Delta_{\cdot j} 
  \Big(n^{-1}\bs{ Y}^\top_{\df_{\mathcal S}(j)}\bs{ Y}_{\df_{\mathcal S}(j)}\Big) \sqrt{n}\bm\Delta_{\cdot j} \Big),\\
\end{equation*}
where $\bm \eta_j = (n^{-1/2}\bs{ Y}_{\df_{\mathcal S}(j)}^\top\bs{ e}_j,\bm 0_{|\df_{\mathcal S}(j)^c|})$ is a $p$-vector.
It suffices to consider 
$\textnormal{Lr}(\mathcal S,\widehat{\bm\Sigma})$, since $\PP(\widehat{\mathcal S}=\mathcal S)\to 1$. 
Under null hypothesis $H_0$, $2\text{Lr} = \sum_{\{j:\df_{\mathcal S}(j)\neq\emptyset\}} T_j$, where by Proposition \ref{proposition:asymptotic} we have 
$T_j = \sum_{r=1}^{|\df_{\mathcal S}(j)|}(\bs{ q}_{j,r}^\top\bs{ e}_j)^2 + o_{\PP}(1)$ with $\bs{ P}_{A_j}-\bs{ P}_{B_j} = \sum_{r=1}^{|\df_{\mathcal S}(j)|} \bs{ q}_{j,r}\bs{ q}_{j,r}^\top$.
Letting $\bs{ Q}_j = \big(\bs{ q}_{j,1},\ldots,\bs{ q}_{j,|\df_{\mathcal S}(j)|}\big)$ 
be an $n\times |\df_{\mathcal S}(j)|$ matrix, then 
\begin{equation*}
  \begin{pmatrix}
    \bs{ Q}_j^\top \bs{ e}_j\\
    \bm \eta_j
  \end{pmatrix} \ \vline \
  \bs{ Y}_{\pa_{\mathcal S}(j)},\bs{ X}
  \sim N\left( \bm 0,
  \begin{pmatrix}
    \bs I_{|\df_{\mathcal S}(j)|} & n^{-1/2} \bs{ Q}_j^\top \bs{ Y}_{\df_{\mathcal S}(j)} \\
    n^{-1/2} \bs{ Y}_{\df_{\mathcal S}(j)}^\top\bs{ Q}_j & n^{-1}\bs{ Y}^\top_{\df_{\mathcal S}(j)}\bs{ Y}_{\df_{\mathcal S}(j)}
  \end{pmatrix} \right),
\end{equation*}
\begin{equation*}
  \bs{ Q}^\top_j\bs{ e}_j \mid \bm \eta_j, \bs{ Y}_{\pa_{\mathcal S}(j)},\bs{ X}
  \sim N\Big(n^{-1/2}\bs{ Y}_{\df_{\mathcal S}(j)}^\top \bs{ Q}_j \bm\eta_j, 
  \bs{ Y}_{\df_{\mathcal S}(j)}^\top (\bs I_n - \bs{ Q}_j\bs{ Q}_j^\top) \bs{ Y}_{\df_{\mathcal S}(j)}\Big).
\end{equation*}
Next, let $Y_k$ be a leaf node of the graph $\mathcal G$.
For fixed $|\mathcal D|>0$, after change of measure,  
\begin{equation*}
  \begin{split}
   & \beta(\bm \theta^\circ,\bm \Delta) \\
  &\geq \liminf_{n\to\infty} \E_{\bm \theta^\circ}
  \left( \I\left( \sum_{j=1}^p \|\bs{ Q}^\top\bs{ e}_j\|^2_2  > \chi^2_{|\mathcal D|,1-\alpha}\right) \exp\Big(L(\bm\theta^{(n)},\bm\Sigma)-L(\bm\theta^\circ,\bm\Sigma)\Big)  \right)\\
  &= \liminf_{n\to\infty} \E_{\bm \theta^\circ} \E
  \left( \I\left( \sum_{j=1}^p \|\bs{ Q}^\top\bs{ e}_j\|^2_2  > \chi^2_{|\mathcal D|,1-\alpha}\right) \exp\Big(L(\bm\theta^{(n)},\bm\Sigma)-L(\bm\theta^\circ,\bm\Sigma)\Big) \ \vline\ \bs{ Y}_{-k},\bs{ X} \right)\\
  &= \liminf_{n\to\infty} \E \left( \PP\left( \|\bm Z_{k} + \bs{ Q}_k^\top\bs{ Y}_{\df_{\mathcal S}(k)} \bm\Delta_{\cdot k}\|_2^2 
  + \sum_{j\neq k}\|\bs{ Q}_j^\top\bs{ e}_j\|_2^2 >\chi^2_{|\mathcal D|,1-\alpha}\ \vline\ \bs{ Y}_{-k},\bs{ X} \right) \right),
  \end{split}
\end{equation*}
where $\bm Z_{k}\sim N(\bm 0,\bs I_{|\df_{\mathcal S}(k)|})$.
By Lemma 3 of \citet{li2020likelihood}, 
$\| \bs{ Q}_k^\top\bs{ Y}_{\df_{\mathcal S}(k)} \bm\Delta_{\cdot k}\|_2^2\geq c_3 \|\bm\Delta_{\cdot k}\|_2^2$ with probability tending to one.
Therefore, a peeling argument yields 
\begin{equation*}
  \beta(\bm \theta^\circ,\bm \Delta)\geq \PP\Big(\|\bm Z + c_3\bm\Delta \|_2^2 \geq \chi^2_{|\mathcal D|,1-\alpha}\Big),
\end{equation*}
where $\bm Z\sim N(\bm 0,\bs I_{|\mathcal D|})$.  Similarly, as $|\mathcal D|\to\infty$, 
\begin{equation*}
  \begin{split}
    \beta(\bm \theta^\circ,\bm H)
    &\geq \PP\Big(\|\bm Z + c_3\bm\Delta\|_2^2> \sqrt{2|\mathcal D|}z_{1-\alpha}+|\mathcal D|\Big)
    \to \PP\Big(Z > z_{1-\alpha} - c_3\|\bm\Delta\|_2^2/\sqrt{2|\mathcal D|} \Big).
  \end{split}
\end{equation*}
This completes the proof.

\subsection{Proof of Proposition \ref{proposition:power-path}} \label{proof:power-path}

By Proposition \ref{proposition:power-edge}, for fixed $|\mathcal D|=|\mathcal H|>0$, $\beta(\bm\theta^\circ,\bm \Delta)$
equals to 
\begin{equation*}
  \begin{split}
     \PP(\text{Pval}_1<\alpha,\ldots,\text{Pval}_{|\mathcal H|}<\alpha)
    \geq 1 - \sum_{k=1}^{|\mathcal H|} \PP(\text{Pval}_k\leq \alpha)
    \geq 1 - |\mathcal H|\PP\left( Z_1^2 \leq \chi^2_{1,1-\alpha} \right), 
  \end{split}
\end{equation*}
where $Z_1\sim N(\delta/\max\limits_{1\leq j\leq p}\Omega_{jj},1)$.
Then $\PP(\widetilde{Z}^2 \leq x) \leq \PP(\widetilde{Z} \leq -\mu +\sqrt{x})
\leq \frac{1}{\sqrt{2\pi}}\frac{e^{-(\mu-\sqrt{x})^2/2}}{\mu-\sqrt{x}}$ for 
$\widetilde{Z}\sim N(\mu,1)$ and any
$0 \leq x<\mu^2$ and $\mu>0$ \citep{small2010expansions}. 
Hence, as $|\mathcal D|=|\mathcal H|\to\infty$, 
\begin{equation*}
  \begin{split}
    \beta(\bm\theta^\circ,\bm \Delta) 
    &\geq 1-|\mathcal H|\PP\left( Z_1^2 \leq \chi^2_{1,1-\alpha} \right)\\
    &\geq 1-\frac{|\mathcal H|}{\sqrt{2\pi}} e^{-\big(\delta\sqrt{\log|\mathcal H|}/\max_{1\leq j\leq p}\Omega_{jj} - \sqrt{\chi^2_{1,1-\alpha}}\big)^2/2}
    \to 1.
  \end{split}
\end{equation*}
This completes the proof.

\subsection{Proof of Theorem \ref{theorem:relax-signal}} \label{proof:relax-signal}

The proof proceeds in two steps to show that \\
(A) $\{ l: |\mathrm V_{lj}|\geq\tau^*_j \}\subseteq \{l:\widehat{\mathrm V}_{lj}\neq 0 \} \subseteq \{l: \mathrm V_{lj} \neq 0\}$ for $1\leq j\leq p$ almost surely, and \\
(B) $\widehat{\mathcal G}_+ = \mathcal G_+$ when $\widehat{\bs V}$ satisfies the property in (A). 

\subsubsection*{Proof of (A)}

This proof is similar to that of Theorem \ref{theorem:consistent-structure-learning} part (A).
To proceed, let $A_j^\circ = \Big\{ l: \mathrm V_{lj}\neq 0 \Big\}$, $A^*_j = \Big\{ l: |\mathrm V_{lj}| > \tau_j^* \Big\}$,
and $A^{[t]}_j = \Big\{ l : |\widetilde{\mathrm V}^{[t]}_{lj}| > \tau_j \Big\}$. 
Then we define the false negative set 
 $\text{FN}^{[t]}_j = A^*_j\setminus A_j^{[t]}$
and the false positive set 
$\text{FP}^{[t]}_j = A_j^{[t]}\setminus A_j^\circ$ for $t\geq 0$.  
Consider 
\begin{equation*}
    \mathscr{E}_j =\Big\{ \|\bs{ X}^\top\widehat{\bm \xi}_j/n\|_\infty\leq 0.5\gamma_j\tau_j \Big\}
\cap \Big\{ \|\widehat{\bs V}^\circ_{\cdot j} - \bs V_{\cdot j}\|_\infty \leq 0.5\tau_j \Big\},
\end{equation*}
where $\widehat{\bm\xi}_j = \bs{ Y}_j - \bs{ X}\widehat{\bs V}^\circ_{\cdot j}$ for $1\leq j \leq p$. 
Again, we shall show that $\text{FN}^{[t]}_j$ and $\text{FP}^{[t]}_j$ are eventually empty
on event $\mathscr{E}_j$ which has a probability tending to one.

Assume $|A_j^\circ\cup A^{[t-1]}_j|\leq 2\kappa^\circ_{\max}$.
Then \eqref{equation:basic-inequality}, \eqref{equation:l1-inequality}, and \eqref{equation:error-bound} remain true on $\mathscr{E}_j$. As a result, $|A_j^\circ \cup A^{[t]}_j|
\leq 2\kappa^\circ_{\max}$ on $\mathscr E_j$ for $t\geq 1$.

To estimate the number of iterations required for termination, it suffices to prove that $|\text{FN}^{[t]}_j| + |\text{FP}^{[t]}_j| < 1$ eventually.
Suppose $|\text{FN}^{[t]}_j|+|\text{FP}_j^{[t]}|\geq 1$. For any $l\in \text{FN}^{[t]}_j\cup\text{FP}_j^{[t]}$, 
by Assumption \ref{assumption:relax-signal}, 
\begin{equation*}
    |\widetilde{\mathrm V}^{[t]}_{lj} - \widehat{\mathrm V}^\circ_{lj}|
  \geq |\widetilde{\mathrm V}^{[t]}_{lj} - V_{lj}|-|\widehat{\mathrm V}^\circ_{lj}-\mathrm V_{lj}|
  \geq \min(\tau_j^* - \tau_j, \tau_j),
\end{equation*}
so $\sqrt{|\text{FN}^{[t]}_j| + |\text{FP}^{[t]}_j|} \leq \|\widetilde{\bs V}_{\cdot j}^{[t]}
-\widehat{\bs V}_{\cdot j}^\circ\|_2 /\min(\tau_j^* - \tau_j, \tau_j)$. 
Letting $\kappa_j^*=|A_j^*|$, by \eqref{equation:error-bound} and the Cauchy-Schwarz inequality, 
\begin{equation*}
    c_1\|\widetilde{\bs V}_{\cdot j}^{[t]} - \widehat{\bs V}_{\cdot j}^\circ\|_2 \leq 
1.5\gamma_j \tau_j \sqrt{|A^\circ_j\triangle A^{[t-1]}_j|} \leq 1.5\gamma_j \tau_j \sqrt{|\text{FN}^{[t-1]}_j| + |\text{FP}^{[t-1]}_j| + (\kappa^\circ_j-\kappa^*_j)}.
\end{equation*}
By the conditions for $(\tau_j,\gamma_j)$ in Theorem \ref{theorem:relax-signal}, we have 
\begin{equation*}
    \begin{split}
        \sqrt{|\text{FN}^{[t]}_j| + |\text{FP}^{[t]}_j|} 
        &\leq \frac{\|\widetilde{\bs V}_{\cdot j}^{[t]}-\widehat{\bs V}_{\cdot j}^\circ\|_2}
  {\min(\tau^*_j-\tau_j,\tau_j)}\\
  &\leq \frac{1.5\gamma_j\tau_j}
  {c_1 \min(\tau^*_j-\tau_j,\tau_j)}\Big(
  \sqrt{|\text{FN}_j^{[t-1]}| + |\text{FP}_j^{[t-1]}|} + \sqrt{\kappa^\circ_j-\kappa^*_j}\Big)\\
  &\leq 0.5 \sqrt{|\text{FN}_j^{[t-1]}| + |\text{FP}_j^{[t-1]}|} + 0.25.
    \end{split}
\end{equation*}
Hence, $\sqrt{|\text{FN}^{[t]}_j| + |\text{FP}^{[t]}_j| }\leq (1/2)^t\sqrt{|A^\circ_j| + |A^{[0]}_j|} + 0.5$.
For $t \geq 1 + \lceil\log \kappa^\circ_j /\log 4\rceil$, we have 
$|\text{FN}_j^{[t]}| + |\text{FP}^{[t]}_j| < 1$ implying that $\text{FN}_j^{[t]} = \emptyset$
and $\text{FP}_j^{[t]} = \emptyset$.

Under the conditions for $(\tau_j,\gamma_j)$ in Theorem \ref{theorem:relax-signal}, using the same argument in Proof of Theorem \ref{theorem:consistent-structure-learning} we have 
\begin{equation*}
  \begin{split}
   \PP\Big( \Big\{l: |\mathrm V_{lj}| > \tau_j^*  \Big\} \subseteq \Big\{l :|\widetilde{\mathrm V}_{l j}| > \tau_j \Big\} \subseteq \Big\{l: \mathrm V_{lj} \neq 0 \Big\};\ 1\leq j\leq p \Big) 
\geq 1 - \sqrt{\frac{2}{\pi}} p q^{-3} n^{-4}.
  \end{split}
\end{equation*}
By Borel-Cantelli lemma, almost surely we have 
\begin{equation*}
    \Big\{l: |\mathrm V_{lj}| > \tau_j^*  \Big\} \subseteq \Big\{l :|\widetilde{\mathrm V}_{l j}| > \tau_j \Big\} \subseteq \Big\{l: \mathrm V_{lj} \neq 0 \Big\}, \quad j=1,\ldots,p,
\end{equation*}
when $n$ is sufficiently large.
Thus, we have $\{l:\widehat{\mathrm V}_{lj} \neq 0 \} = \{l:|\widetilde{\mathrm V}_{lj} |>\tau_j \}$ whenever $\kappa_j=|\{ l: \widetilde{\mathrm V}_{lj} >\tau_j \}|$; $j=1,\ldots,p$.
The desired result follows.

\subsubsection*{Proof of (B)}
This is the same as the Proof of Theorem \ref{theorem:consistent-structure-learning} part (B).

\subsection{Proof of Theorem \ref{theorem:consistent-structure-learning-U}} \label{proof:consistent-structure-learning-U}

By Theorem \ref{theorem:consistent-structure-learning}, it suffices to consider the event $\{\widehat{\mathcal G}_+=\mathcal G_+\}$.
Then with $\mathbf X$ being replaced by $(\mathbf Y_{\cdot,\an_{{\mathcal G}_+}(j)},\mathbf X_{\cdot, \inter_{{\mathcal G}_+}(j)})$, this proof is almost the same as that of Theorem \ref{theorem:consistent-structure-learning} part (A) and thus is omitted.

\section{Supplements for Simulations} \label{section:supplement-simulation}

\subsection{Implementation Details}\label{section:implementation}

DP-LR and LR: 
\begin{itemize}
  \item For Algorithm \ref{algorithm:data-perturbation}, we fix the Monte Carlo sample size $M = 500$. Our limited numerical experience suggests that this choice appears 
adequate for our purpose.

  \item For Algorithms \ref{algorithm:nodewise-regression} and \ref{algorithm:peeling},
  we choose $\tau_j \in \{ 0.05,0.1,0.15 \}$ and  $\gamma_j = \exp(\gamma_j')$ with 
  \begin{equation*}
    \gamma_j' \in \Big\{ \log(\max_{l,j}|\bs{{X}}_{\cdot l}^\top\bs{{Y}}_{\cdot j}|),\ldots,0.05\log(\max_{l,j}|\bs{{X}}_{\cdot l}^\top\bs{{Y}}_{\cdot j}|) \Big \}
    \quad \text{(100 equally spaced values)}.
  \end{equation*}
  Then BIC is used to estimate tuning parameters $\kappa_j\in\{1,\ldots,30\}$; $j=1,\ldots,p$. 
\end{itemize}
2SPLS \citep{chen2018two}: 
\begin{itemize}
  \item We use the R package \texttt{BigSEM} for 2SPLS. 
  {In our experiments, the $\lambda$ sequence of 2SPLS \citep{chen2018two} is set in the same way as the $\gamma$ sequence of the proposed methods. We use the default settings for other parameters.}
\end{itemize}

\subsection{Additional Simulations} \label{section:additional-simulation}

This section supplements Section \ref{section:structure-learning-simulation} by including additional numerical experiments. 

\subsubsection{Random graphs with different sparsity}

\begin{figure}[H]
  \centering
  \includegraphics[width=0.55\textwidth]{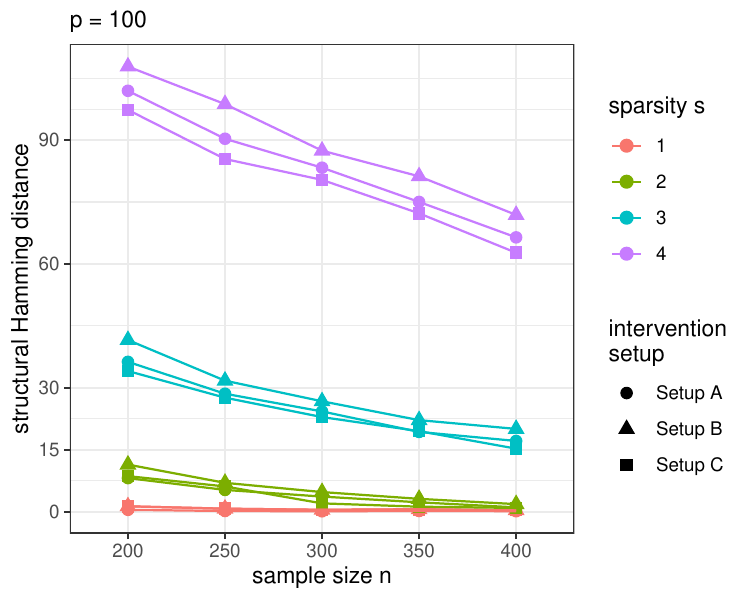}
  \caption{SHDs for the reconstructed DAG by the peeling algorithm.}
  \label{figure:structure-learning-s}
\end{figure}

We examine the proposed method for structure learning of DAGs with different sparsity.
For $\bs U$, the upper off-diagonal entries $\mathrm U_{kj}$; $k<j$ are sampled independently from $\{0,1\}$ according to Bernoulli$(s/p)$; $s=1,2,3,4$, while other entries are zero. The rest of the settings remain the same as in Section \ref{section:structure-learning-simulation}. 
Figure \ref{figure:structure-learning-s} displays the results.
As expected, the performance improves when the sample size $n$ increases and the DAG becomes sparser (controlled by $s$). 

\subsubsection{SHD transition curves of structure learning}

\begin{figure}[H]
  \centering
  \includegraphics[width=.9\textwidth]{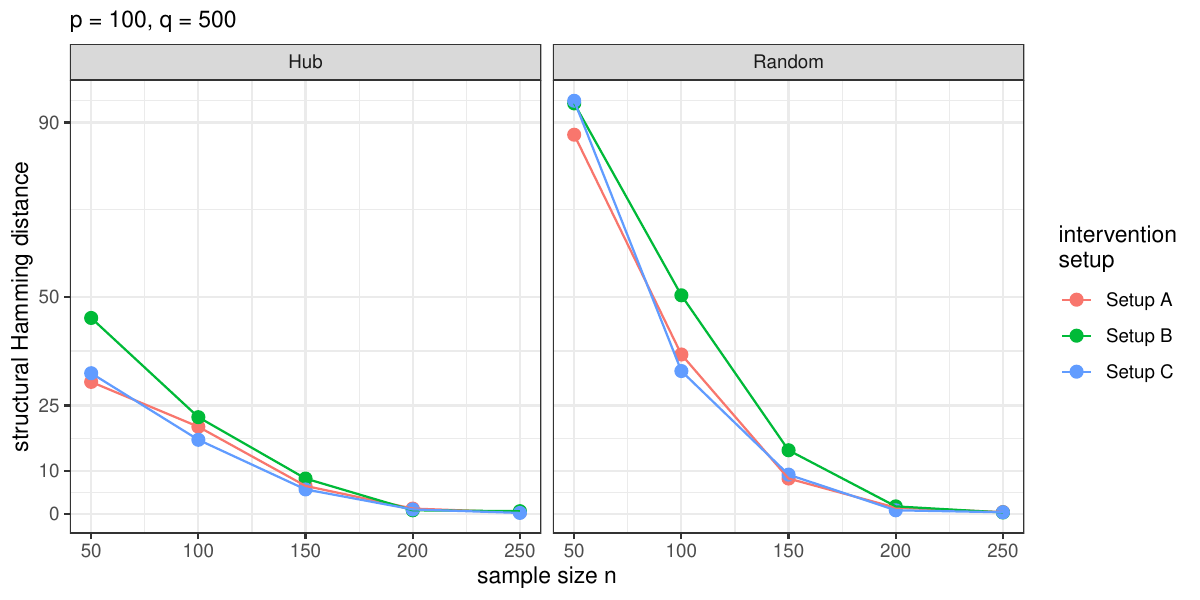}
  \caption{SHDs for the reconstructed DAG by the peeling algorithm. The experiment settings are the same as the ones in Section \ref{section:structure-learning-simulation}.}
  \label{figure:structure-learning-transition-curve}
\end{figure}

We consider different sample sizes $n = 50, 100, 150, 200$ to further examine how the proposed method depends on $n$. Figure \ref{figure:structure-learning-transition-curve} displays the SHD transition curves of the peeling algorithm.

\subsubsection{Structure learning with different numbers of interventions}

\begin{figure}[H]
  \centering
  \includegraphics[width=.9\textwidth]{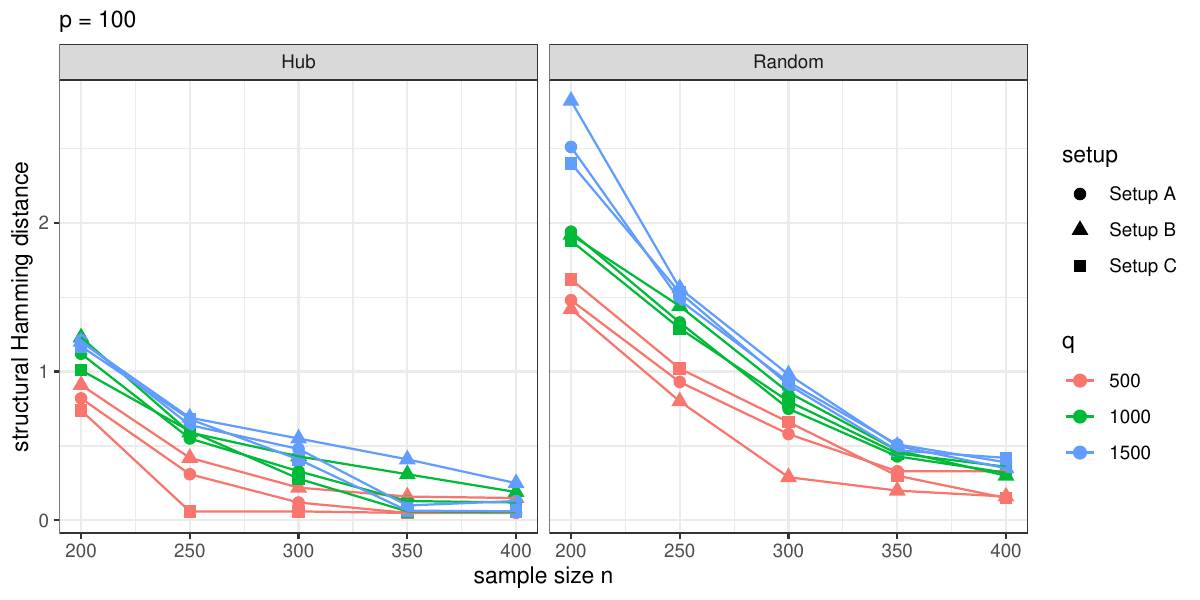}
  \caption{SHDs for the reconstructed DAG by the peeling algorithm. The experiment settings are the same as the ones in Section \ref{section:structure-learning-simulation}.}
  \label{figure:structure-learning-q}
\end{figure}

Finally, we investigate how the number of interventions $q$ affects the learning outcomes.
Figure \ref{figure:structure-learning-q} displays the results when $q = 500, 1000, 1500$. It suggests that the proposed method performs reasonably well at a moderate sample size when many unknown interventions are present.

\vskip 0.2in
\bibliography{ref}

\end{document}